\documentclass[journal ]{new-aiaa}
\usepackage[utf8]{inputenc}
\usepackage{textcomp}

\usepackage{graphicx}
\usepackage{amsmath,empheq}
\usepackage{amsfonts}
\usepackage[version=4]{mhchem}
\usepackage{siunitx}
\usepackage{longtable,tabularx}
\usepackage{subcaption}
\usepackage{multirow}
\DeclareRobustCommand{\rev1}[1]{\textcolor{black}{#1}}
\title{On-Manifold Low-Thrust \rev1{Rephasing} of Quasi-Periodic Orbits}

\author{Ian M. Down\footnote{Research Engineer, Department of Aerospace Engineering; idown@tamu.edu (Corresponding Author)} and Manoranjan Majji\footnote{Professor, Director of the Land, Air, and Space Robotics Laboratory, Department of Aerospace Engineering, AIAA Associate Fellow}}
\affil{Texas A\&M University, College Station, Texas, 77843}
\author{Kathleen C. Howell\footnote{Hsu Lo Distinguished Professor, School of Aeronautics and Astronautics, AIAA Fellow}}
\affil{Purdue University, West Lafayette, Indiana, 47907}

\begin{document}

\maketitle

\begingroup
\renewcommand\thefootnote{}\footnotetext{\rev1{Originally presented at the 2025 AAS/AIAA Spaceflight Mechanics Conference in Kauai, Hawaii on January 21$^\text{st}$ (25-451).}}
\endgroup

\begin{abstract}
A bi-level optimal control framework is introduced to solve the low-thrust rephasing problem on quasi-periodic invariant tori in multi-body environments where deviations away from the torus during maneuver are considered unsafe or irresponsible. It is shown for a large class of mechanical systems that conformity to the torus manifold during periods of non-zero control input is infeasible. The most feasible trajectories on the torus surface are generated through the minimization of fictitious control input in the torus space using phase space control variables mapped via the torus function. These reference trajectories are then transitioned to the phase space both through a minimum tracking error homotopy and minimum time patched solutions. Results are compared to torus agnostic low-thrust transfers using measures of fuel consumption, cumulative torus error, and coast time spent on the torus during maneuver. Modifications to the framework are made for the inclusion of quasi-periodically forced dynamical systems. Lastly, minimum time recovery trajectories with free final torus conditions expose the disparity between the proposed framework and torus agnostic approaches. Examples are drawn from the circular and elliptical restricted three-body problems.
\end{abstract}

\section*{Nomenclature}

{\renewcommand\arraystretch{1.0}
\noindent\begin{longtable*}{@{}l @{\quad=\quad} l@{}}

% Normal Letters
$\boldsymbol{C}$ & Fourier coefficient matrix \\
$\boldsymbol{D}$ & discrete Fourier transform matrix \\
$\boldsymbol{e}$ & instantaneous coordinate torus error vector \\
$E$ & cumulative torus error \\
$\boldsymbol{f}$ & state dynamics \\
$\boldsymbol{F}$ & boundary value problem constraint vector \\
$\boldsymbol{g}$ & forcing phase dynamics \\
$\boldsymbol{h}$ & torus dynamics \\
$\mathcal{H}$ & Hamiltonian \\
$J$ & cost function \\
$m$ & forcing frequency dimension \\
$n$ & phase space dimension \\
$p$ & torus dimension \\
$\boldsymbol{q}$ & coordinate component of state vector, \rev1{DU} \\
$\dot{\boldsymbol{q}}$ & coordinate rate component of state vector, \rev1{DU/TU} \\
$s$ & switching function \\
$t$ & time, \rev1{TU} \\
$\boldsymbol{u}_q$ & continuous velocity control vector, \rev1{DU/TU} \\
$\boldsymbol{u}_{\dot{q}}$ & continuous acceleration control vector, \rev1{DU/TU$^2$} \\
$\boldsymbol{x}$ & state vector \\
$\boldsymbol{X}$ & Fourier dataset matrix \\

% Greek Letters
$\boldsymbol{\gamma}$ & coordinate map component of torus function, \rev1{DU} \\
$\dot{\boldsymbol{\gamma}}$ & coordinate rate map component of torus function, \rev1{DU/TU} \\
$\boldsymbol{\Gamma}$ & torus function \\
$\Delta V$ & propellant use measure, \rev1{DU/TU} \\
$\epsilon$ & torus approximation error \\
$\varepsilon$ & tracking error homotopic parameter \\
$\boldsymbol{\theta}$ & torus angles, \rev1{rad} \\
$\boldsymbol{\vartheta}$ & forcing phase angles, \rev1{rad} \\
$\lambda$ & costate \\
$\boldsymbol{\mu}$ & system parameter set \\
$\boldsymbol{\nu}$ & terminal constraint multiplier \\
$\rho$ & throttle smoothing parameter \\
$\phi$ & terminal penalty function \\
$\boldsymbol{\chi}$ & boundary value problem free variable vector \\
$\boldsymbol{\psi}$ & terminal constraint vector \\
$\boldsymbol{\omega}$ & fundamental torus frequency vector, \rev1{rad/TU}\\
$\boldsymbol{\Omega}$ & torus space control frequency vector, \rev1{rad/TU} \\

\end{longtable*}}

\section{Introduction} \label{sec:intro}

\lettrine{Q}{uasi-periodic orbits} (QPOs) and the quasi-periodic invariant tori (QPIT) they reside on are expanding the trajectory design space for missions in cislunar space and other multi-body orbital regimes, particularly near libration points. Recent advancements in trajectory design methodologies have been made to efficiently incorporate these higher-dimensional structures into the preliminary design process \cite{mccarthy2021leveraging,henry2023quasi,henry2024fully}. As bounded structures in the solution space with multiple phasing angles, QPITs are attractive structures on which to place spacecraft in constellation or formation to enable scientific observation or surveillance capabilities \cite{barden1998fundamental,duering2012uncontrolled,baresi2017spacecraft}. Additionally, many in-plane three-body periodic orbits in the circular problem are surrounded by QPITs with additional out-of-plane motion that allow for better observational geometry of the space \cite{down2024cislunar}. In the conventional sense, periodic orbits in simplified multi-body models like the circular restricted three-body problem (CR3BP) are used as primary design structures. They are then transitioned to QPOs in intermediate models before transitioning to ephemeris, where near-quasi-periodic trajectories are achieved for a given epoch and time horizon through optimization \cite{park2024assessment}. Ideally, the end result is a trajectory that shares both geometric and stability characteristics of the originally selected periodic orbit \cite{zimovan2023baseline}. With their foreseen potential in formation and constellation mission concepts, QPITs are now being examined as primary design structures in simplified models before individual QPO transition to ephemeris \cite{baresi2023transition}.

A key feature of resilient multi-spacecraft systems is the ability for reconfiguration or rephasing. In other words, depart and return to the same primary design structure at an epoch and corresponding phase that would not otherwise be possible without control input. Whether an entire constellation needs to be shifted because of coverage loss from a damaged sensor, or a single spacecraft needs to maneuver for favorable observation conditions of a high-priority target, reconfiguration represents an agile, enabling technology that can ultimately lead to the robust satisfaction of high-level mission requirements. One primary example of rephasing spacecraft for changing observational needs/priorities is seen in the station-relocation of geostationary satellites \cite{zhao2016initial,wu2023atlas}. Maneuvers like these are made compactly possible through the employment of low-thrust engines. This method of propulsion has also been shown to supply the necessary control input for multi-body transfers and station-keeping for a variety of spacecraft classes \cite{mingotti2009low,bosanac2018trajectory}.

For spacecraft dominated by only one gravitational force, reconfiguration is a relatively safe procedure. Low-thrust maneuvers cannot dramatically change a trajectory's geometry in a short amount of time because the solution space is dense with neutrally stable periodic orbits. For trajectories in cislunar space and other multi-body regimes, particularly those around libration points, the same cannot be said. The dynamics in these regions of space are chaotic in nature: a small trajectory perturbation can have large implications on near future geometries. On one hand, high sensitivities are very beneficial. A small usage of propellant can lead to enormous trajectory changes if particular directions are excited, linking hundreds of thousands of kilometers of space almost for free \cite{koon2000heteroclinic}. On the other hand, this requires a full understanding of the dynamics so that transfers and station-keeping maneuvers may be executed with sufficient accuracy to ensure an eventually safe arrival at the intended destination. For reconfiguration on the surfaces of QPITs in multi-body environments, it is thus important to consider deviations away from the design reference torus in maneuver design.

Designing low-thrust reconfiguration maneuvers that remain close to the QPIT in the phase space represent safe trajectories that implicitly avoid the direct excitation of unstable directions. This is potentially advantageous for several reasons. First, 
avoiding the excitation of unstable directions is beneficial in the event of a spacecraft system anomaly mid-maneuver. Strictly minimum energy/fuel maneuvers take advantage of nearby hyperbolic structures to minimize propellant usage. If a maneuver of this kind is aborted in between the desired boundary conditions, departure from the QPIT may occur very quickly. Planning maneuvers close to the QPIT can increase the time until departure, thereby increasing time available for system troubleshooting and odds of recovery. Second, on a similar note, proximity to the QPIT during reconfiguration determines how quickly a spacecraft can immediately return to natural motion on the primary design structure, even if it has not reached the desired boundary conditions. This could be an important consideration for space surveillance systems. When a spacecraft is maneuvering, its attitude is constrained to the demands of the orbital guidance system, and observational targets cannot be tracked. For shifting mission priorities mid-reconfiguration, the system may require an immediate return to nominal observation under natural dynamics on the QPIT. With a reconfiguration maneuver that minimizes the distance away from the QPIT, this return to operation can occur in a minimum amount of time with minimum propellant expenditure than is otherwise possible. Lastly, staying on/near the QPIT can lead to smaller spacecraft state uncertainty growth during maneuver because unstable directions are largely avoided. This is important for space systems that track and estimate the states of other objects in space. Of course, these considerations all come at the cost of excess propellant use compared to strictly minimum propellant transfers. The concept of safety through maneuver planning on the surface of QPITs was recently approached by Takubo \textit{et al}. \cite{takubo2025safe}. There, the authors used local quasi-periodic structures around periodic orbits in restricted multi-body problems to design fuel-optimal relative transfers that guarantee passive safety. However, this method is only applicable at distances away from periodic orbits where the first-order variational space well describes the local motion, which is extremely limiting \cite{down2023relative}.

This paper introduces a novel optimal control framework that minimizes phase space trajectory deviation from a QPIT through its torus function for safe and agile spacecraft rephasing \cite{down2025quasi,downdissertation}. Through a modified invariance condition, it is shown that zero manifold deviation during any maneuver is infeasible for a large class of mechanical systems. Instead, several optimal control problems are solved in the torus space that minimize a fictitious velocity control input, thereby resulting in the most realizable trajectory when considering phase space limitations (i.e. only acceleration level control input). At the torus level, minimum velocity energy and minimum velocity throttle problems are solved to generate reference trajectories. The minimum throttle problem determines a series of coast arcs where it is favorable to obey the natural torus dynamics. Minimum energy solutions are transitioned to the phase space by a tracking error homotopy, while minimum throttle solutions are transitioned by sequential minimum time patches centered around the torus space switching times. Several boundary conditions related to mission scenarios are discussed and numerically examined. Modifications to the framework for tori in quasi-periodically forced dynamical systems \rev1{are discussed and an example in the periodically forced elliptical restricted three body problem (ER3BP) is provided.} Finally, minimum time torus recovery solutions are shown. Numerical examples are provided for 2-dimensional tori in the CR3BP and ER3BP. However, all methods introduced generalize to $p$-dimensional tori in quasi-periodically forced systems. Furthermore, the methodology developed is globally applicable to all tori in the phase space. Quantitative performance measures of $\Delta V$ usage, total torus deviation, and time spent on the torus are used to score the proposed trajectory design methods compared to traditional solutions to the minimum fuel trajectory optimization problem. 

The remainder of the paper is organized as follows. Section \ref{sec:qpit} introduces the necessary mathematical background and numerical approximation of QPITs. Section \ref{sec:modinv} then develops a modified torus invariance condition to establish connections between control in both the torus and phase spaces. Next, Sec. \ref{sec:bilevel} uses this invariance condition as the foundation on which to build the bi-level optimal control framework aimed at generating rephasing trajectories that minimize deviation from a torus in autonomous systems. Section \ref{sec:mods} discusses modifications to the framework in order to accommodate tori within quasi-periodically forced dynamical systems. Section \ref{sec:recovery} poses the minimum time to recovery problem using the torus as a free terminal manifold. And lastly, Sec. \ref{sec:conclusion} discusses limitations and alternative directions this methodology can take in the trajectory design process, and concludes with final remarks.

% David says write minimum-time instead of minimum time
\section{Quasi-Periodic Invariant Tori} \label{sec:qpit}

Quasi-periodic invariant tori are known to exist in simplified models of the three and four-body problem \cite{olikara2012numerical,mccarthy2023four}. They can be found in the vicinity of lower dimensional tori like fixed points (0-dimensional tori) and periodic orbits (1-dimensional tori) that have a complex center subspace. They can also be lifted from lower dimensional tori as forcing frequencies are added to a dynamical system to increase model fidelity. 

Let a given $n$-dimensional nonlinear, quasi-periodically forced Hamiltonian dynamical system in the absence of control be described by its dynamics recast into autonomous phase space form as
\begin{align}
    \dot{\boldsymbol{x}} &= \boldsymbol{f}(\boldsymbol{x},\boldsymbol{\vartheta},\boldsymbol{\mu}),\quad  \boldsymbol{x}\in \mathbb{R}^n,\quad\boldsymbol{\mu}\in\mathbb{R}^r \label{eq:firstdynamics}\\
    \dot{\boldsymbol{\vartheta}} &= \boldsymbol{g}(\boldsymbol{\vartheta},\boldsymbol{\mu}),\quad  \boldsymbol{\vartheta}\in \mathbb{T}_m \nonumber
\end{align}
where $\boldsymbol{\mu}$ are a set of constant system parameters, $\boldsymbol{\vartheta}$ are the set of independent forcing phase angles, \rev1{and $\boldsymbol{g}$ is a smooth function that is conjugate to a set of constant, independent frequencies \cite{arnol2013mathematical}}. Special cases of this class of dynamical systems include autonomous systems ($m=0$) like the CR3BP, and periodically forced systems ($m=1$) like the ER3BP. Within the flow of these systems may exist a variety of different $p$-dimensional invariant tori $\mathbb{T}_p$, where $\mathbb{T}_p=\mathbb{S}^1\times\cdots\times\mathbb{S}^1$, and $\mathbb{S}^1$ is the circle. \rev1{These structures have an associated $\rev1{m\leq }\,p\leq\bar{n}+m$ independent, fundamental frequencies of motion $\boldsymbol{\omega}$ and thus angles $\boldsymbol{\theta}$, where $\bar{n}=n/2$ \cite{broer1996quasi}. Non-isolated invariant tori typically lie in the center manifold of a lower-dimensional invariant torus in astrodynamics systems.} A torus is classified as a QPIT when two or more of its frequencies are incommensurate, or non-resonant, and the torus is invariant under the dynamical system's flow. A QPO is simply a trajectory \rev1{bounded} to the surface of a QPIT. For $m>0$, $\boldsymbol{\vartheta}$ must be a subset of $\boldsymbol{\theta}$ for the invariant torus to exist. Equivalently, the torus' fundamental frequency set must contain frequencies that are resonant with all forcing frequencies found in the system\footnote{An invariant torus' fundamental frequency set is broken down into internal and external frequencies. External torus frequencies are resonant with the dynamical system's forcing frequencies (if present), while internal torus frequencies are not. If the system is autonomous, all torus frequencies are internal.}. For an arbitrary set of initial angles $\boldsymbol{\theta}(t_0)$ on a QPIT, a first return would take infinite time such that $\boldsymbol{\theta}(t_0)=\boldsymbol{\theta}(t_0+\infty)$. The resulting trajectory would densely cover the torus' surface in the process.

An invariant torus embedded in a dynamical system is described by its torus map or function $\boldsymbol{\Gamma}(\boldsymbol{\theta})$, which maps $p$ torus angles into the $n$-dimensional phase space.
\begin{equation}
    \boldsymbol{\theta} = 
    \begin{bmatrix}
        \theta_1 & \cdots & \theta_p
    \end{bmatrix}^\text{T}\in\mathbb{T}_p,\quad \theta_i\in[0,2\pi)
\end{equation}
This map is unique when fully parametrized, and is in general sufficiently smooth and bijective such that $\boldsymbol{\Gamma}^{-1}$ exists and is also sufficiently smooth. In other words, $\boldsymbol{\Gamma}$ is a diffeomorphism. 
\begin{equation}
    \boldsymbol{\Gamma} : \mathbb{T}_p \rightarrow \mathbb{R}^n
\end{equation}
For the torus to be invariant within the flow of a dynamical system, it must obey the field invariance condition. This condition equates the vector fields of both the torus and the dynamical system through the torus function. For this, the time derivative of the torus angles $\dot{\boldsymbol{\theta}}=\boldsymbol{h}(\boldsymbol{\theta})$ associated with the map must be known. For the subset of $\boldsymbol{\theta}$ corresponding to $\boldsymbol{\vartheta}$, this is explicitly defined in the dynamics of Eq. \ref{eq:firstdynamics}. For any remaining torus angles, a choice of $\dot{\theta}_i$ represents a choice of torus parametrization. Though it is not necessary, $\dot{\theta}_i=\omega_i$ is often selected as the constant fundamental torus frequency such that the corresponding torus function is the fundamental torus function. 
\begin{equation}
    \dot{\theta}_i = \omega_i \implies \theta_i(t) = \omega_i(t - t_0) + \theta(t_0) \label{eq:linearFrequency}
\end{equation}
This choice alone does not fully parametrize the torus function for tori with $p>m$ because initial torus phases $\theta_i(t_0)$ must also be chosen relative to some state on the torus in the phase space for the remaining $p-m$ torus angles. The field invariance condition can then be stated as
\begin{equation}
    \frac{\text{d}}{\text{d}t}\left\{\boldsymbol{\Gamma}(\boldsymbol{\theta})\right\}=\left[\frac{\partial \boldsymbol{\Gamma}(\boldsymbol{\theta})}{\partial \boldsymbol{\theta}}\right]\dot{\boldsymbol{\theta}}=
    \left[\frac{\partial \boldsymbol{\Gamma}(\boldsymbol{\theta})}{\partial \boldsymbol{\theta}}\right]\boldsymbol{h}(\boldsymbol{\theta}) = \boldsymbol{f}(\boldsymbol{\Gamma}(\boldsymbol{\theta}),\boldsymbol{\vartheta},\boldsymbol{\mu
    }) \label{eq:invar1}
\end{equation}
Equation \ref{eq:invar1} represents flow invariance over the vector field as a set of partial differential equations. Invariance may also be expressed within the context of a stroboscopic map through untwisting an invariant curve. This reduction is useful for the computation of these structures \cite{baresiSUMMARY}.

\subsection{Torus Function Approximation}
All methods for the numerical computation of invariant tori involve some flavor of discretization scheme. After computation, it is advantageous to create a continuous representation of the structure that can be evaluated at any angle set $\boldsymbol{\theta}$. In other words, an approximation of the torus function $\boldsymbol{\Gamma}(\boldsymbol{\theta})$ is sought. This is done using a multivariate Fourier series basis through the discrete Fourier transform with uniform sampling, shown below for a 2-dimensional torus -- though this generalizes to $p$-dimensional tori.

Let $N_i\in\mathbb{Z}^+$ be the number of discrete points on the torus corresponding to the \rev1{torus} angle $\theta_i$. Then the uniform angle set is defined as
\rev1{\begin{equation}
    \boldsymbol{\theta}_i =
    \begin{bmatrix}
        \theta_{i,1}& \cdots & \theta_{i,N_i}
    \end{bmatrix}\in\mathbb{R}^{1\times N_i},\quad
    \theta_{i,j} = \frac{2\pi(j-1)}{N_i}\label{eq:anglevecforfit} 
\end{equation}}
\noindent From a chosen computational method, states corresponding to each angle set $\left\{\theta_{1,i},\theta_{2,j} \right\}$ are collected and concatenated according to Eq. \ref{eq:datacon}.
\begin{equation}
    \boldsymbol{X} =
    \begin{bmatrix}
        \boldsymbol{x}_{1,1} & \cdots & \boldsymbol{x}_{1,N_2} & \boldsymbol{x}_{2,1} & \cdots & \boldsymbol{x}_{N_1,N_2}
    \end{bmatrix}\in\mathbb{R}^{n\times N_1N_2}
    \label{eq:datacon}
\end{equation}
The Fourier coefficients $\boldsymbol{C}$ are then computed in matrix form by
\begin{align}
    \boldsymbol{C} &= \left(\boldsymbol{D}_1\otimes\boldsymbol{D}_2\right)\cdot\boldsymbol{X}^\text{T},\quad \boldsymbol{D}_i=\frac{1}{N_i}\text{e}^{-\text{i}\cdot\boldsymbol{k}_i^\text{T}\boldsymbol{\theta}_i} \\
    \boldsymbol{k}_i &= \begin{bmatrix}
        -\frac{N_i-1}{2} & \cdots & -1 & 0 & 1 & \cdots & \frac{N_i-1}{2}
    \end{bmatrix} \quad \text{for} \,\, \text{odd} \,\, N_i \label{eq:kodd}\\
    \boldsymbol{k}_i&= \begin{bmatrix}
        -\frac{N_i}{2} & \cdots & -1 & 0 & 1 & \cdots & \frac{N_i}{2}-1
    \end{bmatrix} \quad \text{for} \,\, \text{even} \,\, N_i\label{eq:keven}
\end{align}
where $\otimes$ is the Kronecker product and $\text{i}=\sqrt{-1}\neq i$. Finally, the torus function can be continuously evaluated over the entire torus space domain by Eq. \ref{eq:toruseval}.
\begin{equation}
    \label{eq:toruseval}
    \boldsymbol{\Gamma}(\theta_1,\theta_2) \approx \boldsymbol{C}^\text{T}\cdot\left(\text{e}^{\text{i}\cdot\boldsymbol{k}_1^\text{T}\theta_1}\otimes\text{e}^{\text{i}\cdot\boldsymbol{k}_2^\text{T}\theta_2}\right)
\end{equation}
This approximation is additionally useful because it allows for the straightforward computation of torus function sensitivity to its angle arguments, which will be needed later in the formulation of the minimum deviation rephasing optimal control problem.

Before a torus function can be used for further mission design and analysis, its accuracy in approximating the entire structure must be verified. To numerically evaluate this property, a measure of invariance error is defined that sums the field invariance error at angle sets in between those used in the Fourier fit. For a given set of $N_i$ uniformly distributed angles as in Eq. \ref{eq:anglevecforfit}, the midpoint set of $(N_i-1)$ uniformly distributed angles is found to be
\rev1{\begin{equation}
    \boldsymbol{\theta}_i^{\text{mid}} =
    \begin{bmatrix}
        \theta_{i,1}& \cdots & \theta_{i,N_i-1}
    \end{bmatrix}\in\mathbb{R}^{1\times (N_i-1)},\quad
    \theta_{i,j} = \frac{\pi(2j-1)}{N_i} 
\end{equation}}
Evaluating the field invariance error at these midpoint angles across a 2-dimensional torus function's angle domain gives the following measure of invariance.
\begin{equation}
    \epsilon = \frac{1}{(N_1-1)(N_2-1)}\sum_{i=1}^{N_1-1}\sum_{j=1}^{N_2-1}\left|\left|\left[\frac{\partial \boldsymbol{\Gamma}(\boldsymbol{\theta}^{\text{mid}}_{ij})}{\partial \boldsymbol{\theta}}\right]\boldsymbol{h}(\boldsymbol{\theta})-\boldsymbol{f}(\boldsymbol{\Gamma}(\boldsymbol{\theta}^{\text{mid}}_{ij}),\boldsymbol{\mu
    })\right|\right|
\end{equation}
\begin{equation}
    \boldsymbol{\theta}^{\text{mid}}_{ij}=
    \begin{bmatrix}
        \theta^{\text{mid}}_{1i} & \theta^{\text{mid}}_{2j}
    \end{bmatrix}^{\text{T}}
\end{equation}
where the analytical torus function sensitivities are given in Appendix A. Again, the error measure generalizes to $p$-dimensional tori. In the numerical computation of QPITs, a solution is generally said to be achieved when the problem's constraint vector $||\boldsymbol{F}||<1\times10^{-10}$ \cite{olikaraPHD}. The same criteria is used here for locating sufficient model orders of the torus function approximation. 

\begin{figure}[htbp!]
     \centering
     \begin{subfigure}{0.382\textwidth}
         \centering
         \includegraphics[width=1\textwidth]{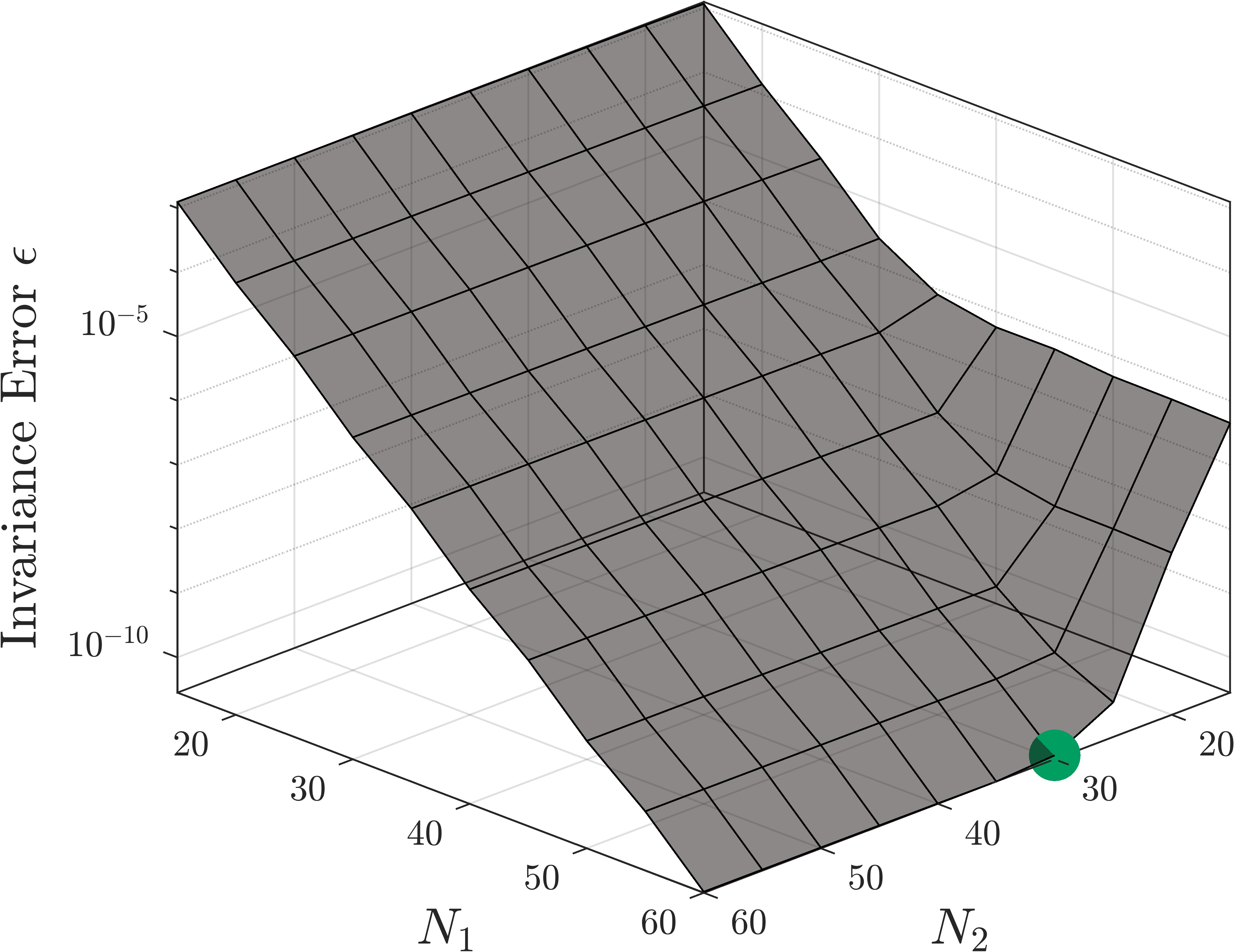}
         \caption{Parametric Error Function}
     \end{subfigure}\hspace{2em}
     \begin{subfigure}{0.315\textwidth}
         \centering
         \includegraphics[width=1\textwidth]{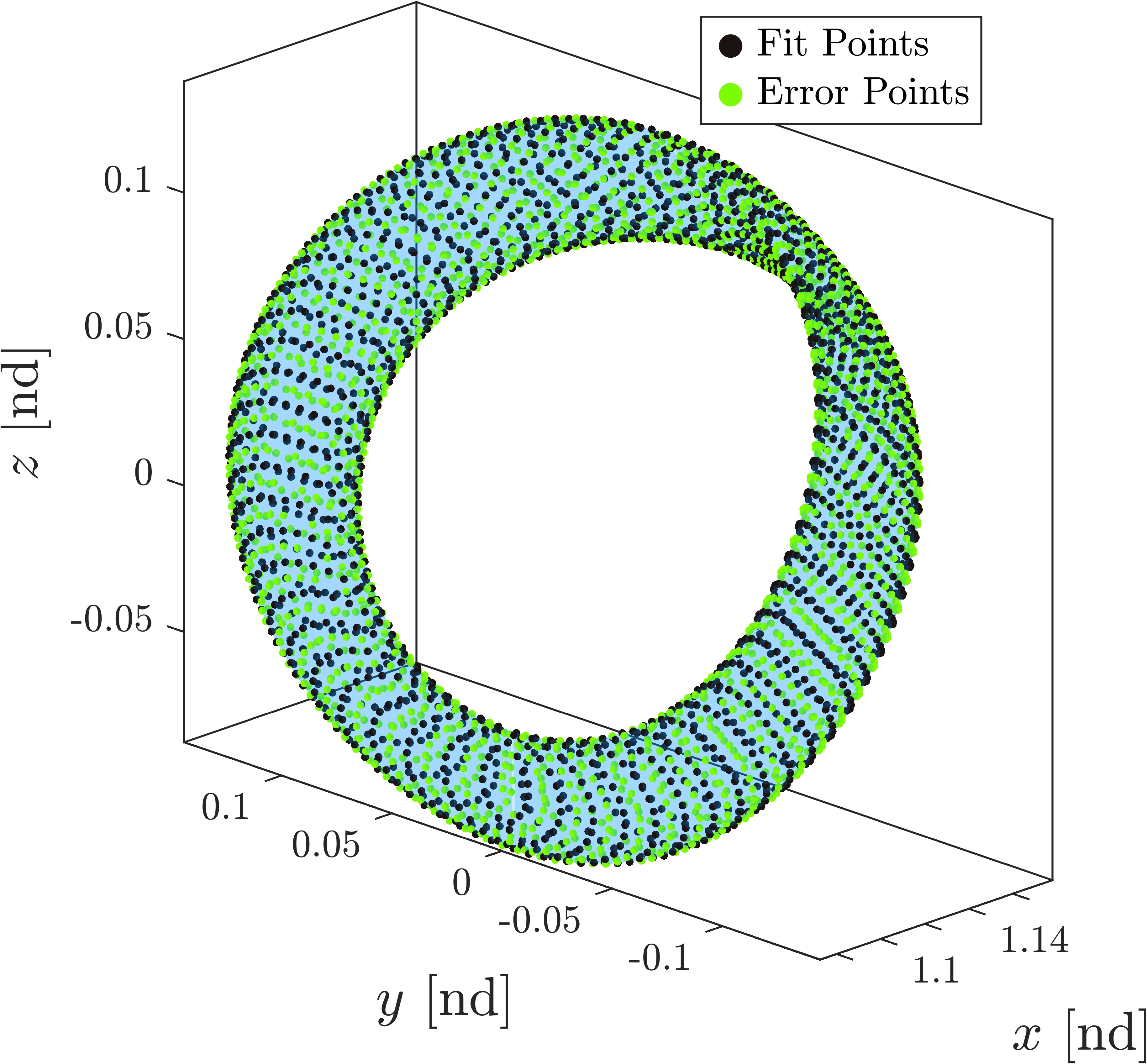}
         \caption{$N_1=60, N_2=30$}
     \end{subfigure}
     \caption{Approximation order analysis for 2-dimensional $L_2$ quasi-halo torus in the Earth-Moon circular restricted problem.}
    \label{fig:approx}
\end{figure}

\begin{figure}[htbp!]
    \centering
    \includegraphics[width=0.95\linewidth]{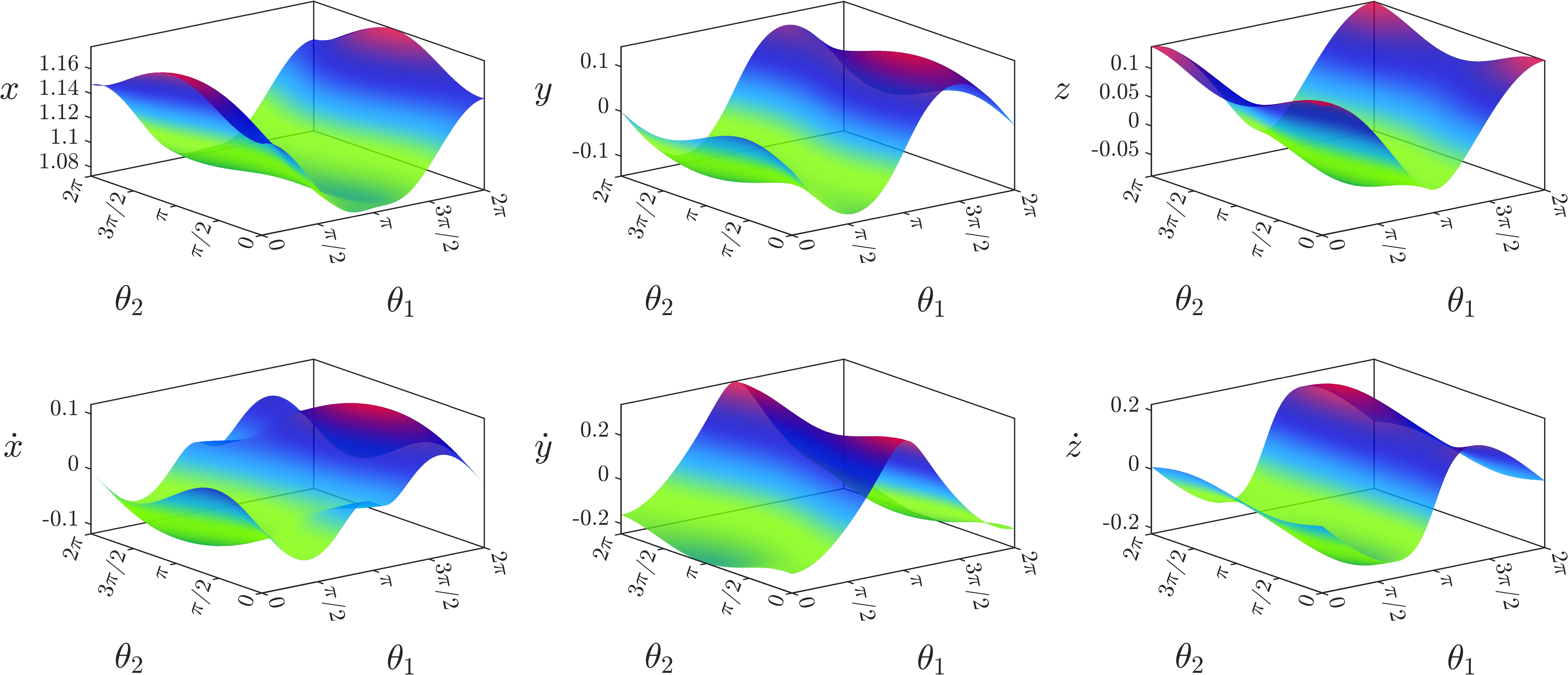}
    \caption{Torus function by component for 2-dimensional $L_2$ quasi-halo torus in the Earth-Moon circular restricted problem.}
    \label{fig:torusfun}
\end{figure}

An application of the Fourier approximation and error measure to determine ideal torus model order is shown in Fig. \ref{fig:approx} for an $L_2$ constant energy quasi-halo torus in the Earth-Moon CR3BP with the following parameters: Jacobi's Constant of 3.098, $\omega_1=1.8922$, and $\omega_2=1.6054$. The CR3BP equations of motion with control acceleration $\boldsymbol{u}=[u_x\,\,u_y\,\,u_z]^\text{T}$are written as
\begin{subequations} \label{eq:CR3BPcontrol}
    \begin{align}
        \Ddot{x}-2\Dot{y}-x & = -\frac{(1-\mu)(x+\mu)}{d_1^3} - \frac{\mu(x-1+\mu)}{d_2^3} +u_x\label{eq:cra} \\
        \Ddot{y}+2\Dot{x}-y & = -\frac{(1-\mu)y}{d_1^3} - \frac{\mu y}{d_2^3}+u_y\label{eq:crb}\\
        \Ddot{z} & = -\frac{(1-\mu)z}{d_1^3} - \frac{\mu z}{d_2^3}+u_z\label{eq:crc}
    \end{align}
\end{subequations}
where $d_1=\left[(x+\mu)^2+y^2+z^2\right]^{\frac{1}{2}}$, $d_2=\left[(x-1+\mu)^2+y^2+z^2\right]^{\frac{1}{2}}$, and $\mu=0.01215$ is the non-dimensional mass parameter of the Earth-Moon system.
From the parametric error function surface, a near linear logarithmic decrease in error is seen for a sufficient number $N_2$ as $N_1$ is increased. Here, $\theta_1$ corresponds to the phase of the underlying halo orbit, while $\theta_2$ corresponds to the winding motion around this orbit. In this case, an ideal model order is reached with half as many $N_2$ as $N_1$. \rev1{In general, more complex torus geometry requires more fit points because additional Fourier frequencies are necessary to accurately capture the structural deviation away from standard circles.} This example is shown for a torus with consistently fast dynamics across the structure. For tori which exhibit both fast and slow dynamics, like those around near-rectilinear halo orbits \rev1{with close perilune passes}, uniform sampling from an equivalent torus computed in a regularized vector field can ensure this means of structural approximation is successful using a tractable amount of fit points \cite{olikaraMASTERS}. This is beyond the scope of the work presented herein. A visualization of the torus function $\boldsymbol{\Gamma}$ created from the approximation in Fig. \ref{fig:approx} is shown in Fig. \ref{fig:torusfun}. Clearly each component tessellates with itself, showcasing the function's $2\pi$ periodicity.

\section{The Modified Invariance Condition}\label{sec:modinv}

The invariance condition in Eq. \ref{eq:invar1} satisfied by a $p$-dimensional torus is expressed with respect to the natural flow of a dynamical system. When a non-trivial control input $\boldsymbol{u}$ is introduced such as the low-thrust of a spacecraft engine, the natural invariance condition is no longer satisfied and motion will in general leave the torus surface in the phase space. To better understand this statement for a large class of mechanical systems where a continuous control input can be applied only at the acceleration level, the following mathematical framework surrounding a modified field invariance condition is developed for \textit{autonomous systems}, i.e. $m=0$. Quasi-periodically forced systems are discussed later in Sec. \ref{sec:mods}.

Let $\boldsymbol{x}$ in a system be decomposed into generalized coordinate $\boldsymbol{q}$ and coordinate rate $\dot{\boldsymbol{q}}$ components.
\begin{equation}
    \boldsymbol{x} =
    \begin{bmatrix}
        \boldsymbol{q} \\ \dot{\boldsymbol{q}}
    \end{bmatrix},\quad\boldsymbol{q},\dot{\boldsymbol{q}} \in \mathbb{R}^{\bar{n}}
\end{equation}
An invariant torus found in this system can also have its torus function decomposed into a coordinate mapping component and coordinate rate mapping component.
\begin{equation}
    \boldsymbol{\Gamma}(\boldsymbol{\theta}) = 
    \begin{bmatrix}
        \boldsymbol{\gamma}(\boldsymbol{\theta}) \\
        \dot{\boldsymbol{\gamma}}(\boldsymbol{\theta})
    \end{bmatrix}
\end{equation}

\rev1{Assume the system is control-affine with control appearing linearly in the dynamics. Without loss of generality, the control map is taken to be the identity matrix.} It is temporarily assumed that the system has physical access to both continuous input at the velocity level $\boldsymbol{u}_q$ and at the acceleration level $\boldsymbol{u}_{\dot{q}}$. The dynamics are now written as
\begin{equation}
    \dot{\boldsymbol{x}} = \boldsymbol{f}(\boldsymbol{x},\boldsymbol{
    \mu}) + \boldsymbol{u},\quad \boldsymbol{u}=
    \begin{bmatrix}
        \boldsymbol{u}_q \\ \boldsymbol{u}_{\dot{q}}
    \end{bmatrix}, \quad\boldsymbol{u}_q,\boldsymbol{u}_{\dot{q}}\in \mathbb{R}^{\bar{n}} \label{eq:DSC}
\end{equation}

With (falsely assumed) independent control available for each state in the phase space, control can be selected to artificially change the torus state while staying on the manifold. However, a connection between control applied in the phase space and artificially changing torus angles must be established to understand how these changes occur. This is captured through the introduction of the continuous control frequency input $\boldsymbol{\Omega}$ in the torus space. The control frequency dimension is locked to the number of internal torus frequencies, which for autonomous systems is the torus dimension $p$. The appended torus dynamics are now
\begin{equation}
    \dot{\boldsymbol{\theta}} = \boldsymbol{\omega} + \boldsymbol{\Omega},\quad\boldsymbol{\Omega}\in\mathbb{R}^p \label{eq:ADC}
\end{equation}
Though the fundamental torus function is used in this development such that the natural torus dynamics are constant, this methodology is generally applicable to any torus parametrization and hence function.

Equating the vector fields of the torus and the system with control now included results in the modified field invariance condition of Eq. \ref{eq:invar2}.
\begin{equation}
    \left[\frac{\partial \boldsymbol{\Gamma}(\boldsymbol{\theta})}{\partial \boldsymbol{\theta}}\right]\left(\boldsymbol{\omega} + \boldsymbol{\Omega}\right) = \boldsymbol{f}(\boldsymbol{\Gamma}(\boldsymbol{\theta}),\boldsymbol{\mu}) +
    \boldsymbol{u}
    \label{eq:invar2}
\end{equation}
Noting the relationship of the invariance condition in Eq. \ref{eq:invar1}, the natural dynamics drop out \rev1{under the assumption that the trajectory lies on the torus manifold. This leaves} behind only the control structure components as
\begin{equation}
    \left[\frac{\partial\boldsymbol{{\gamma}}(\boldsymbol{\theta})}{\partial \boldsymbol{\theta}}\right]\boldsymbol{\Omega} = \boldsymbol{u}_q, \quad \left[\frac{\partial\dot{\boldsymbol{{\gamma}}}(\boldsymbol{\theta})}{\partial \boldsymbol{\theta}}\right]\boldsymbol{\Omega} = \boldsymbol{u}_{\dot{q}} \label{eq:controlCon}
\end{equation}
Equation \ref{eq:controlCon} gives two relationships that must be satisfied for controlled trajectories to remain on the original invariant torus during rephasing. For a large class of mechanical systems which includes astrodynamic models, control cannot be continuously applied at the velocity level, i.e. $\boldsymbol{u}_q=\boldsymbol{0}$. For an arbitrary (non-zero) control frequency $\boldsymbol{\Omega}$, this demands that
\begin{equation}
    \text{dim}\left\{\text{null}\left(\left[\frac{\partial\boldsymbol{{\gamma}}(\boldsymbol{\theta})}{\partial \boldsymbol{\theta}}\right]\right)\right\} \geq 1,\quad \left[\frac{\partial\boldsymbol{{\gamma}}(\boldsymbol{\theta})}{\partial \boldsymbol{\theta}}\right] \in\mathbb{R}^{\bar{n}\times p}
\end{equation}
but by definition, $\boldsymbol{\gamma}(\boldsymbol{\theta})$ is a sufficiently smooth, locally bijective map that is a sub-component of the globally bijective diffeomorphism $\boldsymbol{\Gamma}(\boldsymbol{\theta})$. As such, it is locally non-singular with a well-defined local inverse in the neighborhood of all angle sets $\boldsymbol{\theta}$. Therefore, the sensitivity of this map with respect to its arguments must be locally non-singular and cannot hold a null space with dimension. Thus, it is concluded that any unnatural deviation in the angle space caused by a control frequency $\boldsymbol{\Omega}$ cannot be realized with the control available in this class of mechanical systems. In other words, acceleration level control authority alone is insufficient in maneuvering a particle along an invariant torus in the phase space such that the natural invariance condition is exactly satisfied at every instance in time. Perfectly "safe" rephasing on QPITs by continued satisfaction of the invariance condition is then not possible, but this perspective gives insight on how to construct reference trajectories along the torus that minimize torus deviation instead of impossibly eliminating it all together. This forms the basis of the bi-level optimal control framework for the manifold-constrained low-thrust maneuvering of quasi-periodic orbits for reconfiguration/rephasing.

The modified invariance condition perspective can also be viewed using the inverse torus function, defined below.
\begin{equation}
    \boldsymbol{\Gamma}^{-1}:\mathbb{R}^n\rightarrow\mathbb{T}_p
\end{equation}
Using the fundamental inverse torus function such that $\dot{\boldsymbol{\theta}}=\boldsymbol{\omega}$, an alternate rendition of the field invariance condition enforces mapped dynamic constants.
\begin{align}
    \frac{\text{d}}{\text{d}t}\left\{\boldsymbol{\theta} = \boldsymbol{\Gamma}^{-1}(\boldsymbol{x})\right\} = \dot{\boldsymbol{\theta}}&=\left[\frac{\partial \boldsymbol{\Gamma}^{-1}(\boldsymbol{x})}{\partial\boldsymbol{x}}\right]\dot{\boldsymbol{x}}\\
    \boldsymbol{\omega}&=\left[\frac{\partial \boldsymbol{\Gamma}^{-1}(\boldsymbol{x})}{\partial\boldsymbol{x}}\right]\boldsymbol{f}(\boldsymbol{x},\boldsymbol{\mu})
\end{align}
Again, adding in continuous control inputs to both the torus and phase space, a modified field invariance condition is expressed,
\begin{equation}
    \boldsymbol{\omega}+\boldsymbol{\Omega}=\left[\frac{\partial \boldsymbol{\Gamma}^{-1}(\boldsymbol{x})}{\partial\boldsymbol{x}}\right]\left(\boldsymbol{f}(\boldsymbol{x},\boldsymbol{\mu})+
    \boldsymbol{u}\right)
\end{equation}
leading to the necessary control relationships in Eq. \ref{eq:invinv} after the natural invariance is removed.
\begin{equation}
    \boldsymbol{\Omega}=\left[\frac{\partial \boldsymbol{\Gamma}^{-1}(\boldsymbol{x})}{\partial\boldsymbol{x}}\right]
    \boldsymbol{u}\label{eq:invinv}
\end{equation}

The substitution of Eq. \ref{eq:invinv} into Eq. \ref{eq:invar2} yields the following relationship between tangent bundles of the torus function and its inverse that must hold over all $\boldsymbol{\theta}$ and $\boldsymbol{x}$ in the bijection,
\begin{align}
    &\left[\frac{\partial \boldsymbol{\Gamma}(\boldsymbol{\theta})}{\partial \boldsymbol{\theta}}\right]\cdot\left[\frac{\partial \boldsymbol{\Gamma}^{-1}(\boldsymbol{x})}{\partial\boldsymbol{x}}\right]
    \boldsymbol{u} = \boldsymbol{u} \\
    &\implies\left[\frac{\partial \boldsymbol{\Gamma}(\boldsymbol{\theta})}{\partial \boldsymbol{\theta}}\right]\cdot\left[\frac{\partial \boldsymbol{\Gamma}^{-1}(\boldsymbol{x})}{\partial\boldsymbol{x}}\right] = \mathbb{I}_{n\times n} \\
    &\implies\left[\frac{\partial \boldsymbol{\Gamma}^{-1}(\boldsymbol{x})}{\partial\boldsymbol{x}}\right]\cdot\left[\frac{\partial \boldsymbol{\Gamma}(\boldsymbol{\theta})}{\partial \boldsymbol{\theta}}\right]=\mathbb{I}_{p\times p}
\end{align}
where $\mathbb{I}$ is the identity matrix. With the identity property established, Eq. \ref{eq:invinv} can be returned to Eq. \ref{eq:controlCon}, and the same logic of local differentiability can be applied to conclude that any unnatural deviation in the torus space caused by a control frequency $\boldsymbol{\Omega}$ cannot be realized with the control available in a mechanical system with only acceleration level input.

With a connection between control in both the torus and phase space established, the goal of constructing a continuous, low-thrust trajectory to and from states on an invariant torus which minimizes its deviation from the manifold can now be split into two levels of successive optimization. The first level solves for the most realizable torus space trajectory by minimizing the mapped fictitious control input $\boldsymbol{u}_q$. The second level then realizes this trajectory with phase space available control input by either the minimization of tracking error, or the generation of patchwork solutions based on torus space switching times.
\section{Bi-Level Optimal Control Framework for Autonomous Dynamical Systems}\label{sec:bilevel}

Consider the general optimal control problem of maneuvering a particle to and from states on an invariant manifold embedded within a dynamical system \rev1{where deviation from the manifold is minimized.} As a thought exercise, let this problem be transcribed with a direct method such that the continuous trajectory is discretized and \rev1{the problem is} solved via nonlinear programming \cite{topputo2014survey}. \rev1{In order to evaluate the cost function at each iteration, the distance from each discrete trajectory point to the manifold must be found to quantify the current trajectory's manifold deviation. For an arbitrary point, there is no closed-form expression that yields the distance to the manifold. Thus, the minimum distance from each discrete trajectory point to the manifold must be found numerically within every iteration of the nonlinear program. This ultimately results in a nested optimization problem that is computationally intractable and limits dynamical insight.}

% Consider the general optimal control problem of maneuvering a particle to and from states on an invariant manifold embedded within a dynamical system that minimizes manifold deviation. As a thought exercise, let this problem be transcribed with a direct method such that the continuous trajectory is discretized and solved via nonlinear programming \cite{topputo2014survey}. Then at every discrete point along the trajectory within an iteration, a sub-optimization problem needs to be solved to find the instantaneous minimum distance to the manifold in order to quantify the total structural deviation, and hence evaluate the over-arching cost function. A nested optimization problem of this nature is computationally intractable and limits any insight gained into the problem.

To better approach the solution of this problem, a bi-level optimization framework is now proposed. The rephasing optimal control problem is first solved in the torus space through the minimization of the mapped fictitious phase space velocity control input $\boldsymbol{u}_q$. This is made possible by a torus' Fourier approximation and the control space mappings of Eq. \ref{eq:controlCon} derived from the modified invariance condition in Sec. \ref{sec:modinv}. A solution to this type of torus space optimal control problem represents the most realizable trajectory on the torus to track in the phase space with physically available control input. The torus space solution is then mapped to the phase space by the torus function to generate a reference trajectory. A phase space solution is achieved either by solving a minimum tracking error problem or generating patched solutions based on torus space switch times, depending on the torus space cost function used to generate the reference trajectory. The remainder of this section introduces the bi-level framework for \textit{autonomous} dynamical systems where all torus frequencies are internal. The framework is discussed in the context of indirect methods of optimal control because the problems at hand lack complex constraints and are shown to perform well with simple initial guess and continuation schemes. Either or both levels of the optimal control problems posed here towards the solution of the minimum torus deviation rephasing problem could be solved via direct methods. A brief treatment of indirect optimal control theory is provided in Appendix B, with a full treatment found in \cite{bryson2018applied}. A comparison between indirect and direct methods is detailed in \cite{betts1998survey}.
% Could say more here about direct vs indirect, see pino dissertation

\subsection{Optimal Control in the Torus Space}

Two methods to generate torus space reference trajectories for minimum torus deviation rephasing are now presented. The first method solves the constrained minimum velocity energy (CMVE) problem, which considers a cost function that is quadratic with respect to $\boldsymbol{u}_q$ mapped from the phase space. The second method solves the constrained minimum velocity throttle (CMVT) problem. This problem's cost function is more closely related to the true $\mathcal{L}_2$ norm of $\boldsymbol{u}_q$ and generates thrust and coast arcs on the torus surface with a bang-off-bang control structure. Both optimal control problems are defined below assuming fixed boundary conditions and times in the torus space, i.e. $\boldsymbol{\theta}(t_0) = \boldsymbol{\theta}_0,\,\, \boldsymbol{\psi}(t_f)=\boldsymbol{\theta}(t_f) - \boldsymbol{\theta}_d$, where $\boldsymbol{\theta}_d$ is the desired final angle set. Additional boundary value problems related to mission applications and their transversality conditions are discussed after both methods are detailed.
% Could discuss trivial minimization of Omega here

\subsubsection{Constrained Minimum Velocity Energy}
The quadratic minimization of $\boldsymbol{u}_q$ is defined by the CMVE problem in Eq. \ref{eq:CMVE}.
\begin{align}
    \text{min}\quad &J = \int_{t_0}^{t_f}\frac{1}{2}\boldsymbol{u}_q^\text{T}\boldsymbol{u}_q\:\text{d}t = \int_{t_0}^{t_f}\frac{1}{2}\boldsymbol{\Omega}^\text{T} \boldsymbol{W}_q(\boldsymbol{\theta})\boldsymbol{\Omega}\:\text{d}t \label{eq:CMVE} \\
    \text{s.t.}\quad &\dot{\boldsymbol{\theta}} = \boldsymbol{\omega}+\boldsymbol{\Omega},\quad \left|\left|\boldsymbol{u}_{\dot{q}}\right|\right|=\left|\left|\left[\frac{\partial \dot{\boldsymbol{\gamma}}(\boldsymbol{\theta})}{\partial \boldsymbol{\theta}}\right]\boldsymbol{\Omega}\right|\right| \leq u_{\dot{q},\text{max}} \nonumber 
\end{align}
When mapped into the control frequency space, the cost function is dynamically weighted by the state-dependent weight matrix $\boldsymbol{W}_q(\boldsymbol{\theta})$ which directly links the phase space and torus space through the torus function sensitivity.
\begin{equation}
    \boldsymbol{W}_q(\boldsymbol{\theta}) = \left[\frac{\partial \boldsymbol{\gamma}(\boldsymbol{\theta})}{\partial \boldsymbol{\theta}}\right]^\text{T}\left[\frac{\partial \boldsymbol{\gamma}(\boldsymbol{\theta})}{\partial \boldsymbol{\theta}}\right]\succ 0,\quad\forall\:\boldsymbol{\theta}\in\mathbb{T}_p
\end{equation}
This weight matrix is positive definite by definition of the torus function as a diffeomorphism. 

Applying the indirect optimal control approach, the problem's Hamiltonian is defined with costates $\boldsymbol{\lambda}_\theta=\left[\lambda_1\:\cdots\:\lambda_p \right]^{\text{T}}$.
\begin{equation}
    \mathcal{H} = \frac{1}{2}\boldsymbol{\Omega}^\text{T} \boldsymbol{W}_q(\boldsymbol{\theta})\boldsymbol{\Omega} + \boldsymbol{\lambda}_{\theta}^\text{T}[\boldsymbol{\omega}+\boldsymbol{\Omega}]
\end{equation}
Note that because the fundamental torus function is used, the only state dependence within $\mathcal{H}$ comes from the dynamic weighting matrix. Applying the stationary condition, the control frequency structure is analytically derived by nature of the quadratic cost function. Despite this optimal control problem occurring exclusively in the torus space, the relationship between $\boldsymbol{\Omega}$ and $\boldsymbol{u}_{\dot{q}}$ from Eq. \ref{eq:controlCon} \rev1{can be used to derive an instantaneous maximum control frequency $\Omega_{\text{max}}$ given a maximum control acceleration $u_{\dot{q},\text{max}}$ obtained from a simplified model of a low-thrust spacecraft engine.} Enforcing this input constraint does not translate directly to trajectories in the phase space because of non-zero $\boldsymbol{u}_q$, but it does maintain dynamical system consistency at both optimization levels, \rev1{important for the eventual transition of torus space solutions to the phase space}. Note that while the maximum acceleration control magnitude is constant, the maximum control frequency magnitude is state-dependent because of the torus function. Rather than include this constraint in the problem's Hamiltonian, Pontryagin's Minimum Principle (PMP) is employed in the application of the stationary condition to instantaneously minimize $\mathcal{H}$ at each time $t$ by using the maximum available control frequency. This results in the following control structure, where $\boldsymbol{p}$ is the weighted primer vector.
\begin{align}
    \boldsymbol{p} &= -\boldsymbol{W}_q(\boldsymbol{\theta})^{-1}\boldsymbol{\lambda}_{\theta},\quad\widehat{\boldsymbol{p}}=\boldsymbol{p}/||\boldsymbol{p}|| \\
    \boldsymbol{\Omega} &= 
    \begin{dcases}
    \boldsymbol{p},& \text{if } \,\,\left|\left|\left[\frac{\partial \dot{\boldsymbol{\gamma}}(\boldsymbol{\theta})}{\partial \boldsymbol{\theta}}\right]\boldsymbol{p}\right|\right| \leq u_{\dot{q},\text{max}}\\
    \Omega_{\text{max}}(\boldsymbol{\theta})\widehat{\boldsymbol{p}},              & \text{otherwise}
    \end{dcases} \\
    \Omega&_{\text{max}}(\boldsymbol{\theta}) = \frac{u_{\dot{q},\text{max}}}{\sqrt{\widehat{\boldsymbol{p}}^\text{T} \boldsymbol{W}_{\dot{q}}(\boldsymbol{\theta})\widehat{\boldsymbol{p}}}},\quad
    \boldsymbol{W}_{\dot{q}}(\boldsymbol{\theta}) = \left[\frac{\partial \dot{\boldsymbol{\gamma}}(\boldsymbol{\theta})}{\partial \boldsymbol{\theta}}\right]^\text{T}\left[\frac{\partial \dot{\boldsymbol{\gamma}}(\boldsymbol{\theta})}{\partial \boldsymbol{\theta}}\right]\succ0\label{eq:maxfreq}
\end{align}

Lastly, the costate dynamics are derived to be a function of the second order torus function sensitivities, which are available for use from the Fourier approximation, and given in Appendix A.
\begin{align}
    &\dot{\boldsymbol{\lambda}}_{\theta} = -\frac{\partial }{\partial \boldsymbol{\theta}}\left\{\frac{1}{2}\boldsymbol{\Omega}^\text{T} \boldsymbol{W}_q(\boldsymbol{\theta})\boldsymbol{\Omega}\right\} = -\left[\frac{\partial \boldsymbol{u}_q}{\partial \boldsymbol{\theta}}\right]^\text{T}\boldsymbol{u}_q \label{eq:CMVEcostates}\\
    &\left[\frac{\partial \boldsymbol{u}_q}{\partial \boldsymbol{\theta}}\right] =
    \begin{bmatrix}
        \boldsymbol{\Omega}^\text{T}\boldsymbol{H}^1 \\ \vdots \\
        \boldsymbol{\Omega}^\text{T}\boldsymbol{H}^{\bar{n}}
    \end{bmatrix},\quad
    H_{j,k}^i =
    \frac{\partial^2\gamma_i}{\partial\theta_j\partial\theta_k},\quad j,k =1,\dots,p \nonumber
\end{align}

\subsubsection{Constrained Minimum Velocity Throttle}
The cost function corresponding to the direct $\mathcal{L}_2$ minimization of $\boldsymbol{u}_q$ mapped into the torus space is given in Eq. \ref{eq:CMVTbad}.
\begin{equation}
    J=\int_{t_0}^{t_f}u_q\:\text{d}t=\int_{t_0}^{t_f}\sqrt{\boldsymbol{u}_q^\text{T}\boldsymbol{u}_q}\:\text{d}t = \int_{t_0}^{t_f}\sqrt{\boldsymbol{\Omega}^\text{T} \boldsymbol{W}_q(\boldsymbol{\theta})\boldsymbol{\Omega}}\:\text{d}t\label{eq:CMVTbad}
\end{equation}
Because of the necessary presence of $\boldsymbol{W}_q(\boldsymbol{\theta})$ under the radical, the control frequency $\boldsymbol{\Omega}$ cannot be determined analytically from the stationary condition, nor can the control frequency magnitude and direction be decoupled like in the traditional minimum-fuel problem \cite{pontryagin2018mathematical}. Still, a thrust/coast solution in the torus space is desired because it identifies optimal switching times when control should be applied, in turn determining passively safe coast arcs on the torus surface. A direct method could be used with Eq. \ref{eq:CMVTbad} to generate a solution in the torus space. To proceed with the indirect approach, an alternative cost function that retains an $\mathcal{L}_2$ structure is designed and displayed in Eq. \ref{eq:CMVT}.
\begin{align}
    \text{min}\quad &J = \int_{t_0}^{t_f}\frac{\Omega}{2}\cdot\widehat{\boldsymbol{\Omega}}^\text{T} \boldsymbol{W}_q(\boldsymbol{\theta})\widehat{\boldsymbol{\Omega}}\:\text{d}t  \label{eq:CMVT}\\
        \text{s.t.}\quad &\dot{\boldsymbol{\theta}} = \boldsymbol{\omega}+\Omega\cdot\widehat{\boldsymbol{\Omega}},\quad \left|\left|\boldsymbol{u}_{\dot{q}}\right|\right|=\left|\left|\left[\frac{\partial \dot{\boldsymbol{\gamma}}(\boldsymbol{\theta})}{\partial \boldsymbol{\theta}}\right]\boldsymbol{\Omega}\right|\right| \leq u_{\dot{q},\text{max}} \nonumber
\end{align}
The control frequency magnitude $\Omega=||\boldsymbol{\Omega}||>0$ now appears linearly, while the control frequency direction $\widehat{\boldsymbol{\Omega}}$ remains quadratically weighted. This allows for the reuse of the weighted primer vector $\boldsymbol{p}$ to determine control direction $\widehat{\boldsymbol{\Omega}} = \widehat{\boldsymbol{p}}$, while PMP must be employed to determine a control switching structure through the instantaneous minimization of $\mathcal{H}$.

The problem's Hamiltonian is written
\begin{equation}
    \mathcal{H} = \frac{\Omega}{2}\cdot\widehat{\boldsymbol{\Omega}}^\text{T} \boldsymbol{W}_q(\boldsymbol{\theta})\widehat{\boldsymbol{\Omega}} + \boldsymbol{\lambda}_{\theta}^\text{T}[\boldsymbol{\omega}+\Omega\cdot\widehat{\boldsymbol{\Omega}}]
\end{equation}
and the following switching function is deduced through the examination of its components dependent on $\Omega$.
\begin{align}
    s(t) &= \frac{1}{2}\widehat{\boldsymbol{\Omega}}^\text{T} \boldsymbol{W}_q(\boldsymbol{\theta})\widehat{\boldsymbol{\Omega}} + \boldsymbol{\lambda}_{\theta}^\text{T}\widehat{\boldsymbol{\Omega}} \\
    \Omega &= 
    \begin{dcases}
    \Omega_{\text{max}}(\boldsymbol{\theta}),& \text{if } s(t)<0\\
    0,              & \text{otherwise}
    \end{dcases} \label{eq:CMVTswitch}
\end{align}
While the switching structure of Eq. \ref{eq:CMVTswitch} is straightforward to implement, its discontinuous nature causes numerical difficulty in the integration of continuous-time dynamical systems. Thus, the hyperbolic tangent smoothing function of Eq. \ref{eq:hyper} is used, where $\rho$ is the smoothing parameter \cite{taheri2018generic}. As $\rho\rightarrow0$, the control approaches the desired on-off structure.
\begin{equation}
    \Omega\approx\frac{\Omega_{\text{max}}(\boldsymbol{\theta})}{2}\left(1 + \tanh{\frac{s(t)}{\rho}}\right) \label{eq:hyper}
\end{equation}

Finally, the costate dynamics derived from the first--order necessary condition take the same qualitative form as those from the CMVE problem with $\boldsymbol{u}^*_q=\left[\frac{\partial \boldsymbol{\gamma}(\boldsymbol{\theta})}{\partial \boldsymbol{\theta}}\right]\widehat{\boldsymbol{\Omega}}$.
\begin{align}
    &\dot{\boldsymbol{\lambda}}_{\theta} = -\frac{\partial }{\partial \boldsymbol{\theta}}\left\{\frac{\Omega}{2}\cdot\widehat{\boldsymbol{\Omega}}^\text{T} \boldsymbol{W}_q(\boldsymbol{\theta})\widehat{\boldsymbol{\Omega}}\right\} = -\Omega\cdot\left[\frac{\partial \boldsymbol{u}^*_q}{\partial \boldsymbol{\theta}}\right]^\text{T}\boldsymbol{u}^*_q\\
    &\left[\frac{\partial \boldsymbol{u}^*_q}{\partial \boldsymbol{\theta}}\right] =
    \begin{bmatrix}
        \widehat{\boldsymbol{\Omega}}^\text{T}\boldsymbol{H}^1 \\ \vdots \\
        \widehat{\boldsymbol{\Omega}}^\text{T}\boldsymbol{H}^{\bar{n}}
    \end{bmatrix},\quad
    H_{j,k}^i =
    \frac{\partial^2\gamma_i}{\partial\theta_j\partial\theta_k},\quad j,k =1,\dots,p \nonumber
\end{align}

\subsubsection{Mission Driven Boundary Constraints}
A torus function plot like that shown in Fig. \ref{fig:torusfun} can help provide geometric context in selecting desired, fixed boundary conditions in the torus space for specified initial and final times. The indirect solution for a 2-dimensional torus must then complete the following boundary value problem with free variables $\boldsymbol{\chi}$ and constraints $\boldsymbol{F}=\boldsymbol{0}$.
\begin{equation}
    \boldsymbol{\chi} =
    \begin{bmatrix}
        \lambda_1(t_0) \\ \lambda_2(t_0)
    \end{bmatrix},\quad
    \boldsymbol{F}=
    \begin{bmatrix}
        \theta_1(t_f) - \theta_{1,d} \\ \theta_2(t_f) - \theta_{2,d}
    \end{bmatrix}
\end{equation}
Still, flexible boundary conditions are appropriate in particular mission scenarios. For example, leaving one terminal angle free may be advantageous when that angle's geometric variation in configuration space is smaller than others on the torus. In other words, the goal behind rephasing may be accomplished by only changing the angle related to an underlying periodic orbit, leaving the location on the terminal invariant curve or surface free. In the indirect approach, an additional transversality condition is then required to capture the allowed final state variation. Let $\theta_i(t_f)$ be free, then the terminal constraint $\lambda_i(t_f)=0$ must be enforced in the boundary value problem. For example,
\begin{equation}
    \boldsymbol{\chi} =
    \begin{bmatrix}
        \lambda_1(t_0) \\ \lambda_2(t_0)
    \end{bmatrix}\quad
    \boldsymbol{F}=
    \begin{bmatrix}
        \theta_1(t_f) - \theta_{1,d} \\ \lambda_2(t_f)
    \end{bmatrix}
\end{equation}

Another set of boundary conditions that appear regularly in astrodynamics relate to the rendezvous problem. Let the desired set of terminal angles be fully parametrized by a fixed set of initial angles and the final time such that the final time alone is free.
\begin{equation}
    \boldsymbol{\theta}_d(t_f) = \boldsymbol{\omega}(t_f - t_0) + \boldsymbol{\theta}_d(t_0),\quad t_f\,\,\text{free}
\end{equation}
Maneuvering to this particular QPO defined by $\boldsymbol{\theta}_d(t_0)$ could be relevant in rephasing relative to another sensor in a constellation or formation. In contrast to conventional astrodynamic problems, leaving time free and not present in the cost function makes this type of problem on a QPIT ill-posed because of the quasi-periodic definition. Assuming $\boldsymbol{\theta}(t_0)\neq\boldsymbol{\theta}_d(t_0)$, there exists some time $t_f<\infty$ such that $\boldsymbol{\theta}(t_f)=\boldsymbol{\theta}_d(t_f)$ because natural motion on the QPIT will eventually satisfy the boundary conditions without the need for control. This results in the global minimum $J=0$ for both CMVE and CMVT. Nonetheless, given mission constraints for an allowable variation in $t_f$, a local minimum can still be located as either an extrema point, or domain boundary. Searching for an extrema point with a newly required transversality condition related to final time variation is found to be computationally difficult because the precision required in the initial guess of final time is too high. Instead, the allowable final time domain is discretized into a finite number of fixed boundary value problems with each previous solution warm-starting the next.

\subsubsection{Numerical Implementation}
Boundary conditions in the torus space thus far have been implicitly defined on the interval $[0,2\pi)$. The question arises of whether to treat $\boldsymbol{\theta}_d\in\mathbb{T}_p$ or $\boldsymbol{\theta}_d\in\mathbb{R}^p$ in efforts to achieve a numerical solution to the boundary value problems posed above. The real-valued costates don't share this ambiguity because they live in the cotangent space of the torus\footnote{Costates represent cost function sensitivity to state variations.}. Assume $\boldsymbol{\theta}_d\in\mathbb{T}_p$ and consider the evaluation of the constraint $F = \theta_1(t_f) - \theta_{1,d}$. If $\theta_{1,d} = 2\pi$ and $\theta_1(t_f)=2\pi-\delta\theta$ for $\delta\theta<<0$, then $F=-\delta\theta\approx0$. But if $\theta_1(t_f)=\delta\theta$, then $F=\delta\theta-2\pi\approx-2\pi$. Both cases are equidistant from the desired torus angle, yet result in drastically different constraint magnitudes, which are ultimately used to iteratively update the solution in a nonlinear root solver. There is the option to numerically force $F\in\mathbb{T}_1$, but this artificially constricts the problem's convergence basin with $F_\text{max}=2\pi$. Because of this discrepancy, the numerical enforcement of boundary conditions must occur with $\boldsymbol{\theta}_d\in\mathbb{R}^p$.

Let $\theta_i(t_f)\in\mathbb{R}$ be the natural torus angle after $t_f$ given by Eq. \ref{eq:linearFrequency} and $\theta_{i,d}(t_f)\in\mathbb{T}_1$ be the desired final angle in the torus domain. Then the real-valued, desired final angle is found by solving the integer programming problem
\begin{equation}
    \min_{\ell\in\mathbb{Z}} \left|2\pi \ell+ \theta_{i,d}(t_f) - \theta_i(t_f)\right|
\end{equation}
where $2\pi \ell+ \theta_{i,d}(t_f)\in\mathbb{R}$ is to be used in the numerical solution to the torus space optimal control problems. This is solved by a simple line search over $\ell$. 

Using a real-valued final torus angle also streamlines the initial guess scheme for solution by multiple shooting -- the primary solution method used in this paper. For a trajectory broken down into $M$ segments, the initial angles of the $j$th segment are chosen by linear interpolation between the given trajectory's initial angles and the real-valued, desired final angles.
\begin{equation}
    \theta_i(t_j) = \theta_i(t_0) + \frac{j-1}{M}\left(\theta_{i,d}(t_f)-\theta_{i}(t_0)\right),\quad j=1,...\,,M
\end{equation}
To complete the initial guess scheme, different values for segmented initial costates vary by problem type. For the CMVE problem, zeros are used to initialize all costates. Convergence with this initial guess scheme was $>95\%$ for all cases with fixed terminal conditions examined in this paper. For the CMVT problem, random numbers are sampled from a standard normal distribution to be used as initial costates for each segment at $\rho=1$. This problem is found to be more sensitive than CMVE, but all cases examined with fixed terminal conditions saw convergence within 10 randomly generated costate vectors.

Within the multiple shooting scheme, a variable-step, variable-order Adams--Bashforth--Moulton PECE integrator is used in place of a more conventional Runge--Kutta method because the state-costate dynamics require the relatively expensive evaluation of the torus function approximation at every step \cite{shampine1997matlab}. A relative and absolute integration tolerance of $1\times10^{-10}$ and $1\times10^{-11}$, respectively, are found to strike a balance between integration accuracy and time. A trust-region nonlinear root solver with dogleg stepping is used to iterate towards a solution with a numerically computed Jacobian. Lastly, natural parameter continuation over $\rho$ is used to refine CMVT solutions towards on-off control structures. Recall from Eq. \ref{eq:maxfreq} that the maximum control frequency is state-dependent at every time step. This makes continuation over $\rho$ more sensitive than conventional trajectory optimization problems, particularly in problems with allowable final state variation or multi-revolutions. All examples presented in this paper were able to reach at least $\rho=2\times10^{-4}$, where switching times can still be identified and the phase space transition procedure can proceed.

\subsubsection{Numerical Example}
\begin{figure}[htbp!]
     \centering
     \begin{subfigure}{0.32\textwidth}
         \centering
         \includegraphics[width=1\textwidth]{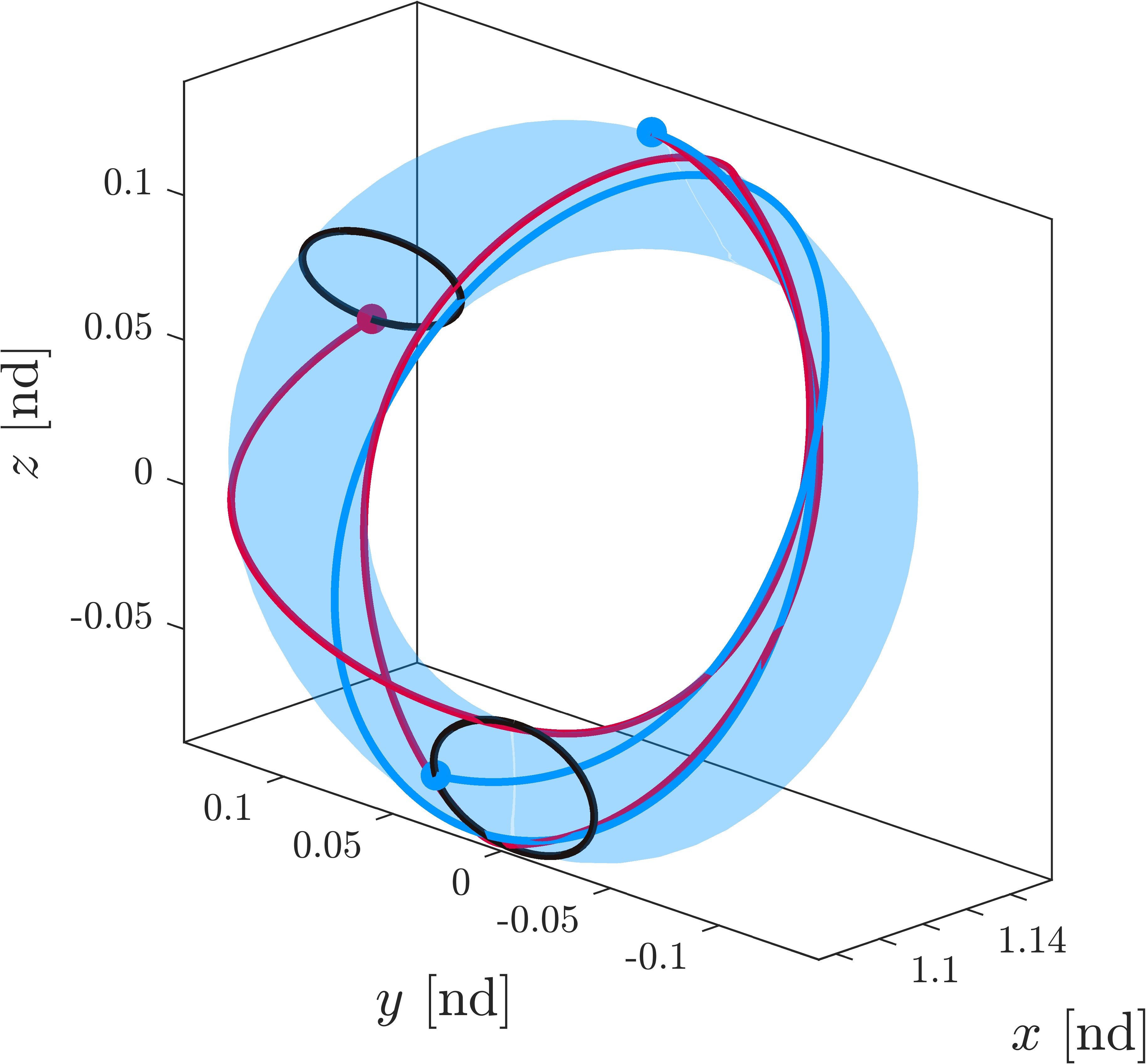}\\
         \includegraphics[width=1\textwidth]{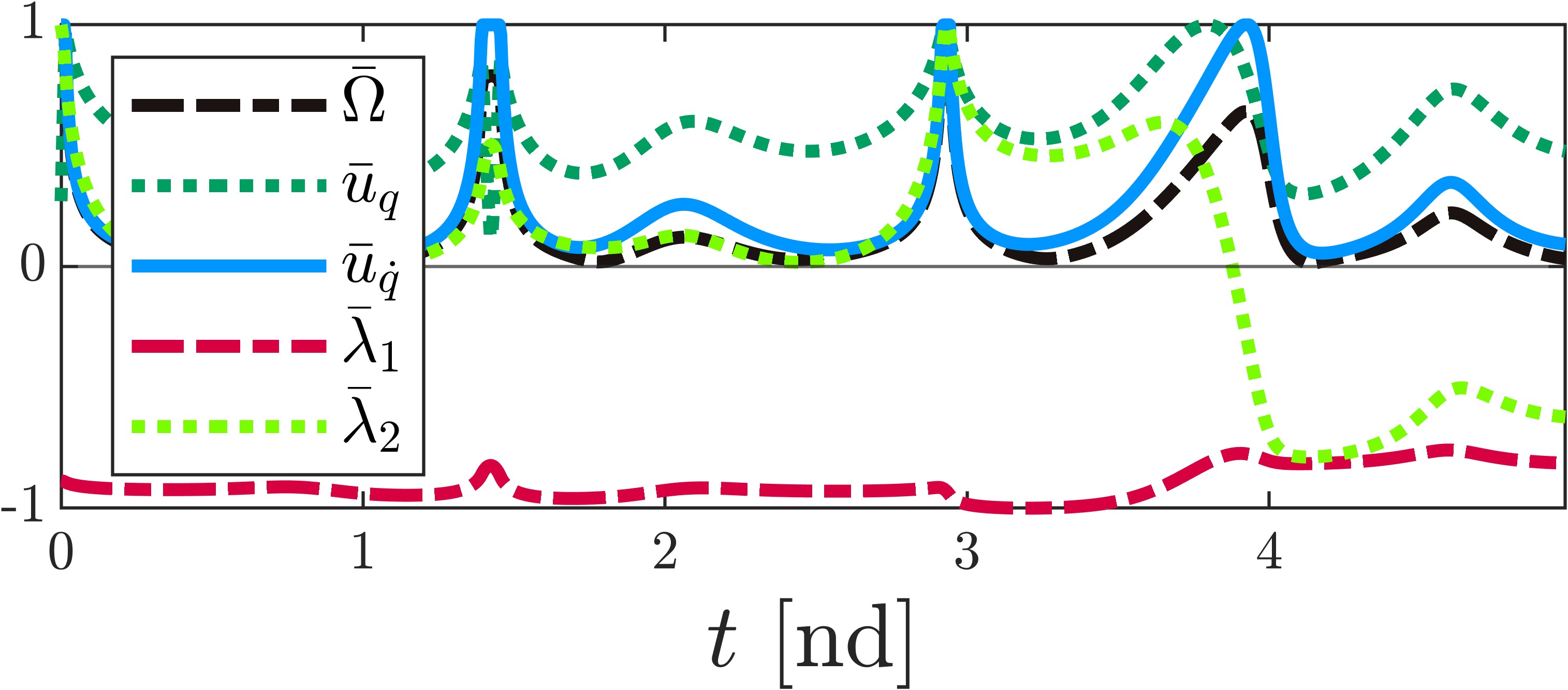}
         \caption{CMVE $\boldsymbol{\theta}(t_f)$ fixed.}\label{fig:leg1}
     \end{subfigure}\hfill
     \begin{subfigure}{0.32\textwidth}
         \centering
         \includegraphics[width=1\textwidth]{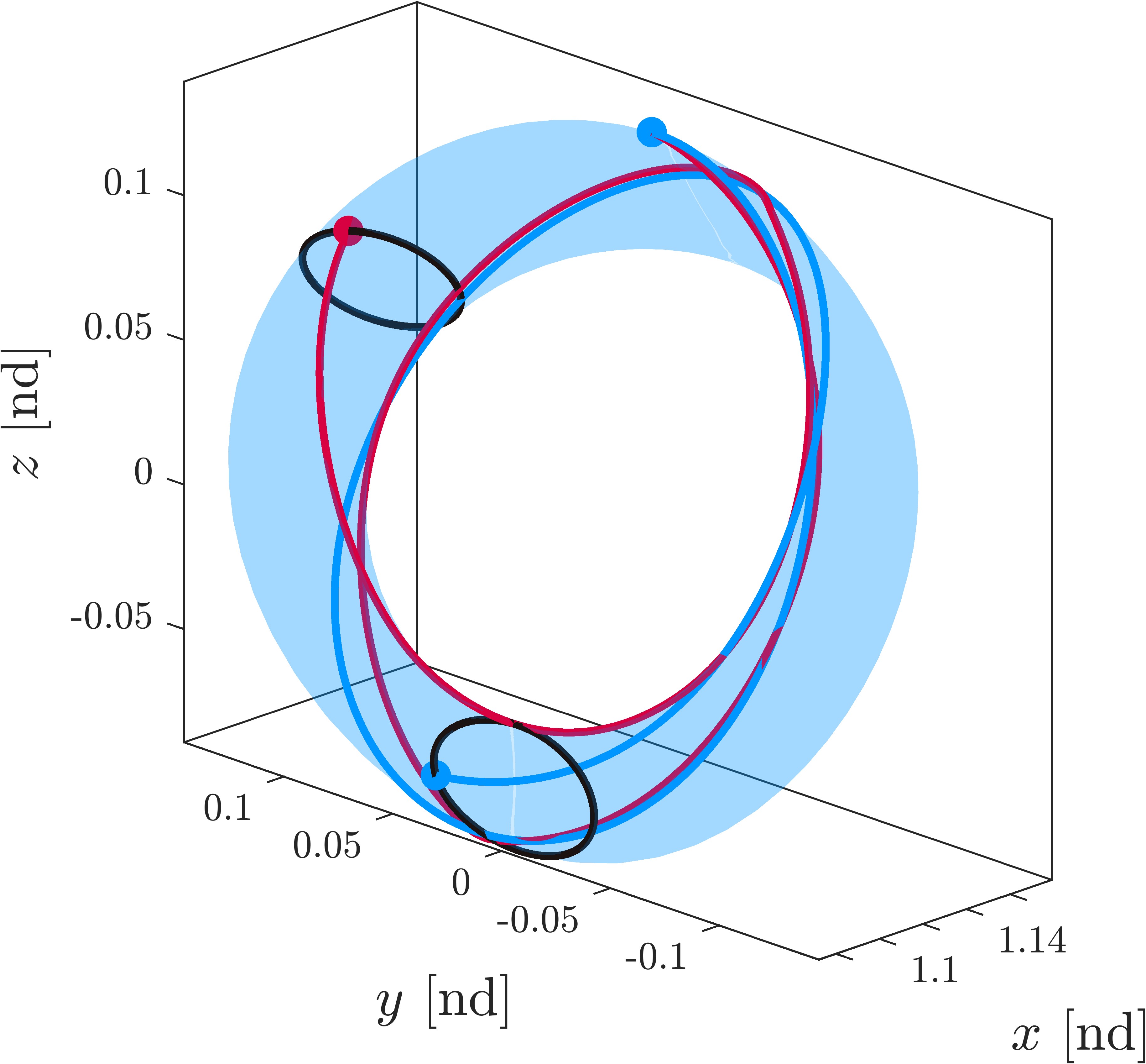}\\
         \includegraphics[width=1\textwidth]{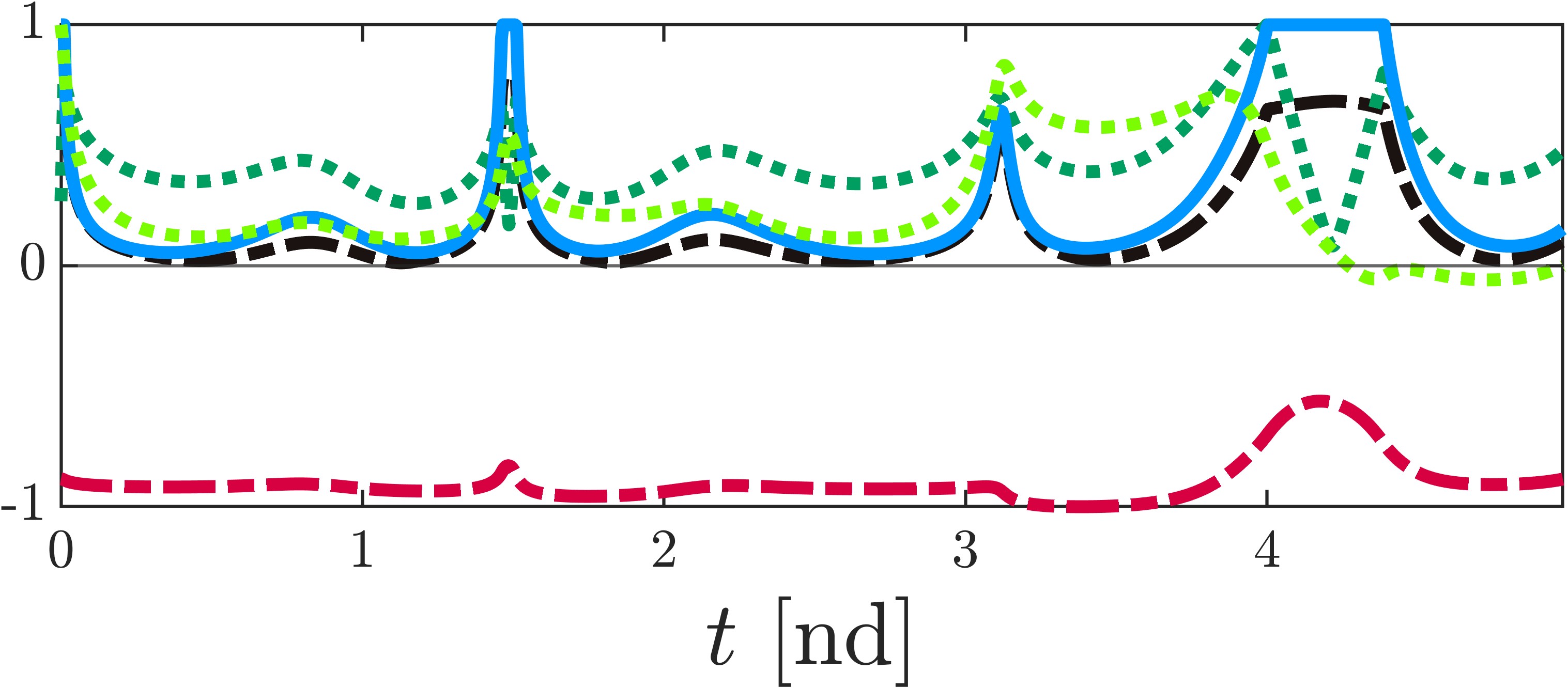}
         \caption{CMVE $\theta_2(t_f)$ free.}
     \end{subfigure}\hfill
     \begin{subfigure}{0.32\textwidth}
         \centering
         \includegraphics[width=1\textwidth]{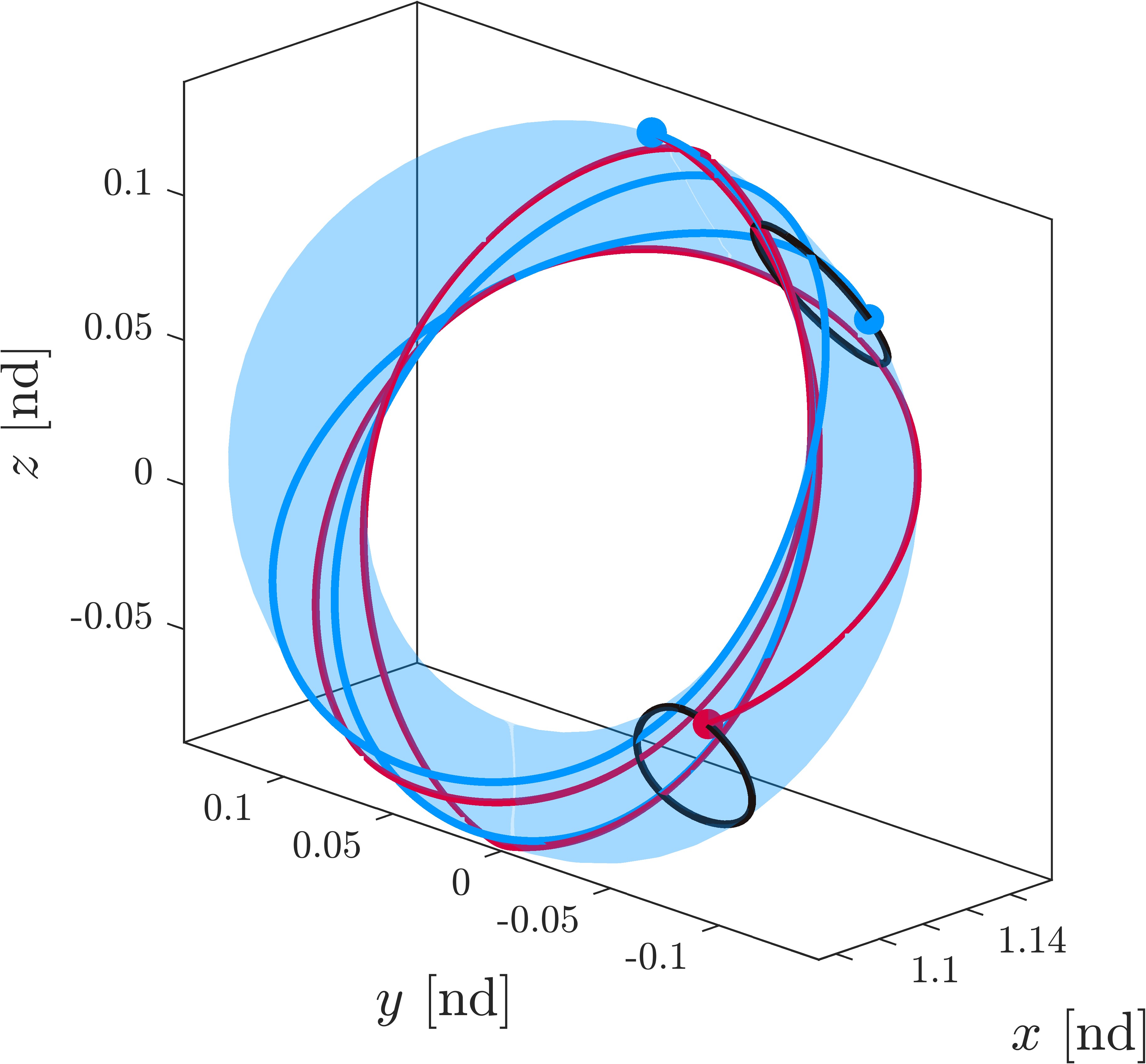}\\
         \includegraphics[width=1\textwidth]{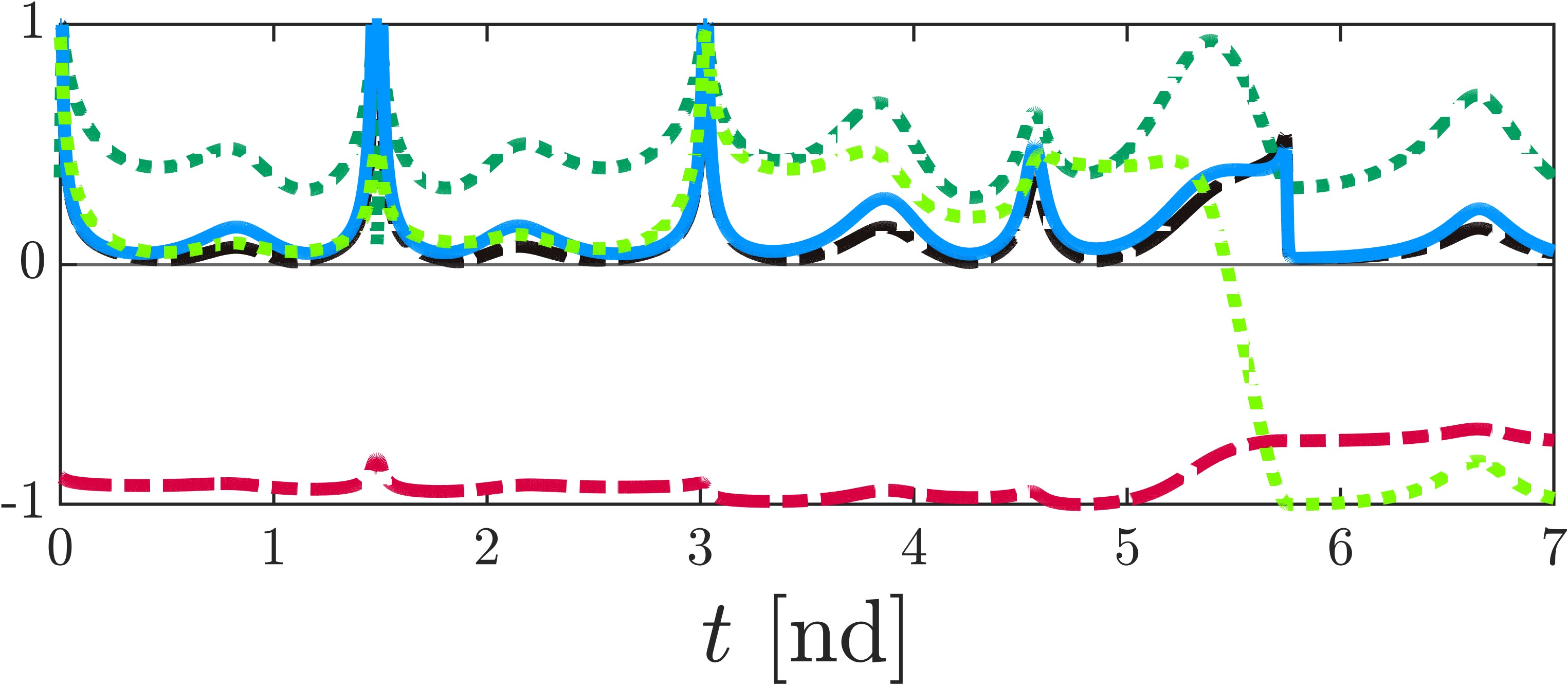}
         \caption{CMVE $t_f$ free.}
     \end{subfigure} \\
     \begin{subfigure}{0.32\textwidth}
         \centering
         \includegraphics[width=1\textwidth]{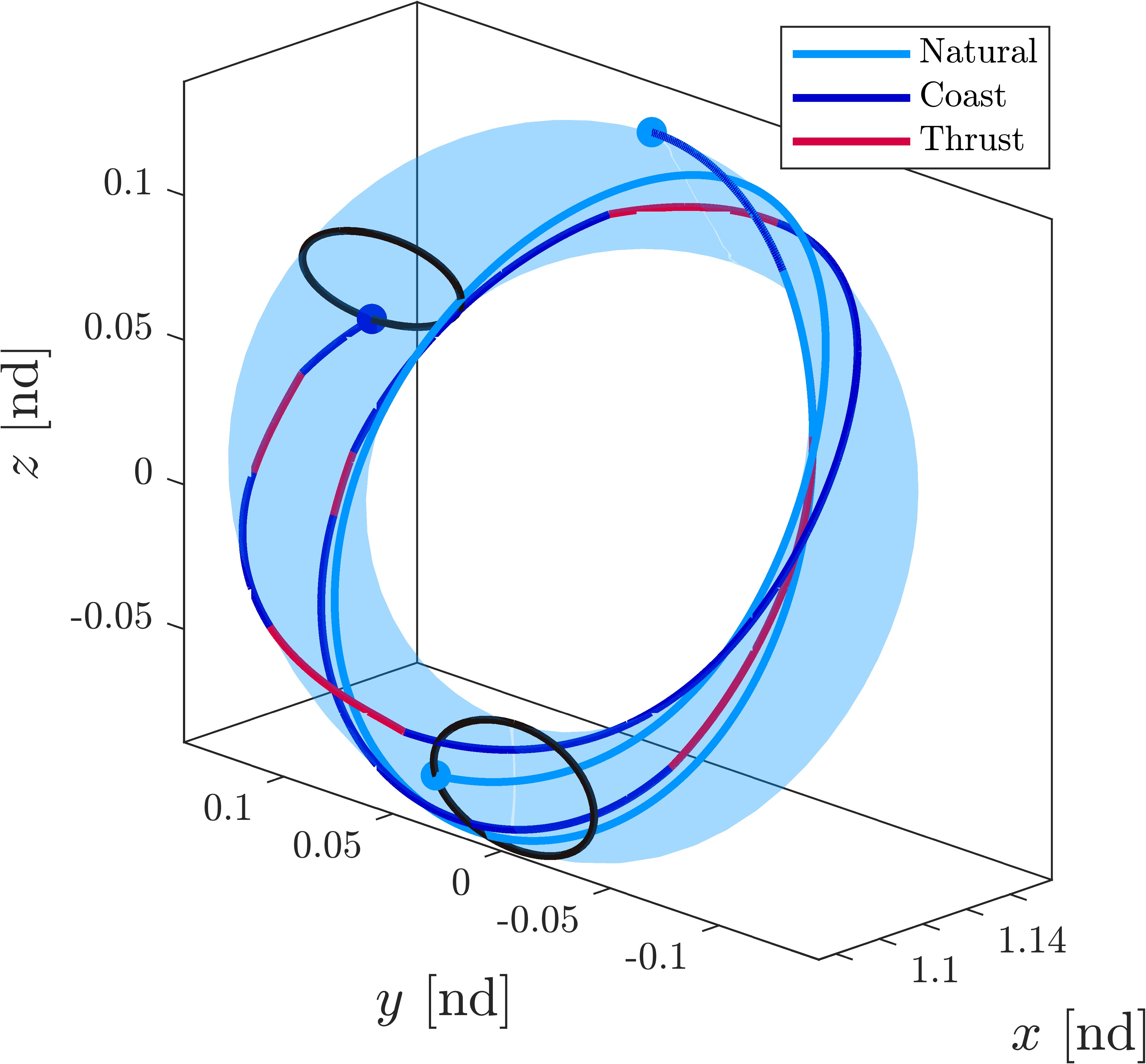}\\
         \includegraphics[width=1\textwidth]{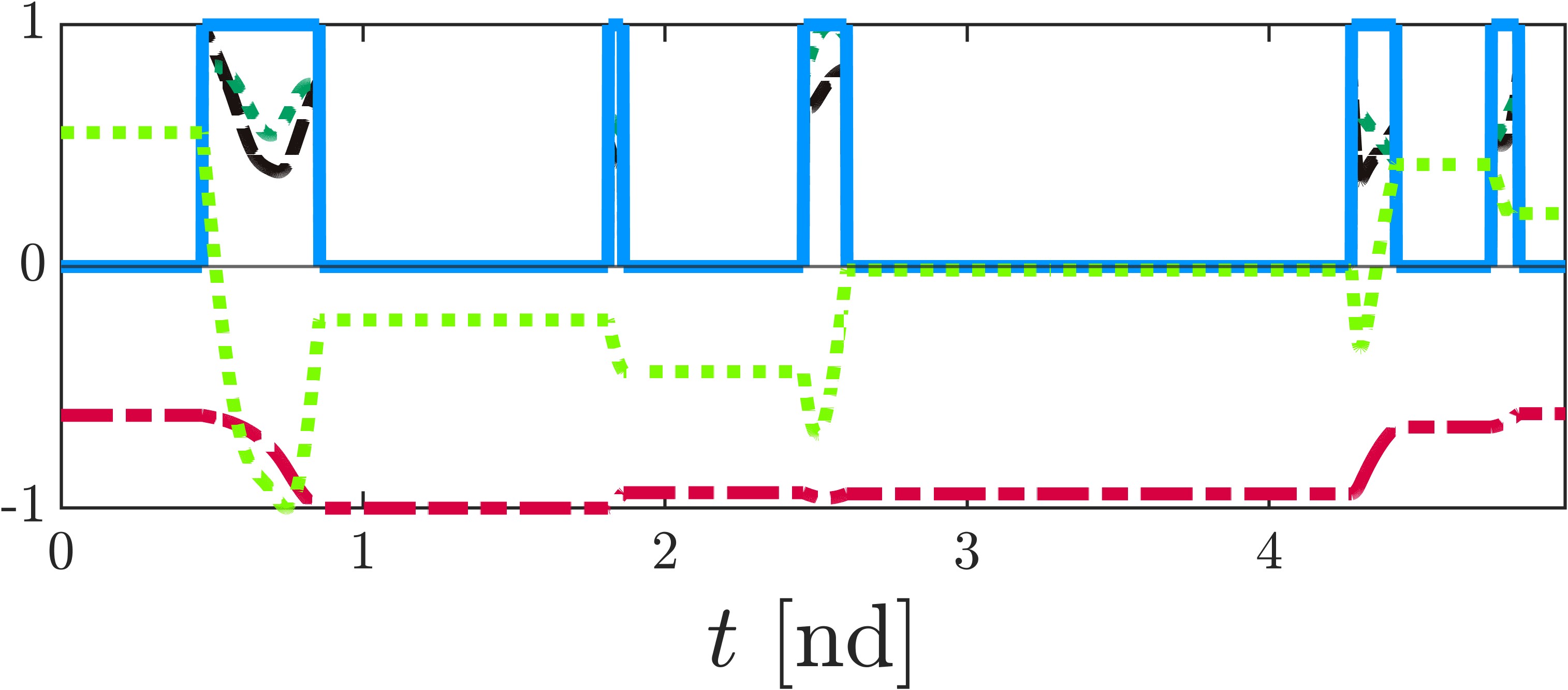}
         \caption{CMVT $\boldsymbol{\theta}(t_f)$ fixed.}\label{fig:leg2}
     \end{subfigure}\hfill
     \begin{subfigure}{0.32\textwidth}
         \centering
         \includegraphics[width=1\textwidth]{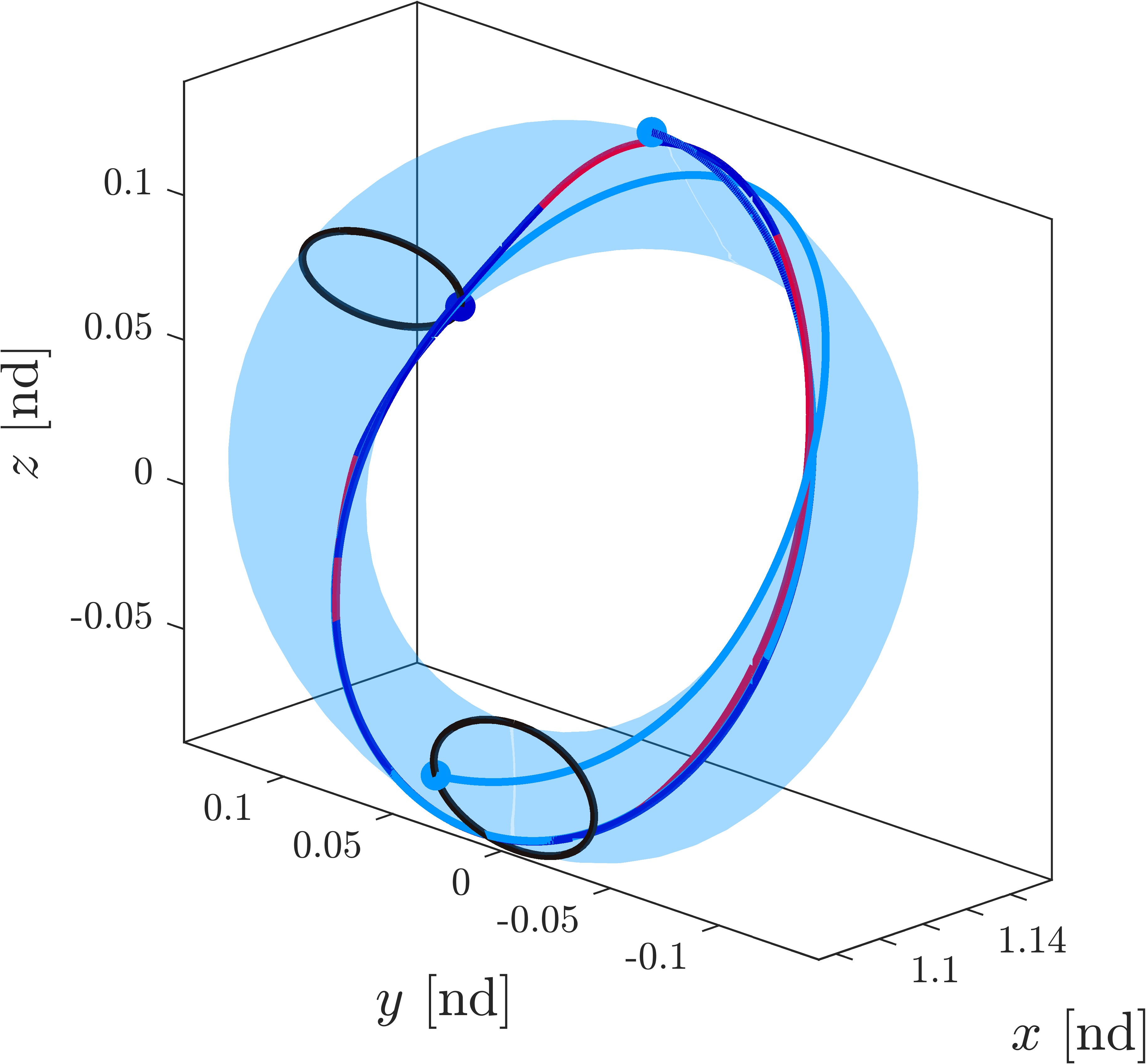}\\
         \includegraphics[width=1\textwidth]{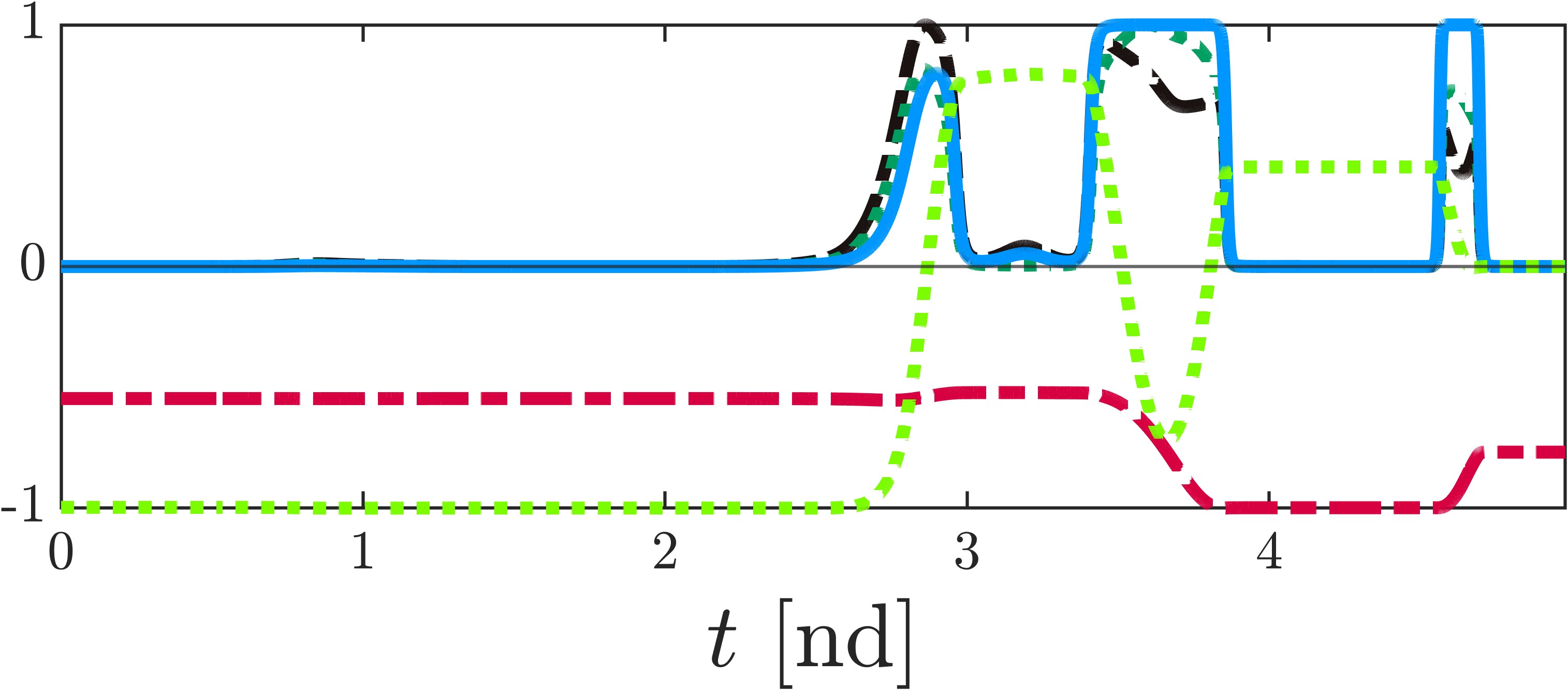}
         \caption{CMVT $\theta_2(t_f)$ free.}
     \end{subfigure}\hfill
     \begin{subfigure}{0.32\textwidth}
         \centering
         \includegraphics[width=1\textwidth]{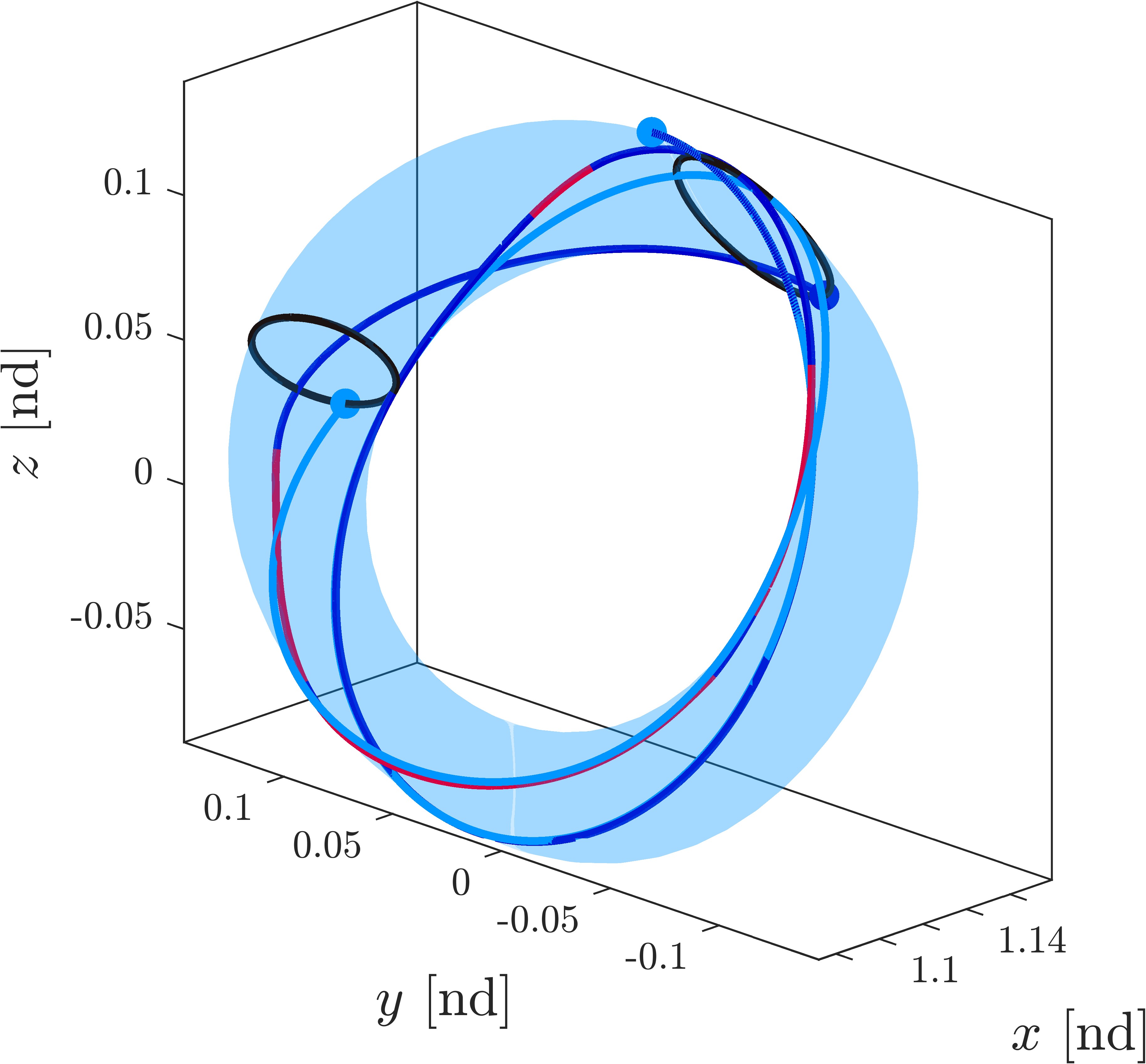}\\
         \includegraphics[width=1\textwidth]{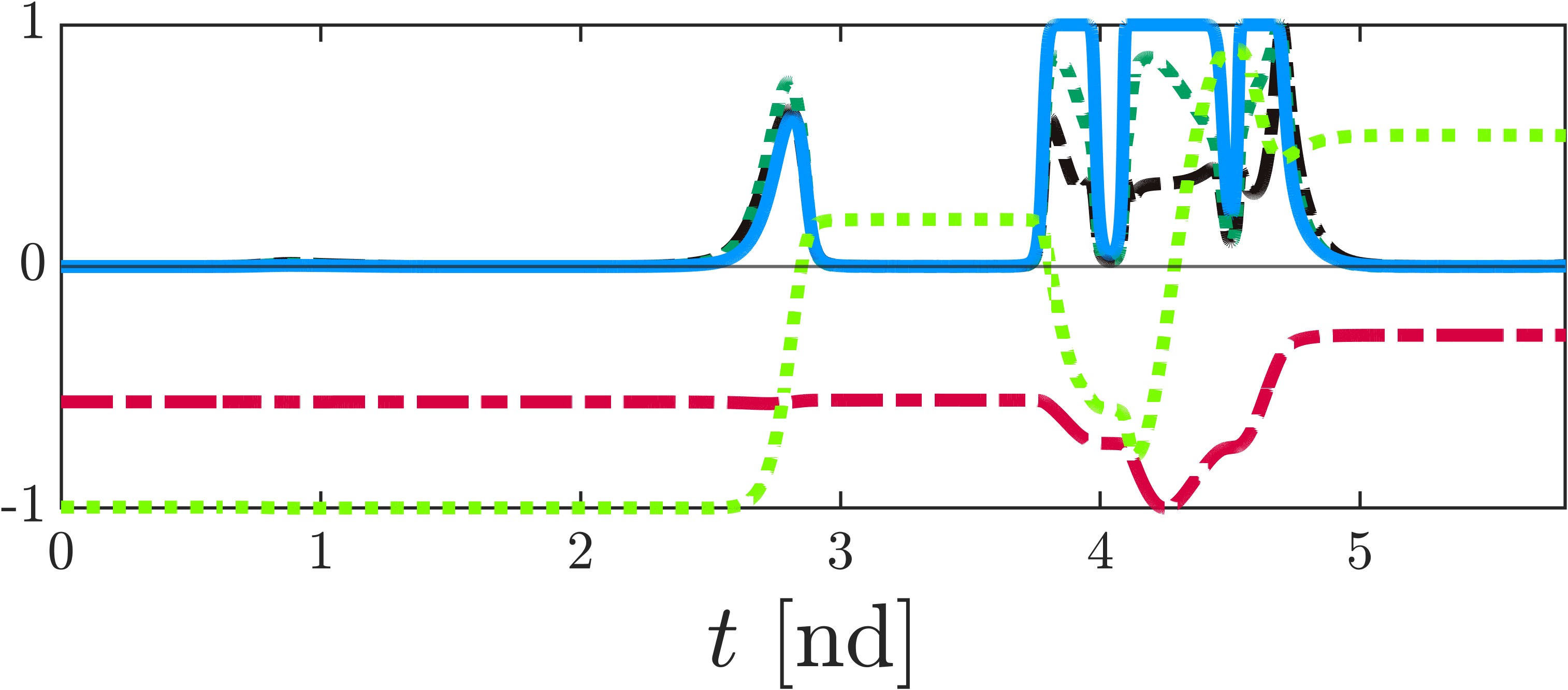}
         \caption{CMVT $t_f$ free.}
     \end{subfigure}
        \caption{Torus space solutions for rephasing with mixed boundary conditions on a 2-dimensional QPIT in the CR3BP. \rev1{Legends from Figs. \ref{fig:leg1} and \ref{fig:leg2} apply to all subplots. Time histories are normalized to their respective maximum absolute values.}}
        \label{fig:L1plots}
\end{figure}

Examples for all three boundary conditions for both the CMVE and CMVT torus space optimal control problems are shown in Fig. \ref{fig:L1plots} for the quasi-halo torus approximated in Fig. \ref{fig:approx}. \rev1{Trajectories from the torus space are visualized in configuration space via the torus function, but do not represent immediately physically realizable solutions because of nonzero control frequency.} Initial torus angles are $\boldsymbol{\theta}=[0,\,0]^\text{T}$ for all examples. Light blue trajectories correspond to the natural torus motion without the introduction of control, while red trajectories correspond to non-zero control frequency, and dark blue to coast arcs during transfer windows. Costates, control frequency magnitude, and mapped phase space control magnitudes are all normalized to $[-1,\,1]$. $M=10$ multiple shooting segments were used in computation. A value of $u_{\dot{q},\text{max}}=0.5$ was used to limit control input, which corresponds to about 1.36 mm/s$^2$ of control acceleration in the Earth-Moon CR3BP. A time window of $t_f\in[3,\,7]$ [nd] was considered for the free final time rendezvous examples. Numerical results of each case appear in Table \ref{tab:level1}.
\begin{table}[htbp!]
\caption{\label{tab:level1} Numerical results for rephasing optimal control problems in the torus space.}
\centering
\begin{tabular}{lccccc}
\hline
Problem & Boundary Conditions & $J\times10^{-3}$ & $\theta_1(t_f)$ & $\theta_2(t_f)$ & $t_f$ \\\hline
CMVE & Fixed $\boldsymbol{\theta}_d(t_f)$& 2.0455 & 5.0256 & 2.4982 & 4.981\\
& Free $\theta_2(t_f)$& 1.2906& 5.0256& 6.1660 & 4.981\\
& Free $t_f$, Fixed $\boldsymbol{\theta}_d(t_0)$& 1.2373& 2.5642& 5.7398& 7.000\\\hline
CMVT& Fixed $\boldsymbol{\theta}_d(t_f)$& 9.8865 & 5.0265& 2.4982& 4.981\\
& Free $\theta_2(t_f)$& 7.1354 & 5.0265& 4.0971& 4.981\\
& Free $t_f$, Fixed $\boldsymbol{\theta}_d(t_0)$& 9.2436 & 0.2794& 3.8014& 5.793\\
\hline
\end{tabular}
\end{table}

Leaving $\theta_2$ and $t_f$ free in the transfers leads to lower cost for both optimal control problems. The cost values for CMVE and CMVT examples with the same boundary conditions are not directly comparable because they have differing non-dimensional units. The fixed boundary condition solution to the CMVT problem was successfully continued down to $\rho=1\times10^{-6}$, resulting in clearly defined switching times. The remaining two CMVT problems saw numerical continuation difficulty at $\rho=1.38\times10^{-4}$ and $\rho=1.75\times10^{-4}$, respectively, but still illicit a switching structure that can be transitioned at those values.

\begin{figure}[htpb!]
    \centering
    \includegraphics[width=0.4\linewidth]{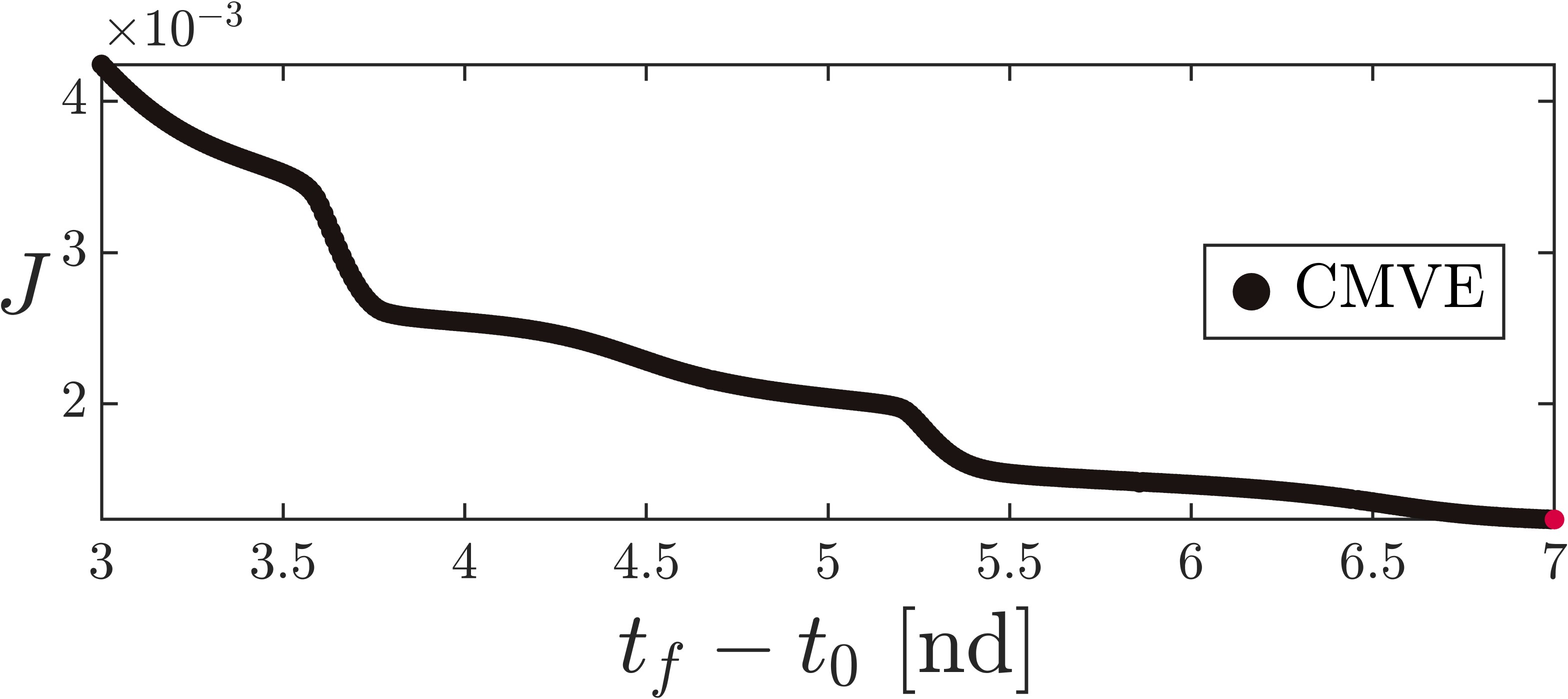}
    \includegraphics[width=0.4\linewidth]{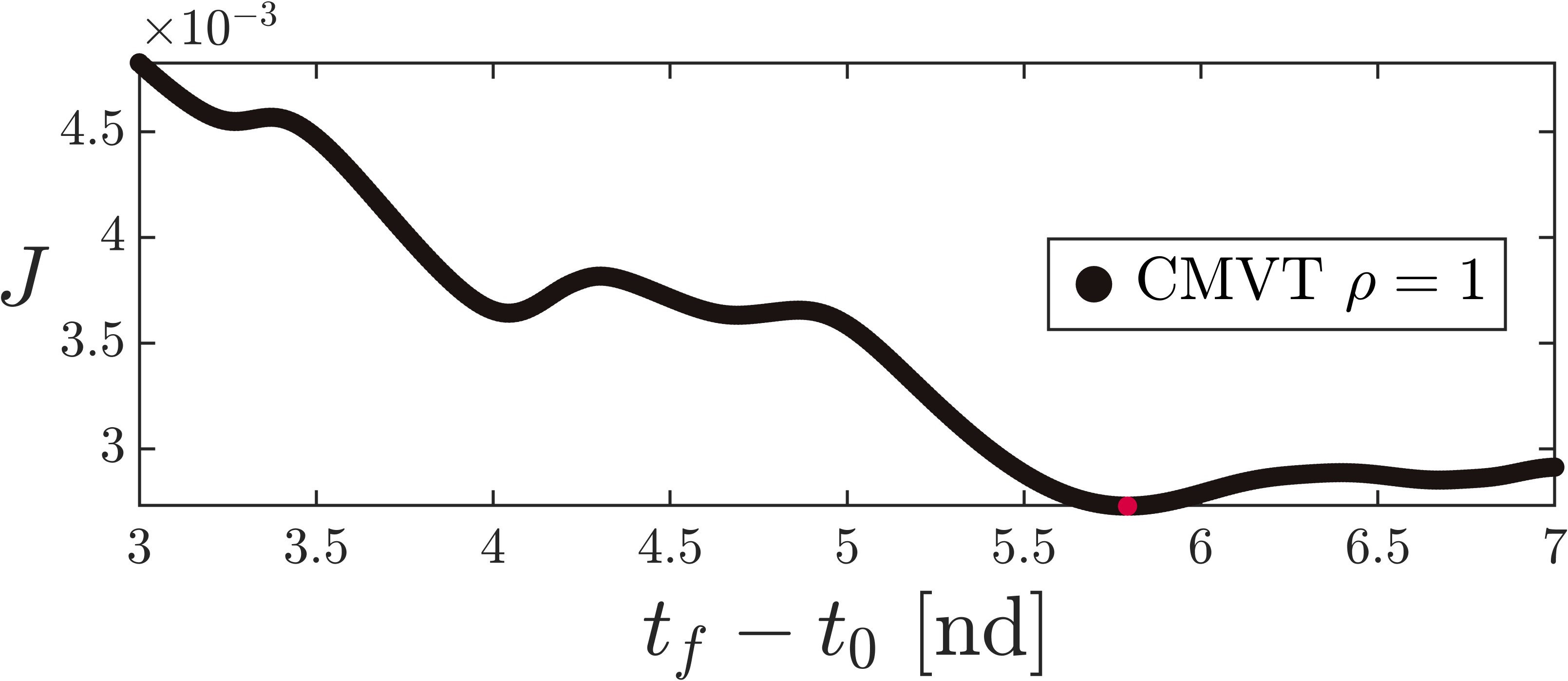}
    \caption{$t_f$ free rendezous rephasing cost.}
    \label{fig:L1costs}
\end{figure}

The cost function sweep over the final time domain for the free final time rendezvous example are shown for both CMVE and CMVT in Fig. \ref{fig:L1costs}. The CMVE solution space is monotonically decreasing, so the final time domain boundary is taken as the set's minimum. In contrast, the CMVT solution space has five local minima. Both solution spaces are smooth, differentiable curves indicating well-behaved problems. 

With torus space solutions obtained, they must now be carried over into the phase space to be evaluated as potential solutions to the rephasing problem.

%%%%%%%%%%%%%%%%%%%%%%%%%%%%%%%%%%%%%%%%%%%%
\subsection{Solution Transition to the Phase Space}
\begin{figure}[htbp!]
     \centering
     \begin{subfigure}{0.49\textwidth}
         \centering
         \includegraphics[width=0.48\textwidth]{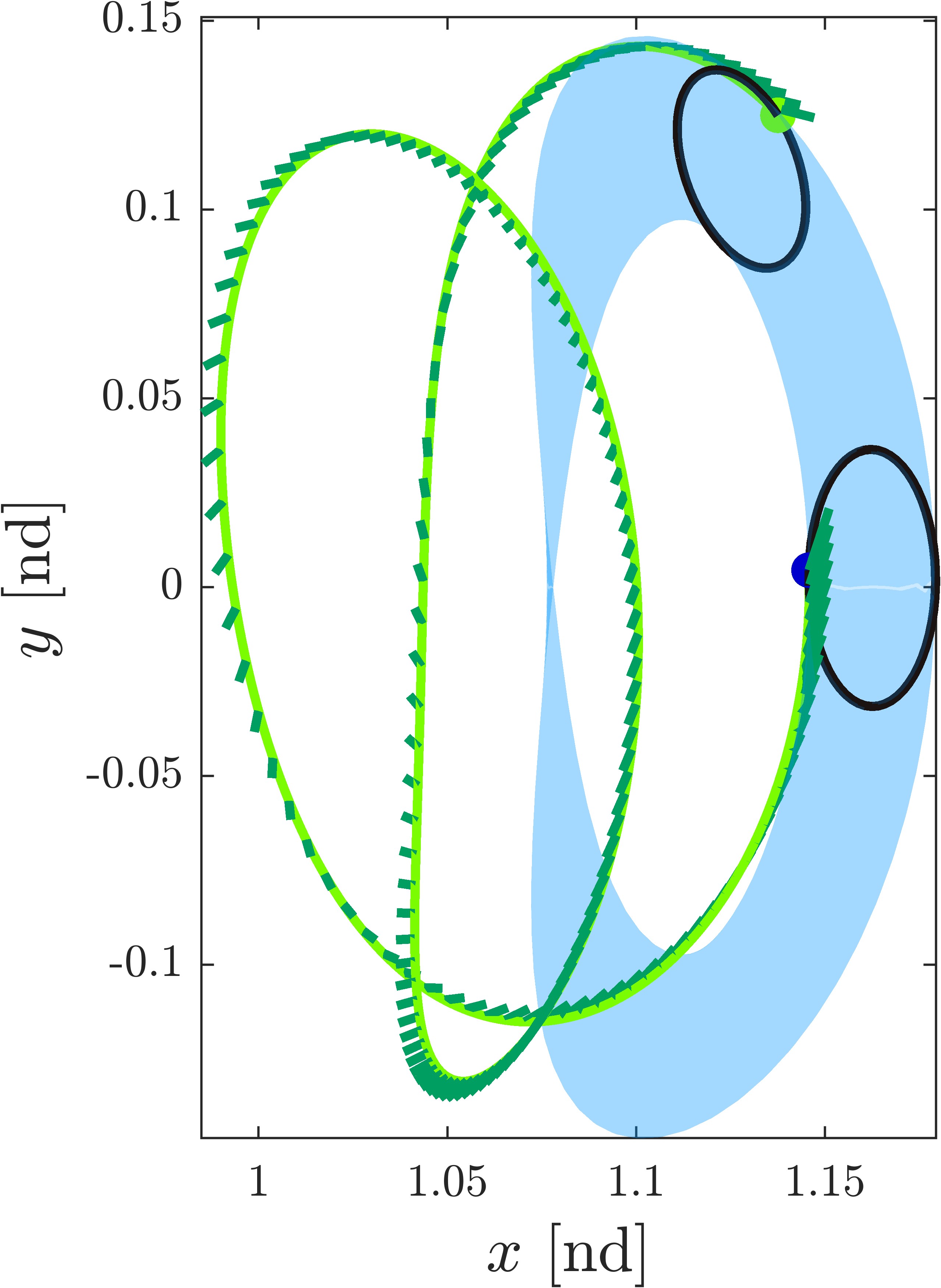}\hfill
         \includegraphics[width=0.463\textwidth]{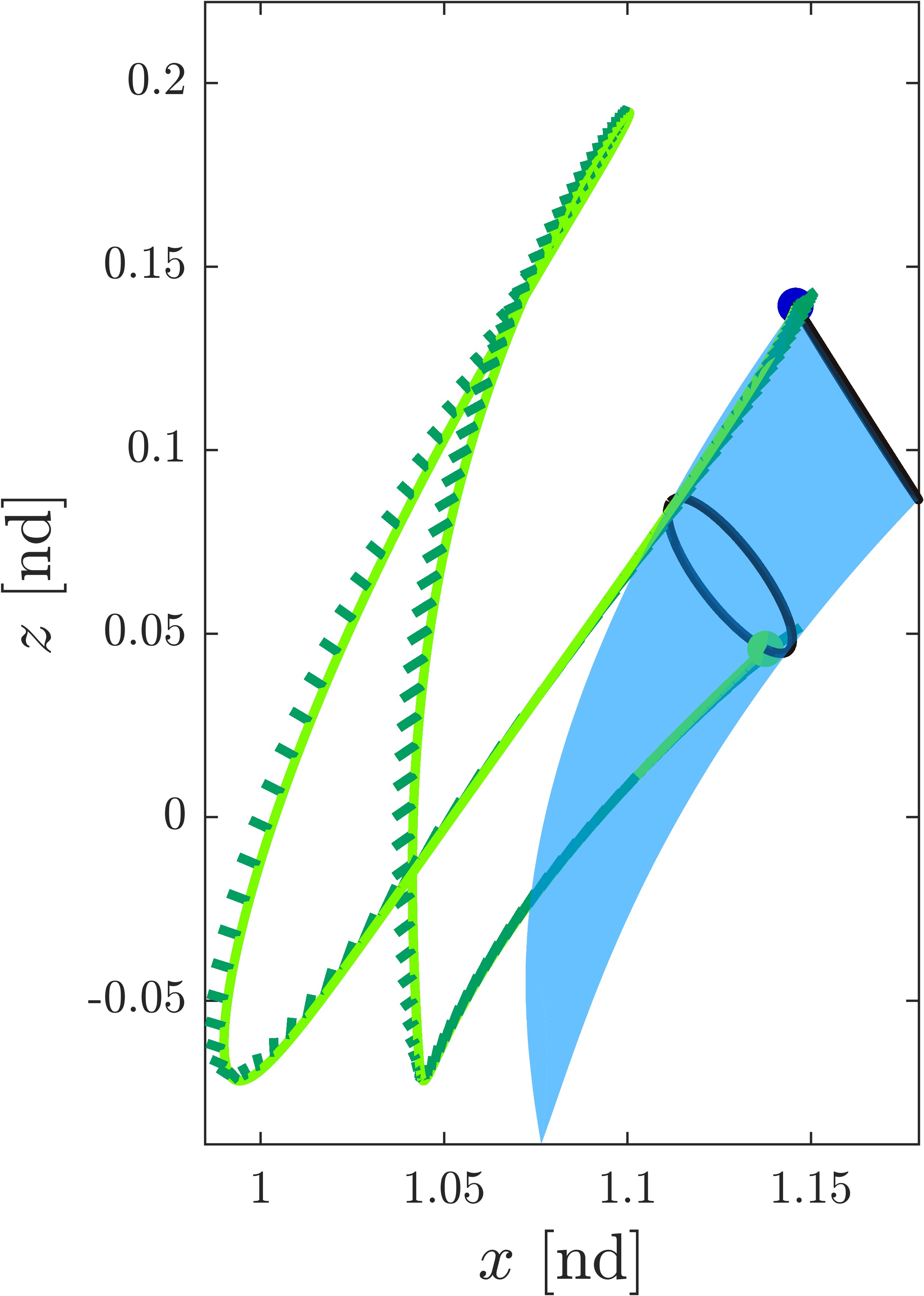}
         \caption{CME Solution}
     \end{subfigure} \hfill
     \begin{subfigure}{0.49\textwidth}
         \centering
         \includegraphics[width=0.48\textwidth]{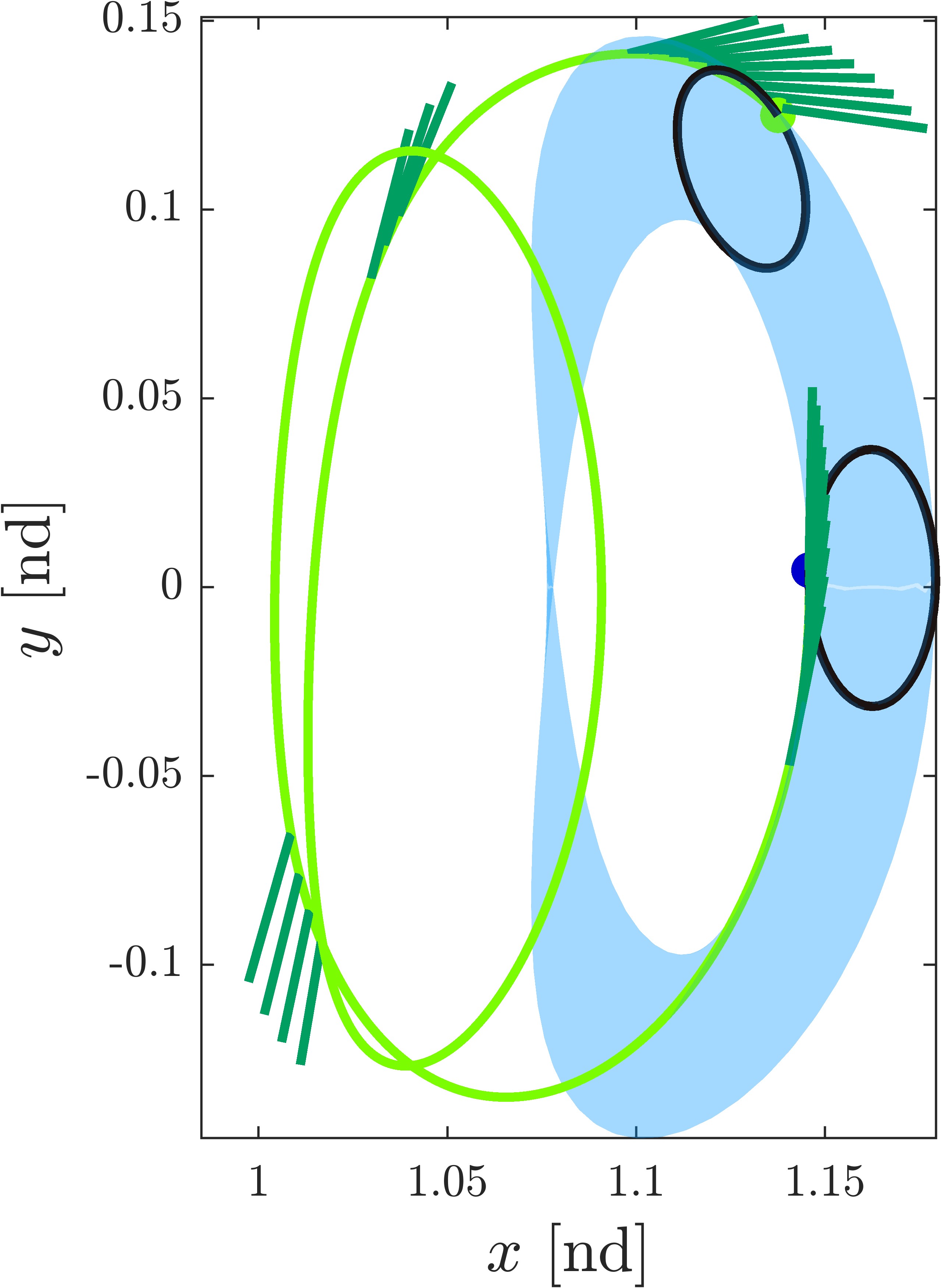}\hfill
         \includegraphics[width=0.463\textwidth]{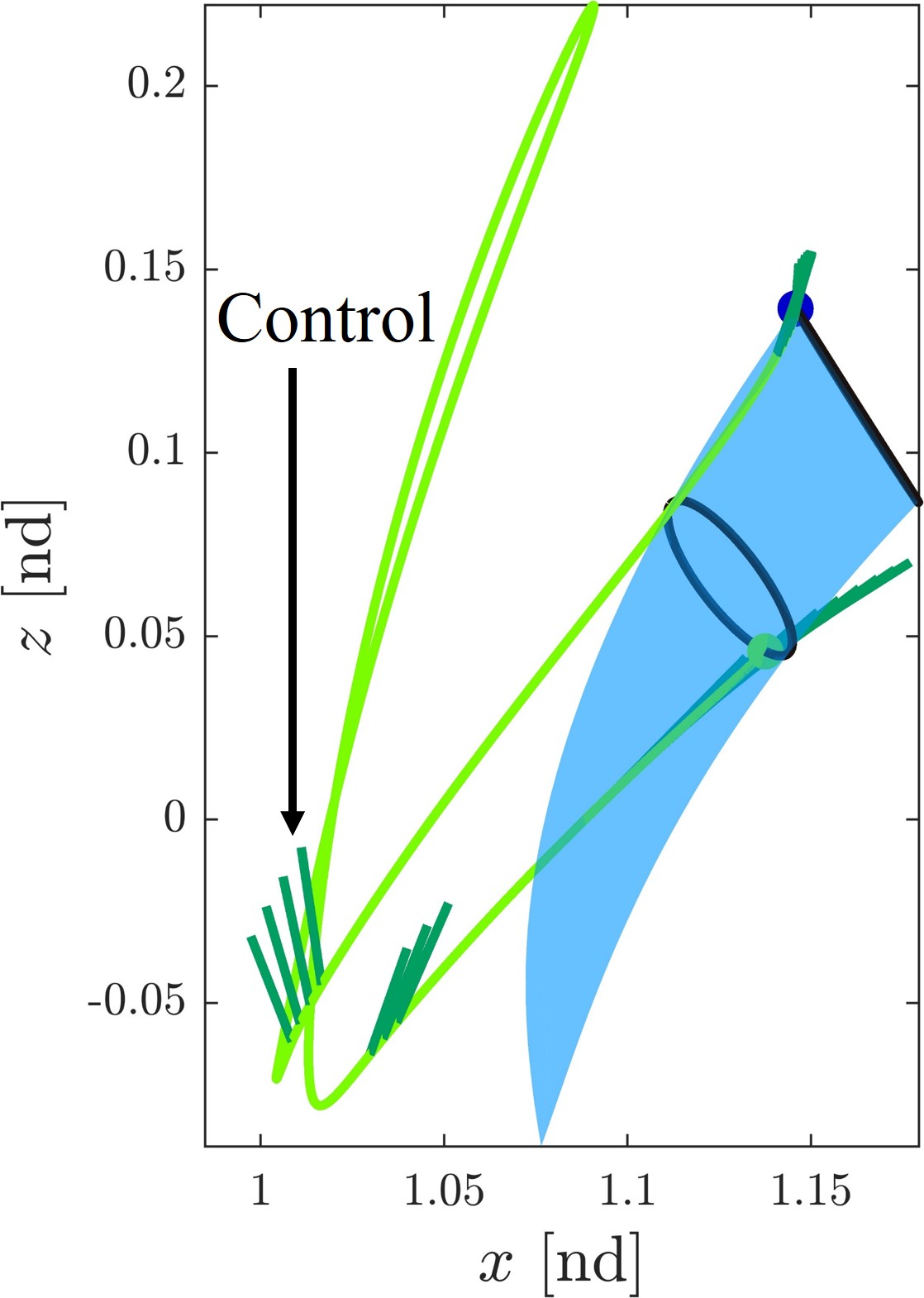}
         \caption{CMT Solution}
     \end{subfigure}
        \caption{Torus agnostic solutions for rephasing with fixed boundary conditions on a 2-dimensional QPIT in the CR3BP.}
        \label{fig:L2torusagnostic}
\end{figure}

Torus space solutions to both the CMVE and CMVT problems in the first level of optimization give reference trajectories $\boldsymbol{\theta}^*(t)$ that can be mapped into the phase space through the torus function. However, the nature of each torus space solution method warrants separate treatment on how these reference trajectories are transitioned to physically available solutions in the phase space. For comparison, two \rev1{classical control} solutions to the rephasing problem are obtained from the constrained minimum energy (CME) and constrained minimum throttle (CMT) problems. CME considers the quadratic cost function $J = \int_{t_0}^{t_f}u_{\dot{q}}^2\:\text{d}t$ while CMT considers the $\mathcal{L}_2$ norm cost function $J = \int_{t_0}^{t_f}u_{\dot{q}}\:\text{d}t$ for $u_{\dot{q}}>0$. Similar to the CMVT problem, the CMT problem produces a bang-off-bang control structure. \rev1{These solutions are referred to as ``torus agnostic" because they don't consider torus deviation by construction.} Both torus agnostic solutions to the fixed boundary condition problem in Fig. \ref{fig:L1plots} are shown in Fig. \ref{fig:L2torusagnostic}. \rev1{Control directions and magnitudes are plotted along the trajectories in dark green.} Note that both exhibit significant deviation from the original torus, with the CMT solution naturally leveraging a nearby halo structure during its long duration coast arc. All transitioned solutions in this section are obtained via indirect, multiple shooting with two segments. Individual patch transitions of the CMVT reference trajectory only require single shooting.

% \begin{figure}[htbp!]
%      \centering
%      \begin{subfigure}{1\textwidth}
%          \centering
%          \includegraphics[width=0.38\textwidth]{Figures/L2_Plots/L2CME_CSiso.jpg}\hfill
%          \includegraphics[width=0.30\textwidth]{Figures/L2_Plots/L2CME_CSbird.jpg}\hfill
%          \includegraphics[width=0.29\textwidth]{Figures/L2_Plots/L2CME_CSside.jpg}
%          \caption{CME Solution}
%      \end{subfigure}
%      \\
%      \begin{subfigure}{1\textwidth}
%          \centering
%          \includegraphics[width=0.38\textwidth]{Figures/L2_Plots/L2CMT_CSiso.jpg}\hfill
%          \includegraphics[width=0.30\textwidth]{Figures/L2_Plots/L2CMT_CSbird.jpg}\hfill
%          \includegraphics[width=0.29\textwidth]{Figures/L2_Plots/L2CMT_CSside.jpg}
%          \caption{CMT Solution}
%      \end{subfigure}
%         \caption{Torus agnostic solutions for rephasing with fixed boundary conditions on a 2-dimensional QPIT in the CR3BP.}
%         \label{fig:L2torusagnostic}
% \end{figure}

\subsubsection{Constrained Minimum Energy Tracking Error Homotopy}
The CMVE solution in the torus space assumes continuous and variable control input. To best track this in the phase space, a solution should also have this control structure. Because the control quantity $\boldsymbol{u}_q$ is non-zero over this reference trajectory, its corresponding phase space coordinate and coordinate rate components can never be simultaneously tracked with zero error given finite control input at the acceleration level. Thus, only the coordinate component of the state is used in the following error state definition.
\begin{equation}
    \boldsymbol{e}(t) = \boldsymbol{q}(t) - \boldsymbol{\gamma}(\boldsymbol{\theta}^*(t)) \label{eq:errordef}
\end{equation}

A new cost function is defined to minimize Eq. \ref{eq:errordef}. This cost function is connected to the standard CME problem via the homotopic parameter $\varepsilon\in[0,1)$ in order to qualitatively match the torus space control structure and assist in solution convergence. The constrained minimum energy tracking error homotopy (CMETEH) problem is now stated as
\begin{align}
    \text{min}\quad &J = \int_{t_0}^{t_f}\frac{1}{2}\left[(1-\varepsilon)u_{\dot{q}}^2 + \varepsilon\boldsymbol{e}^\text{T}\boldsymbol{e}\right]\:\text{d}t \\
    \text{s.t.}\quad &\dot{\boldsymbol{x}}=\boldsymbol{f}(\boldsymbol{x},\boldsymbol{\mu})+\boldsymbol{B}\boldsymbol{u}_{\dot{q}},\quad \left|\left|\boldsymbol{u}_{\dot{q}}\right|\right| \leq u_{\dot{q},\text{max}} \nonumber \\
    &\boldsymbol{x}(t_0) = \boldsymbol{\Gamma}(\boldsymbol{\theta}^*(t_0))\nonumber \\
    &\boldsymbol{x}(t_f) = \boldsymbol{\Gamma}(\boldsymbol{\theta}^*(t_f)) \nonumber
\end{align}
where $\boldsymbol{B}=\left[\boldsymbol{0}_{\bar{n}\times\bar{n}}\,\,\,\,\mathbb{I}_{\bar{n}\times\bar{n}}\right]^\text{T}$. A value $\varepsilon=0$ corresponds to the CME solution, while $\varepsilon=1$ corresponds to the minimum tracking error solution. 

The problem's Hamiltonian is written as
\begin{equation}
    \mathcal{H} = \frac{1}{2}\left[(1-\varepsilon)u_{\dot{q}}^2 + \varepsilon\boldsymbol{e}^\text{T}\boldsymbol{e}\right] + \boldsymbol{\lambda}_{x}^\text{T}\left[\boldsymbol{f}(\boldsymbol{x})+\boldsymbol{B}\boldsymbol{u}_{\dot{q}}\right]
\end{equation}
with $\boldsymbol{\lambda}_x^\text{T}=[\boldsymbol{\lambda}_q^\text{T}\,\,\,\,\boldsymbol{\lambda}_{\dot{q}}^\text{T}]$. The control structure and costate dynamics resulting from the indirect problem's necessary conditions are
\begin{equation}
    \widehat{\boldsymbol{u}}_{\dot{q}} = -\widehat{\boldsymbol{\lambda}}_{\dot{q}}
\end{equation}
\begin{equation}
    u_{\dot{q}} = 
    \begin{dcases}
    \frac{1}{(1-\varepsilon)}||\boldsymbol{\lambda}_{\dot{q}}||,& \text{if } \frac{1}{(1-\varepsilon)}||\boldsymbol{\lambda}_{\dot{q}}||\leq u_{{\dot{q}},\text{max}}\\
    u_{{\dot{q}},\text{max}},              & \text{otherwise}
    \end{dcases}
\end{equation}
\begin{equation}
    \dot{\boldsymbol{\lambda}}_{x} = -\left[\frac{\partial \boldsymbol{f}}{\partial \boldsymbol{x}}\right]^\text{T}\boldsymbol{\lambda}_x +
    \varepsilon\begin{bmatrix}
        \boldsymbol{e} \\ \boldsymbol{0}_{\bar{n}}
    \end{bmatrix} \label{eq:costateCMETEH}
\end{equation}

\begin{figure}[htbp!]
     \centering
     \begin{subfigure}{0.216\textwidth}
         \centering
         \includegraphics[width=1\textwidth]{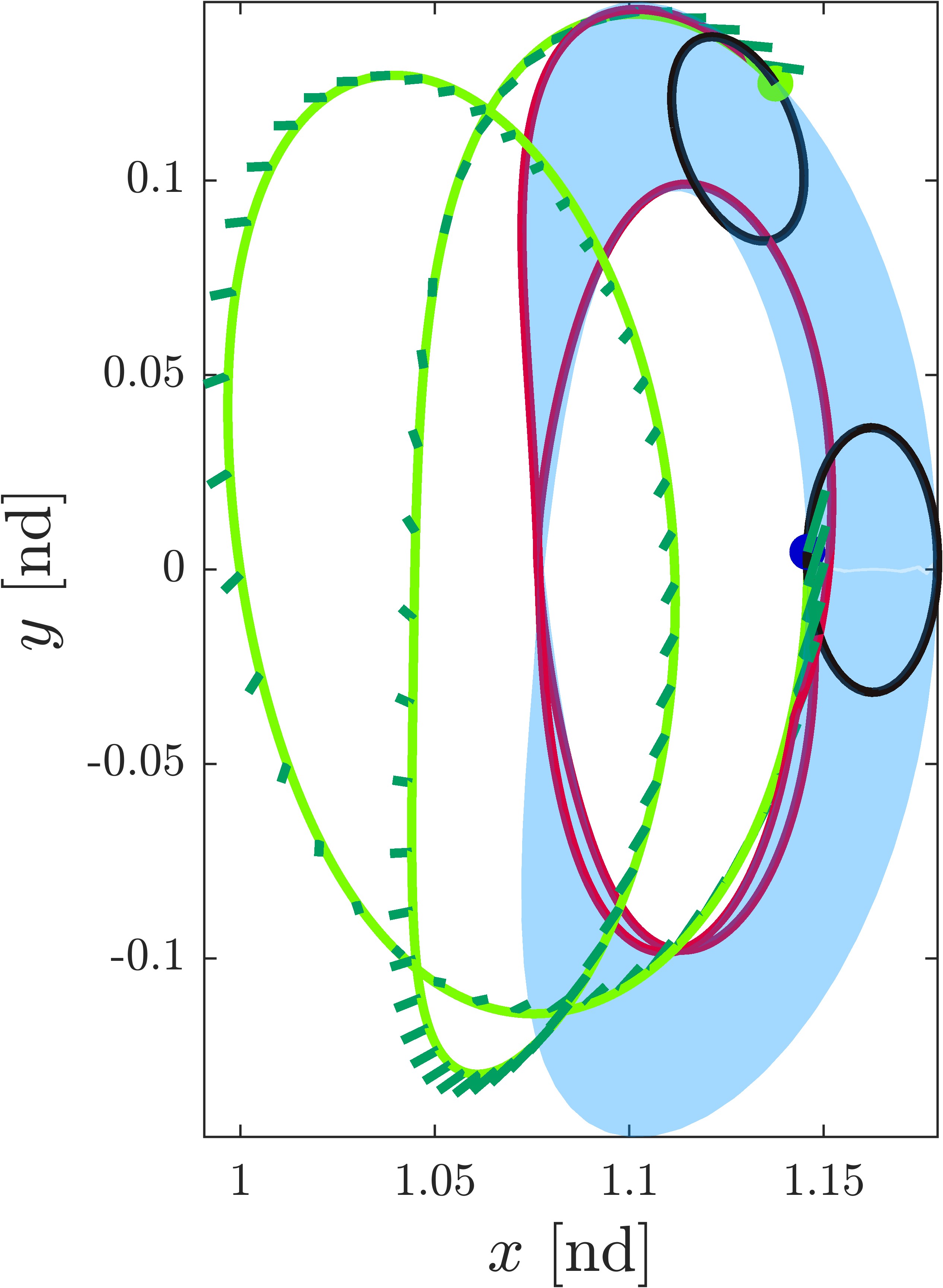}
         \caption{$\varepsilon=0.1$}
     \end{subfigure}\hfill
     \begin{subfigure}{0.203\textwidth}
         \centering
         \includegraphics[width=1\textwidth]{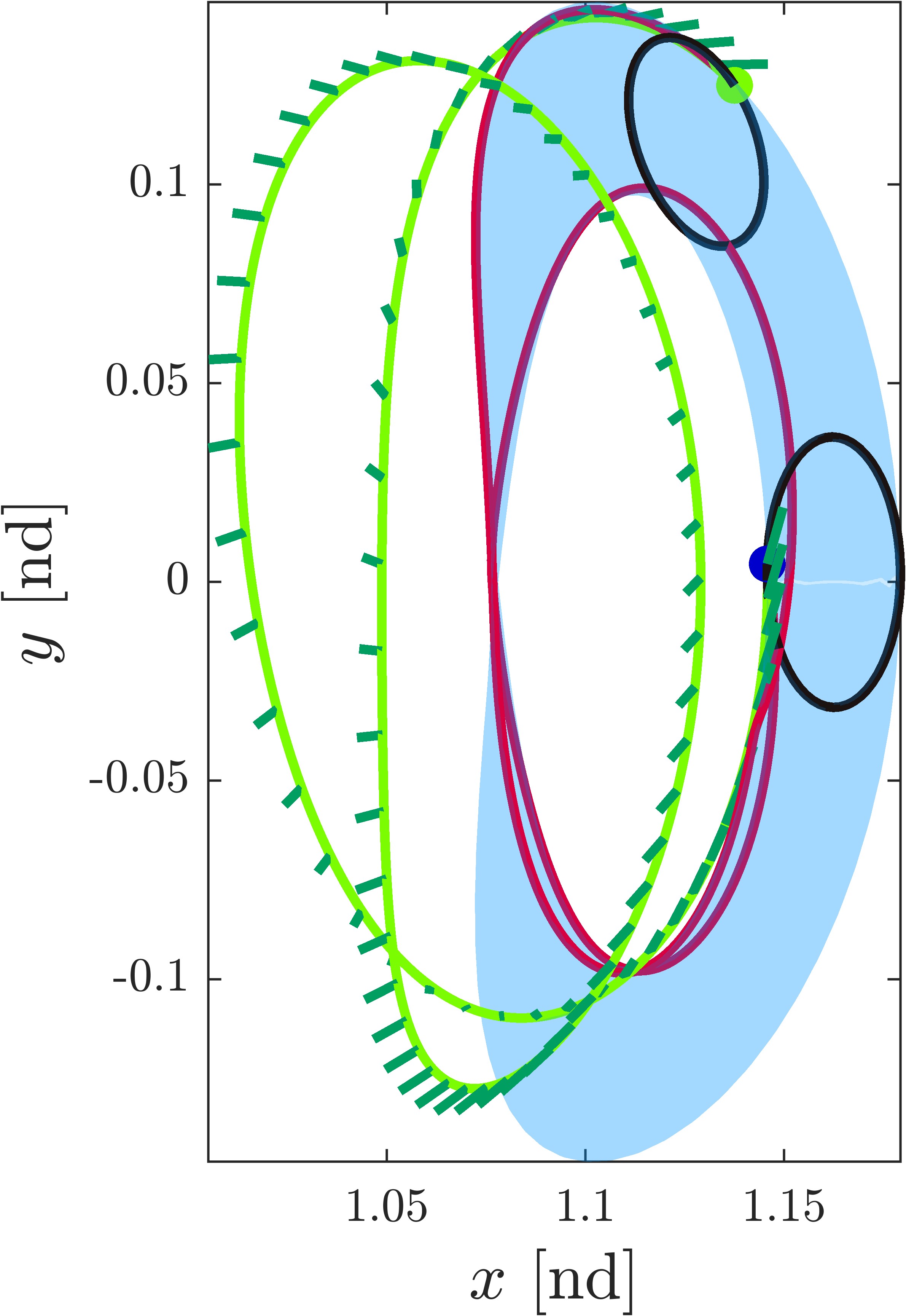}
         \caption{$\varepsilon=0.3$}
     \end{subfigure}\hfill
     \begin{subfigure}{0.190\textwidth}
         \centering
         \includegraphics[width=1\textwidth]{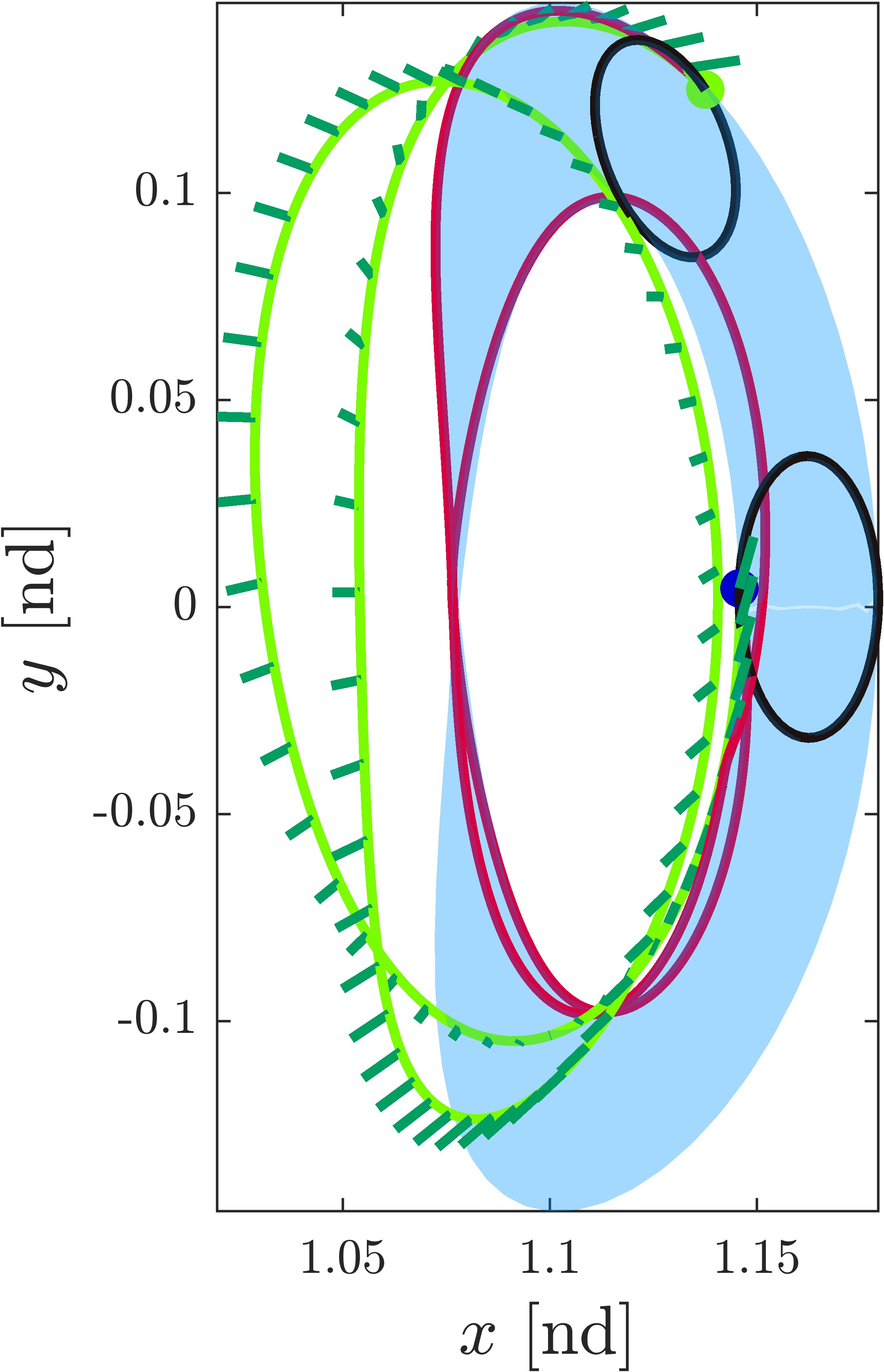}
         \caption{$\varepsilon=0.5$}
     \end{subfigure}\hfill
     \begin{subfigure}{0.178\textwidth}
         \centering
         \includegraphics[width=1\textwidth]{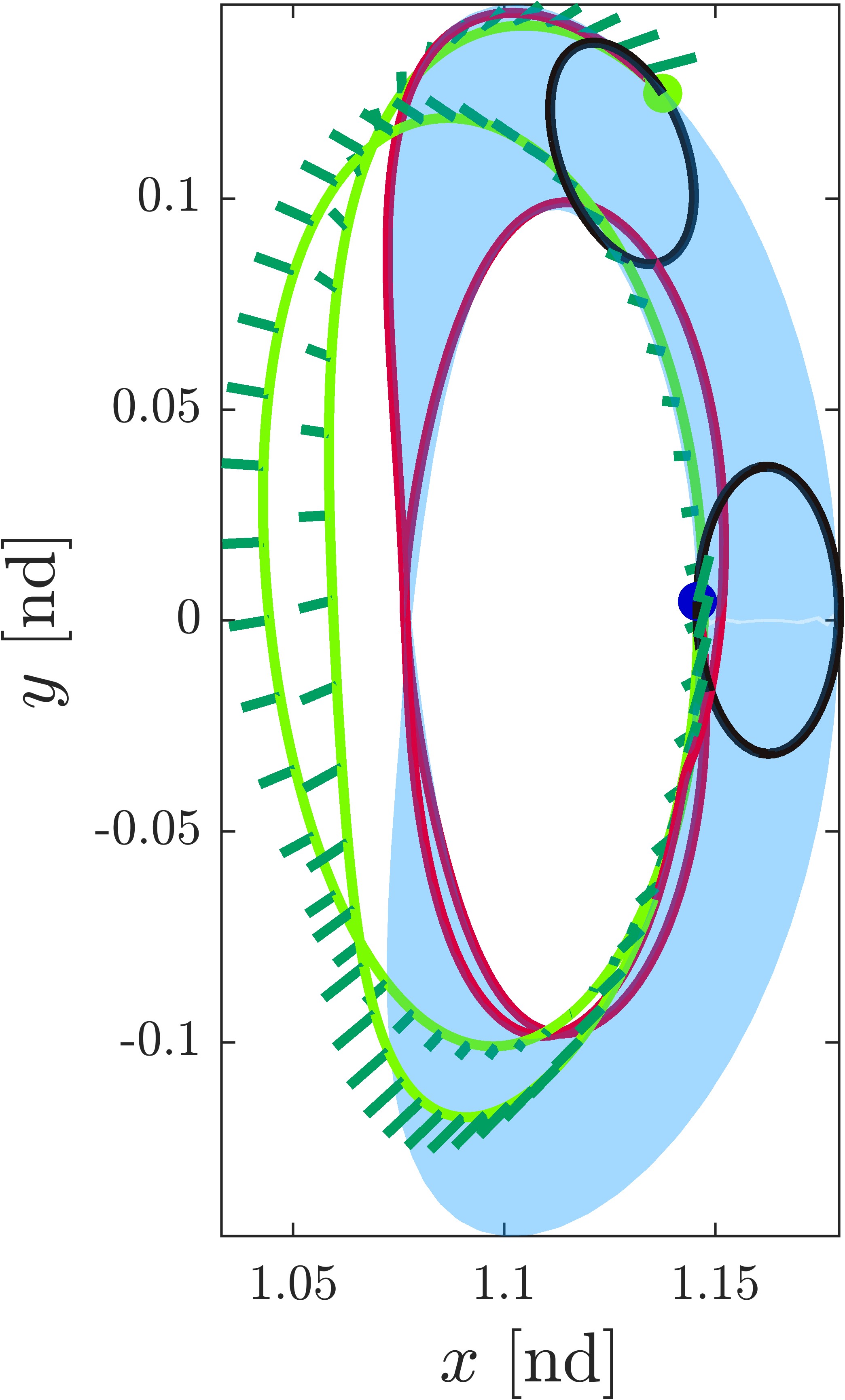}
         \caption{$\varepsilon=0.7$}
     \end{subfigure}\hfill
     \begin{subfigure}{0.1555\textwidth}
         \centering
         \includegraphics[width=1\textwidth]{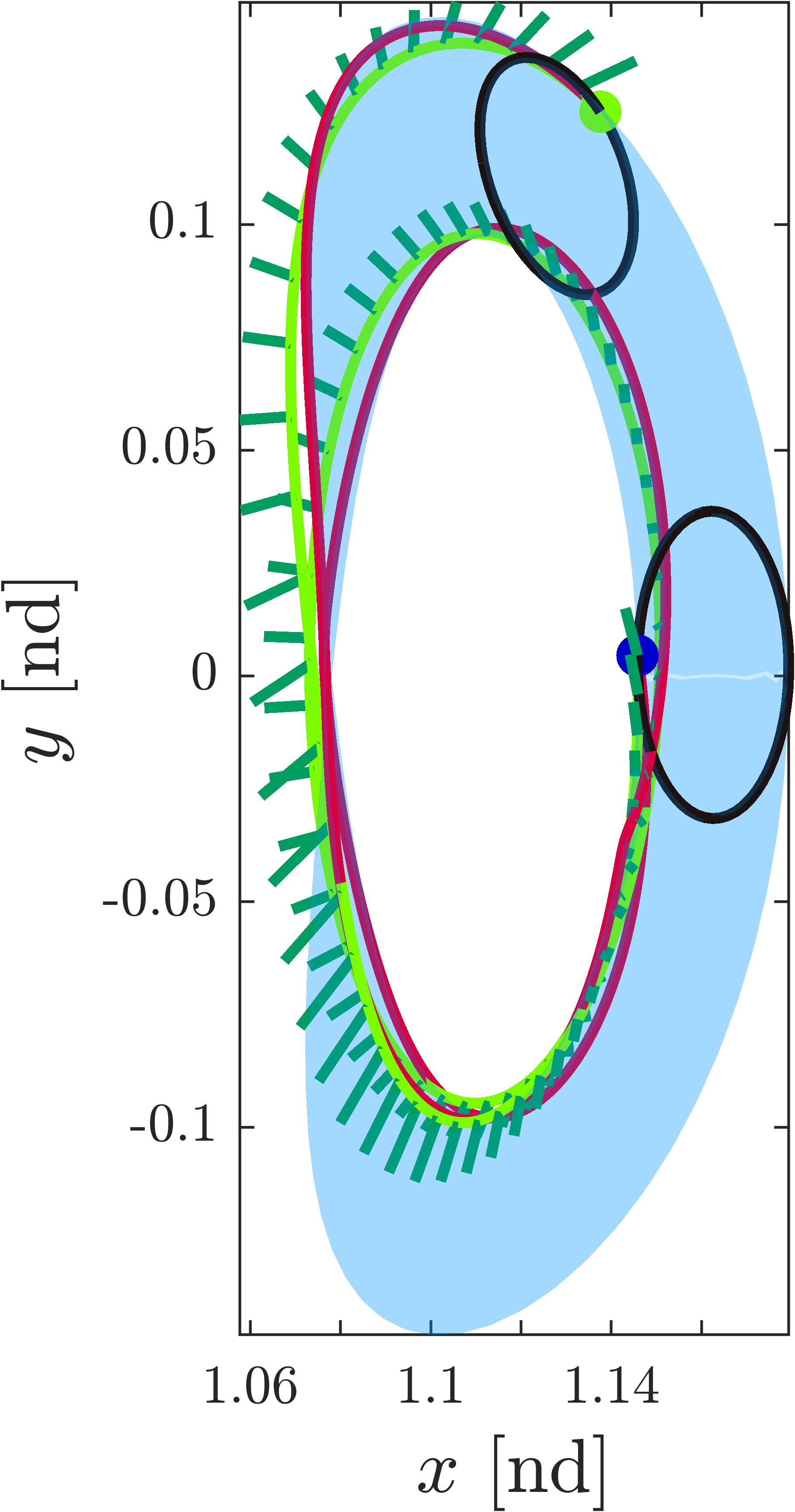}
         \caption{$\varepsilon=0.99$}
     \end{subfigure}
        \caption{CMETEH solutions using the CMVE solution in the torus space.}
        \label{fig:L2homotopy}
\end{figure}

A singularity exists at $\varepsilon=1$ in the determination of the control input magnitude. In other words, for a true minimum tracking error solution, control should be set to the maximum available at all times. It is also seen in Eq. \ref{eq:costateCMETEH} that including tracking error in the cost function affects the coordinate costate dynamics. In solving this boundary value problem via shooting, the reference trajectory is propagated alongside the phase space states and costates using Eqs. \ref{eq:CMVE} and \ref{eq:CMVEcostates}. The solution process begins with $\varepsilon=0$, which takes advantage of the minimum energy quadratic cost to exhibit regular convergence using initial costates guesses of zero. Once this CME solution is obtained, solutions are continued over the homotopic parameter towards $\varepsilon=1$. As $\varepsilon$ is increased, the solution approaches adherence to the torus surface. This is illustrated in Fig. \ref{fig:L2homotopy}.
% \begin{figure}[htbp!]
%      \centering
%      \begin{subfigure}{0.18\textwidth}
%          \centering
%          \includegraphics[width=1\textwidth]{Figures/L2_Plots/L2CMETEH_CSbirdp1.jpg}
%          \caption{$\varepsilon=0.1$}
%      \end{subfigure}\hfill
%      \begin{subfigure}{0.169\textwidth}
%          \centering
%          \includegraphics[width=1\textwidth]{Figures/L2_Plots/L2CMETEH_CSbirdp3.jpg}
%          \caption{$\varepsilon=0.3$}
%      \end{subfigure}\hfill
%      \begin{subfigure}{0.158\textwidth}
%          \centering
%          \includegraphics[width=1\textwidth]{Figures/L2_Plots/L2CMETEH_CSbirdp5.jpg}
%          \caption{$\varepsilon=0.5$}
%      \end{subfigure}\hfill
%      \begin{subfigure}{0.149\textwidth}
%          \centering
%          \includegraphics[width=1\textwidth]{Figures/L2_Plots/L2CMETEH_CSbirdp7.jpg}
%          \caption{$\varepsilon=0.7$}
%      \end{subfigure}\hfill
%      \begin{subfigure}{0.138\textwidth}
%          \centering
%          \includegraphics[width=1\textwidth]{Figures/L2_Plots/L2CMETEH_CSbirdp9.jpg}
%          \caption{$\varepsilon=0.9$}
%      \end{subfigure}\hfill
%      \begin{subfigure}{0.131\textwidth}
%          \centering
%          \includegraphics[width=1\textwidth]{Figures/L2_Plots/L2CMETEH_CSbirdp99.jpg}
%          \caption{$\varepsilon=0.99$}
%      \end{subfigure}
%         \caption{CMETEH solutions using the CMVE solution in the torus space.}
%         \label{fig:L2homotopy}
% \end{figure}

\begin{figure}[htbp!]
     \centering
     \begin{subfigure}{1\textwidth}
         \centering
         \includegraphics[width=0.36\textwidth]{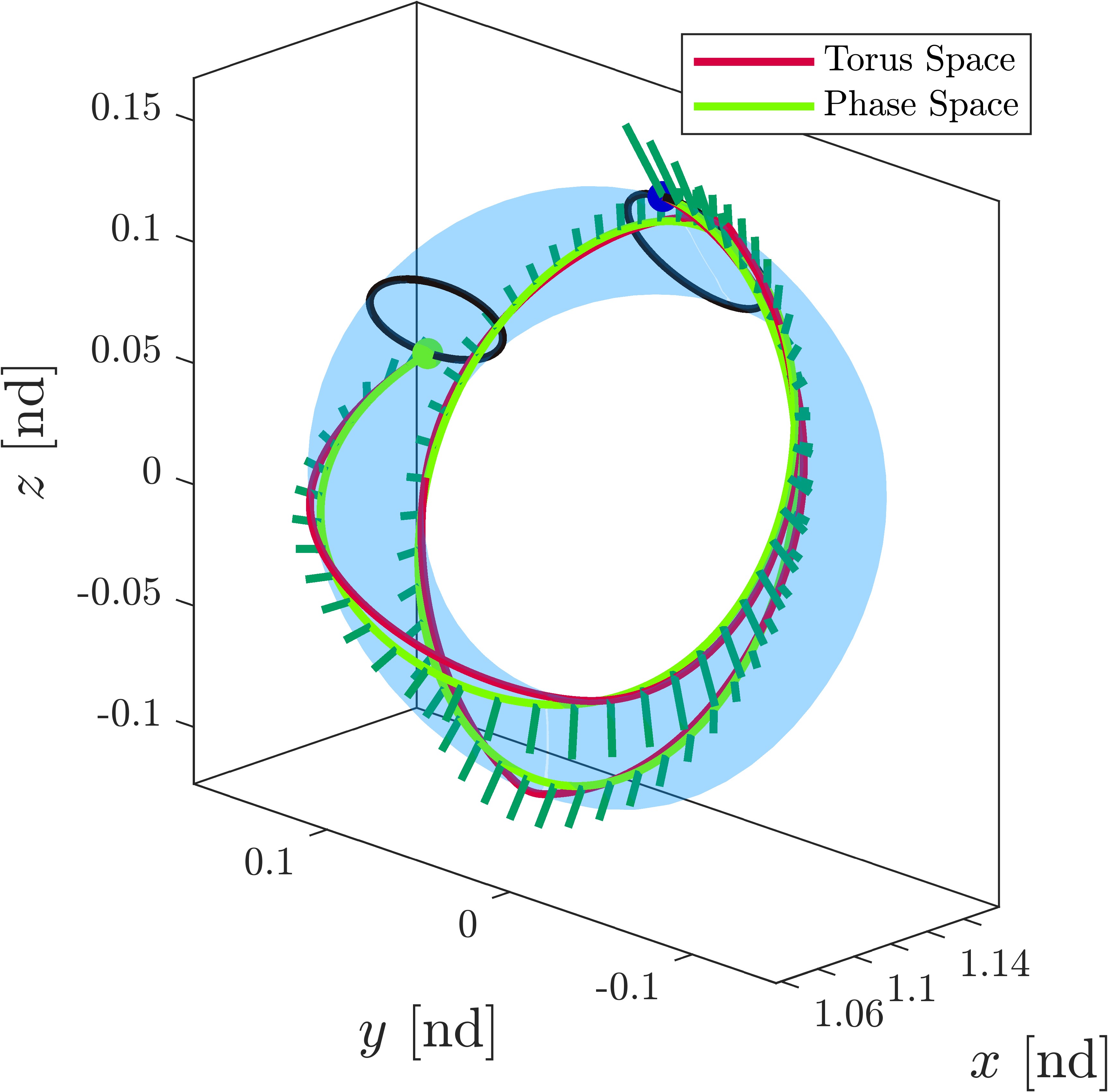}
     \end{subfigure}
     \caption{CMETEH solution with $\varepsilon=0.9968$ to the rephasing problem on a 2-dimensional QPIT in the CR3BP.}
        \label{fig:L2CMETEH}
\end{figure}

Near $\varepsilon=1$, problem sensitivity impedes the continuation towards full minimum tracking. In other words, too small of a step size in $\varepsilon$ is required to truly reach a solution that utilizes all available control to track the reference trajectory. This held true for both natural parameter continuation and pseudo-arc length continuation. For all cases, a step size $\delta\varepsilon<1\times10^{-5}$ was required around parameter values $\varepsilon=0.995$. The final value used in the continued $L_2$ quasi-halo example was $\varepsilon=0.9968$ and the geometric results on top of the CMVE reference trajectory are shown in Fig. \ref{fig:L2CMETEH}. Quantitative performance of this solution is discussed in the later section.
% \begin{figure}[htbp!]
%      \centering
%      \begin{subfigure}{1\textwidth}
%          \centering
%          \includegraphics[width=0.47\textwidth]{Figures/L2_Plots/L2CMETEH_CSiso.jpg}\hfill
%          \includegraphics[width=0.23\textwidth]{Figures/L2_Plots/L2CMETEH_CSbird.jpg}\hfill
%          \includegraphics[width=0.245\textwidth]{Figures/L2_Plots/L2CMETEH_CSside.jpg}
%      \end{subfigure}
%      \caption{CMETEH solution with $\varepsilon=0.9968$ to the rephasing problem on a 2-dimensional QPIT in the CR3BP.}
%         \label{fig:L2CMETEH}
% \end{figure}

\subsubsection{Constrained Minimum Time Patches}
The CMVT torus space solution returns a series of thrust and coast arcs on the torus surface. Unlike the CMVE solution, thrust is set to a maximum whenever on. With non-zero velocity control input during these maximum thrust arcs, tracking their trajectory or even just the arc's boundary conditions in the phase space is not possible because there is no available control margin. Still, it's desired to preserve the coast arc structure generated from the CMVT solution because any time spent on the torus naturally during transfer results in zero torus deviation. To do so, constrained minimum time patches (CMTP) are now used to connect coast arcs in the phase space generated from the CMVT torus space solution. This is different from the bang-off-bang solution provided by the CMT problem seen in Fig. \ref{fig:L2torusagnostic} because any coast arcs are now constrained to the torus surface. 

Let $t_{0,t}^i$ and $t_{f,t}^i$ be the initial and final times for the $i$th thrust arc obtained from the CMVT solution. Then the CMTP problem to transition this thrust arc to the phase space is given as
\begin{align}
    \text{min}\quad &J = \int_{t_{0,t}^i}^{t_f}\text{d}t,\quad t_f\,\,\text{free}\\
    \text{s.t.}\quad &\dot{\boldsymbol{x}}=\boldsymbol{f}(\boldsymbol{x},\boldsymbol{\mu})+\boldsymbol{B}\boldsymbol{u}_{\dot{q}},\quad \left|\left|\boldsymbol{u}_{\dot{q}}\right|\right| \leq u_{\dot{q},\text{max}} \nonumber \\
    &\boldsymbol{x}(t_{0,t}^i) = \boldsymbol{\Gamma}(\boldsymbol{\theta}^*(t_{0,t}^i))\nonumber \\
    &\boldsymbol{x}(t_f) = \boldsymbol{\Gamma}(\boldsymbol{\omega}(t_f - t_{f,t}^i) + \boldsymbol{\theta}^*(t_{f,t}^i)) \nonumber
\end{align}
with Hamiltonian
\begin{equation}
    \mathcal{H} = 1 + \boldsymbol{\lambda}_{x}^\text{T}\left[\boldsymbol{f}(\boldsymbol{x})+\boldsymbol{B}\boldsymbol{u}_{\dot{q}}\right]
\end{equation}
The resulting control magnitude and direction are $u_{\dot{q}}=u_{\dot{q},\text{max}}$ and $\widehat{\boldsymbol{u}} = -\widehat{\boldsymbol{\lambda}}_{\dot{q}}$. The costate dynamics are found to be the same as Eq. \ref{eq:costateCMETEH} barring the torus error term. The addition of the free final time variable warrants the enforcement of the following indirect transversality condition, expressed in Eq. \ref{eq:CMTPHf}.
\begin{equation}
    \boldsymbol{\lambda}_x^\text{T}(t_f)\left[\frac{\partial \boldsymbol{\Gamma}(\boldsymbol{\theta}(t_f))}{\partial \boldsymbol{\theta}}\right]\boldsymbol{\omega} + \mathcal{H}(t_f) =0
    \label{eq:CMTPHf}
\end{equation}

\begin{figure}[htbp!]
     \centering
     \begin{subfigure}{1\textwidth}
         \centering
         \includegraphics[width=0.36\textwidth]{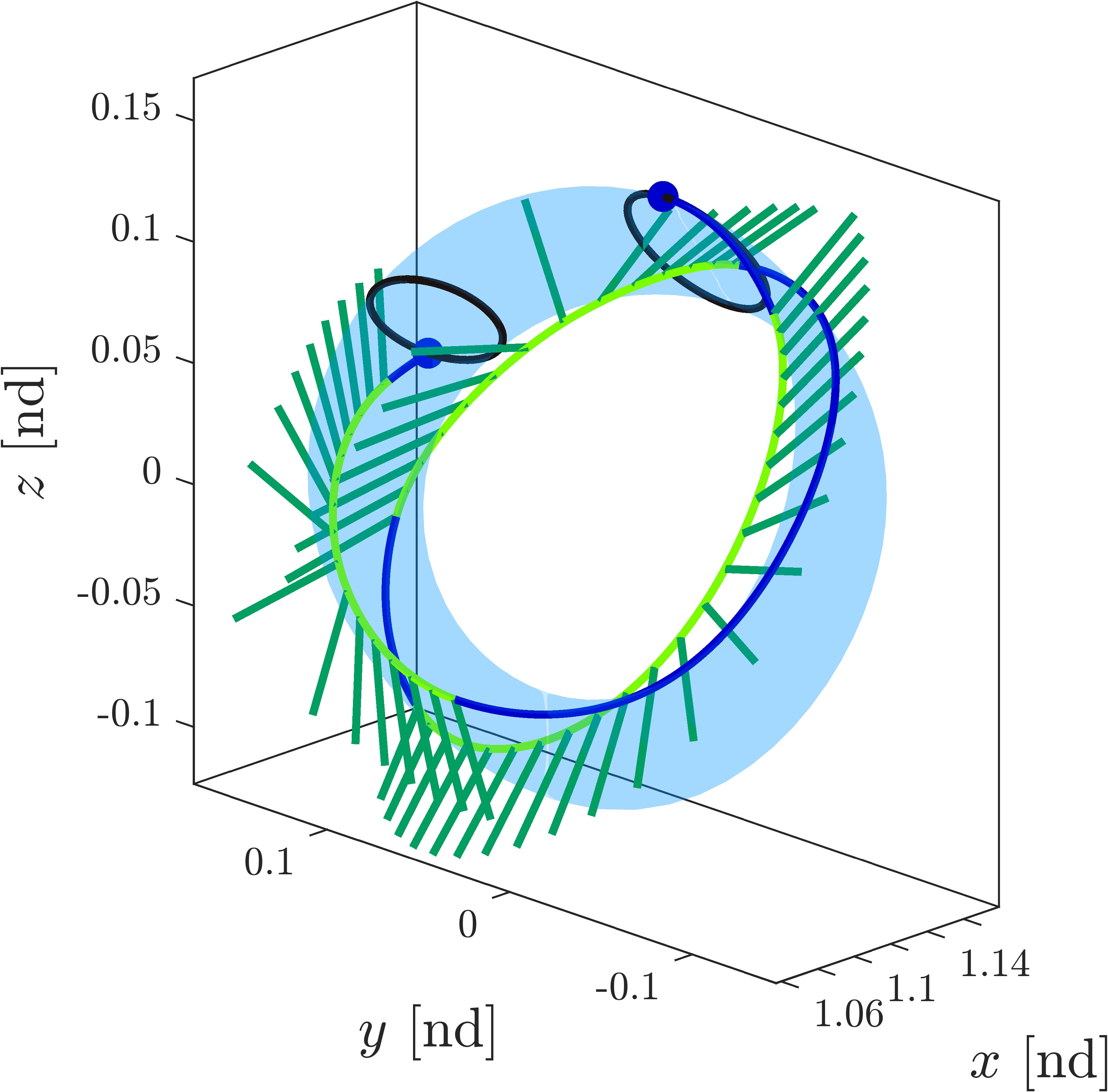}
     \end{subfigure}
        \caption{CMTP solution to the rephasing problem on a 2-dimensional QPIT in the CR3BP.}
        \label{fig:L2CMTP}
\end{figure}

In the CMVT solution, there is a possibility that two thrust arcs in the torus space lie too close together such that a minimum time patch to the first takes more time than the coast separating the two allows. In that case, the thrust arcs are grouped and treated as one with $\boldsymbol{x}(t_{0,t}^i) = \boldsymbol{\Gamma}(\boldsymbol{\theta}^*(t_{0,t}^i))$ and $\boldsymbol{x}(t_f) = \boldsymbol{\Gamma}(\boldsymbol{\omega}(t_f - t_{f,t}^{i+1}) + \boldsymbol{\theta}^*(t_{f,t}^{i+1}))$. This happens to be the case in transitioning the CMVT solution to the continued $L_2$ quasi-halo example, whose control history is shown in Fig. \ref{fig:L1plots}. Thrust arcs two and three, as well as four and five are both grouped into CMTP solutions such that the final transitioned trajectory has three thrust arcs and four coast arcs, as shown in Fig. \ref{fig:L2CMTP}.
% \begin{figure}[htbp!]
%      \centering
%      \begin{subfigure}{1\textwidth}
%          \centering
%          \includegraphics[width=0.47\textwidth]{Figures/L2_Plots/L2patch_CSiso.jpg}\hfill
%          \includegraphics[width=0.23\textwidth]{Figures/L2_Plots/L2patch_CSbird.jpg}\hfill
%          \includegraphics[width=0.245\textwidth]{Figures/L2_Plots/L2patch_CSside.jpg}
%      \end{subfigure}
%         \caption{CMTP solution to the rephasing problem on a 2-dimensional QPIT in the CR3BP.}
%         \label{fig:L2CMTP}
% \end{figure}

\subsection{Performance Metrics}
Three quantitative measures are introduced to better compare the newly presented, manifold-conscious solutions to traditional, torus-agnostic solutions. These include: total $\Delta V$ required for the transfer, cumulative torus error $E$, and time spent on the torus $t_{\text{on}}$ without control during rephasing. For this simplified comparison of acceleration level continuous control input, $\Delta V$ is computed as
\begin{equation}
    \Delta V = \int_{t_0}^{t_f}u_{\dot{q}}\:\text{d}t,\quad u_{\dot{q}}>0
\end{equation}
using quadrature with dense, uniform temporal spacing.

Cumulative torus error measures the total deviation away from the structure during the transfer. Only the coordinate component of error, as in Eq. \ref{eq:errordef}, is used. In the evaluation of torus agnostic solutions, the idea of a reference trajectory is absent. Hence, the following minimization problem is formed and solved via nonlinear programming for a discretized trajectory. For every discrete state $\boldsymbol{x}$ on a phase space trajectory solution, the set of angles $\Breve{\boldsymbol{\theta}}$ is found whose map back to the configuration space yields the minimum $\mathcal{L}_2$ norm of the difference. In other words, the minimum distance from the discrete state to the torus surface in configuration space is found. The problem is stated as
\begin{equation}
    \min_{\breve{\boldsymbol{\theta}}\in\mathbb{T}_p} ||\boldsymbol{e}||,\,\,
    \boldsymbol{e} = \boldsymbol{q}-\boldsymbol{\gamma}(\breve{\boldsymbol{\theta}})
\end{equation}
and the cumulative error measure is then
\begin{equation}
    E = \int_{t_0}^{t_f}||\boldsymbol{e}(t)||\:\text{d}t
    \label{eq:cumerr}
\end{equation}
which is solved via quadrature for the same discretized solution used in the computation of $\Delta V$.

Time spent on the torus is simply computed by either summing initial/terminal coast arcs in a torus agnostic solution, or summing all coast arcs in the CMTP realization of the CMVT problem. Time spent on the torus during the overall planned rephasing transfer presents an advantage from a mission operations perspective. When on the torus, nominal operations (i.e. observation) may be resumed within the specified transfer window. Additionally, in the operations scenario, time is given to assess the maneuver's accuracy and correct/prepare for the next. Finally, total maneuver abort during coast arcs on the torus requires no additional recovery maneuvers to return to nominal operation, further emphasizing the idea of passive safety through manifold proximity.

\begin{figure}[htbp!]
    \centering
    \includegraphics[width=0.4\linewidth]{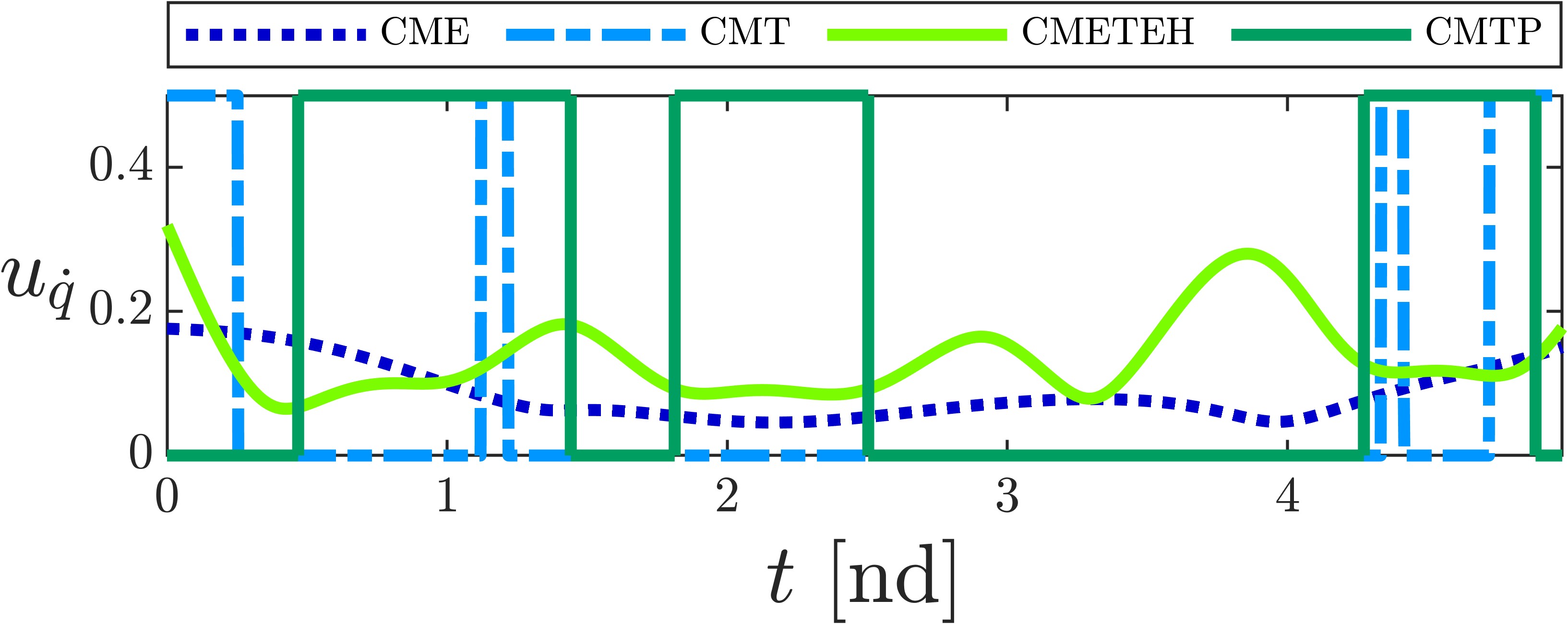}
    \caption{Control time histories for different solutions to the $L_2$ quasi-halo torus rephasing problem.}
    \label{fig:L2control}
\end{figure}

Results for the rephasing trajectories for the $L_2$ quasi-halo example with fixed boundary conditions are presented in Table \ref{tab:level2} after their transition to the phase space.
\begin{table}[htbp!]
\caption{\label{tab:level2} rephasing performance metrics with $\varepsilon=0.9968$ used in the CMETEH solution.}
\centering
\begin{tabular}{llccc}
\hline
Torus Space & Phase Space & $\Delta V$ & $E$ & $t_{\text{on}}$ \\\hline
None & CME & 0.4365 & 0.2017 & 0\\
None & CMT& 0.3455 & 0.2620 & 0\\
\hline
CMVE & CMETEH & 0.6760 & 0.0204 & 0\\
CMVT & CMTP & 1.1385 & 0.0111 & 2.7037\\
\hline
\end{tabular}
\end{table}
The torus agnostic CMT solution results in the least fuel consumption, but also the largest cumulative torus error. In contrast, the CMTP solution yields the least torus error, while spending more than half of the transfer time under natural coast on the torus surface. This comes at the cost of more than three times the amount of fuel required. The CMETEH solution strikes an attractive balance with only double the amount of fuel required than the CMT solution for an order of magnitude less torus deviation. However, this solution type requires a variable-thrust engine. The control magnitude time histories of all four solutions are shown in Fig. \ref{fig:L2control}.
Note that although the CMETEH solution was stopped at $\varepsilon=0.9968$ because of numerical continuation step-size requirements for solution convergence, the required control magnitude is not close to the maximum available. This suggests that lower total torus error can be achieved by decreasing problem sensitivity, which is possible with further domain segmentation in the multiple shooting scheme. Overall, these performance metrics demonstrate the success of the proposed bi-level optimization framework in minimizing torus deviation during rephasing.
\section{Modification for Quasi-Periodically Forced Dynamical Systems}\label{sec:mods}

Thus far, the modified invariance condition and bi-level optimal control framework have been introduced and employed for tori in autonomous systems. This methodology is now extended to the general quasi-periodically forced system of Eq. \ref{eq:firstdynamics}. \rev1{Recall that the system in Eq. \ref{eq:firstdynamics} is in its conjugate form.} In multi-body astrodynamics, intermediate models with a tractable amount of forcing frequencies may accurately capture the motion of spacecraft in particular regimes without directly working with full ephemeris models. For example, the \rev1{periodically forced} ER3BP is known to well represent motion around the Earth-Moon $L_1$ and $L_2$ libration points with just one forcing frequency of motion present \cite{park2025}. Because of this, it may be advantageous to begin designing rephasing maneuvers in this general class of systems.

In order to construct a modified invariance condition like that in Eq. \ref{eq:invar2} for a torus in a quasi-periodically forced system, its torus dimension must be larger than the system's forcing phase dimension, i.e. $p>m$. For $p=m$, $\boldsymbol{\theta}=\boldsymbol{\vartheta}$ and each angle set $\boldsymbol{\theta}$ is tied to a particular epoch of the dynamical system, leaving no degree-of-freedom in the torus space for an appended control frequency. \rev1{In other words, only internal torus frequencies, if they exist, can be controlled.} Assuming $p>m$, the torus angles are further decomposed as
\begin{equation}
    \boldsymbol{\theta}^\text{T} = 
    \begin{bmatrix}
        \tilde{\boldsymbol{\theta}}^\text{T} & \boldsymbol{\vartheta}^\text{T}
    \end{bmatrix},\quad\tilde{\boldsymbol{\theta}}\in\mathbb{T}_{p-m}
\end{equation}
where $\tilde{\boldsymbol{\theta}}$ are the angles not tied to the system's phase. Then, the fundamental torus dynamics of the free torus space can be written $\dot{\tilde{\boldsymbol{\theta}}}=\widetilde{\boldsymbol{\omega}}+\widetilde{\boldsymbol{\Omega}}$, and the following corresponding decomposed modified invariance condition can be constructed using the same procedure outlined in Sec. \ref{sec:modinv}.
\begin{equation}
    \left[\frac{\partial \boldsymbol{\Gamma}(\boldsymbol{\theta})}{\partial \tilde{\boldsymbol{\theta}}}\right]\left(\widetilde{\boldsymbol{\omega}}+\widetilde{\boldsymbol{\Omega}}\right) +  \left[\frac{\partial \boldsymbol{\Gamma}(\boldsymbol{\theta})}{\partial \boldsymbol{\vartheta}}\right]\boldsymbol{g}(\boldsymbol{\vartheta},\boldsymbol{\mu}) = \boldsymbol{f}(\boldsymbol{\Gamma}(\boldsymbol{\theta}),\boldsymbol{\vartheta},\boldsymbol{\mu}) +
    \boldsymbol{u}
    \label{eq:invartilde}
\end{equation}
Natural invariance terms fall out, leaving behind the control mappings of Eq. \ref{eq:controlContilde}.
\begin{equation}
    \left[\frac{\partial\boldsymbol{{\gamma}}(\boldsymbol{\theta})}{\partial \tilde{\boldsymbol{\theta}}}\right]\widetilde{\boldsymbol{\Omega}} = \boldsymbol{u}_q, \quad \left[\frac{\partial\dot{\boldsymbol{{\gamma}}}(\boldsymbol{\theta})}{\partial \tilde{\boldsymbol{\theta}}}\right]\widetilde{\boldsymbol{\Omega}} = \boldsymbol{u}_{\dot{q}} \label{eq:controlContilde}
\end{equation}

The CMVE and CMVT torus space optimal control problems defined in Eqs. \ref{eq:CMVE} and \ref{eq:CMVT}, respectively, can now be formulated for quasi-periodically forced dynamical systems using the control mappings of Eq. \ref{eq:controlContilde}. The problem dimension has now changed from $p$ to $p-m$, as only $\tilde{\boldsymbol{\theta}}$ can be manipulated. Note, however, that the sensitivities in Eq. \ref{eq:controlContilde} depend on $\boldsymbol{\theta}$, and thus so do the CMVE and CMVT cost functions. This means the torus space optimal control problems are now time-dependent. Once torus space reference trajectories are computed in the reduced space, the same methods of transition are used to obtain physically realizable rephasing trajectories.

\subsection{Numerical Example}
To demonstrate the proposed methodologies on tori in dynamical system's with forcing frequencies, a torus in the elliptical restricted three body problem is examined, \rev1{which represents the special case of $m=1$.} The ER3BP equations of motion with control expressed in the rotating-pulsating frame appear below with now an additional parameter $e$ as the primaries' eccentricity.
\begin{subequations} \label{eq:ER3BPc}
    \begin{align}
        x''-2y'& = \frac{1}{1+e\cos f}\left[x-\frac{(1-\mu)(x+\mu)}{d_1^3} - \frac{\mu(x-1+\mu)}{d_2^3}+u_x\right]\label{eq:crac} \\
        y''+2x' & = \frac{1}{1+e\cos f}\left[y-\frac{(1-\mu)y}{d_1^3} - \frac{\mu y}{d_2^3}+u_y\right]\label{eq:crbc}\\
        z'' & = \frac{1}{1+e\cos f}\left[-ez\cos f -\frac{(1-\mu)z}{d_1^3} - \frac{\mu z}{d_2^3}+u_z\right]\label{eq:crcc}
    \end{align}
\end{subequations}
The ER3BP is formulated with the primaries' true anomaly $f$ as the independent variable such that $\frac{\text{d}}{\text{d}f}=(\cdot)'$ and the system forcing frequency is 1. Control is parametrized inside the bracketed terms so that any optimal control solutions may be examined without frame pulsation. At $e=0$, the ER3BP exactly reduces to the CR3BP. 

\begin{figure}[htbp!]
     \centering
     \begin{subfigure}{0.175\textwidth}
         \centering
         \includegraphics[width=1\textwidth]{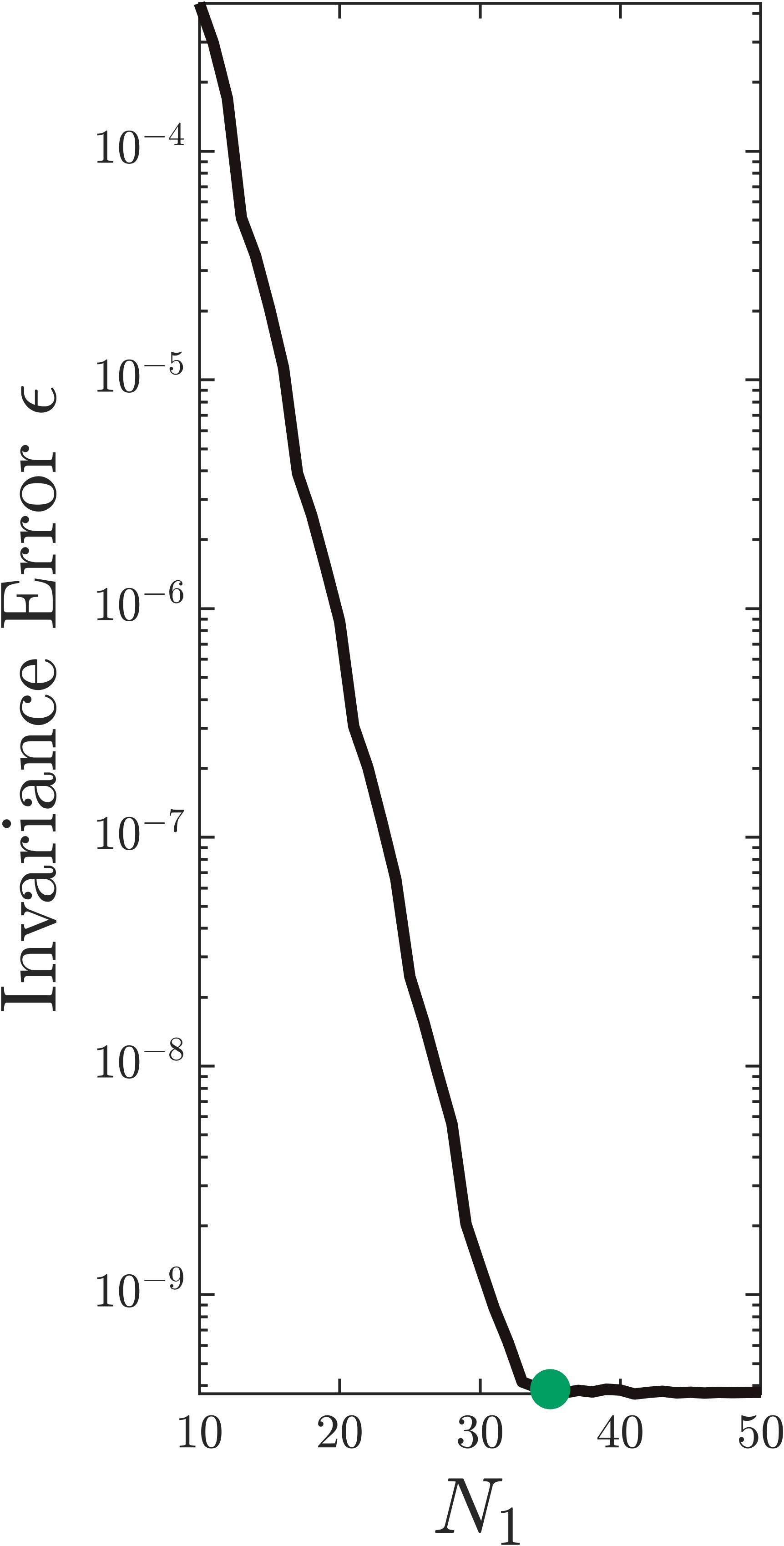}
         \caption{Error Function}
     \end{subfigure}\hspace{1em}
     \begin{subfigure}{0.27\textwidth}
         \centering
         \includegraphics[width=1\textwidth]{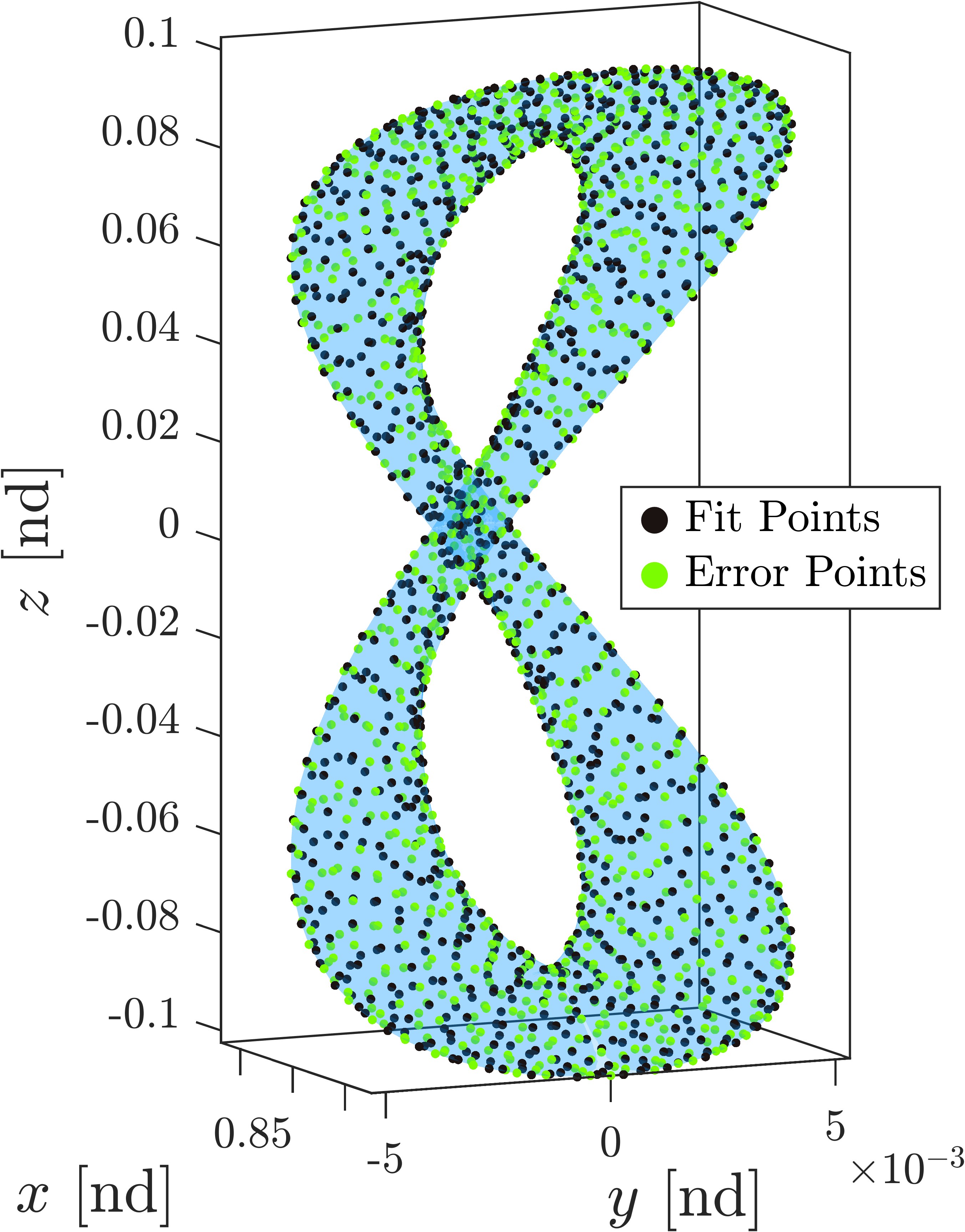}
         \caption{$N_1=35, N_2=30$}
     \end{subfigure}
        \caption{Approximation order analysis for 2-dimensional $L_1$ quasi-vertical torus in the elliptical restricted problem.}
        \label{fig:ER3BPtorusfit}
\end{figure}

A non-resonant $L_1$ vertical orbit in the Earth-Moon CR3BP is transitioned into a 2-dimensional torus in the ER3BP up to a primary system eccentricity of $e=0.4$. The torus frequencies are $\omega_1=2.1981$ and $\omega_2=1$ with $\theta_2=f$ such that only $\theta_1$ is free to manipulate. With $N_2=30$ fixed, Fourier approximation errors are shown in Fig. \ref{fig:ER3BPtorusfit} as $N_1$ is varied.
An ideal model order is reached at $N_1=35, N_2=30$. Note that the error cannot be reduced further unless $N_2$ is increased.

\begin{figure}[htbp!]
     \centering
     \begin{subfigure}{0.27\textwidth}
         \centering
         \includegraphics[width=1\textwidth]{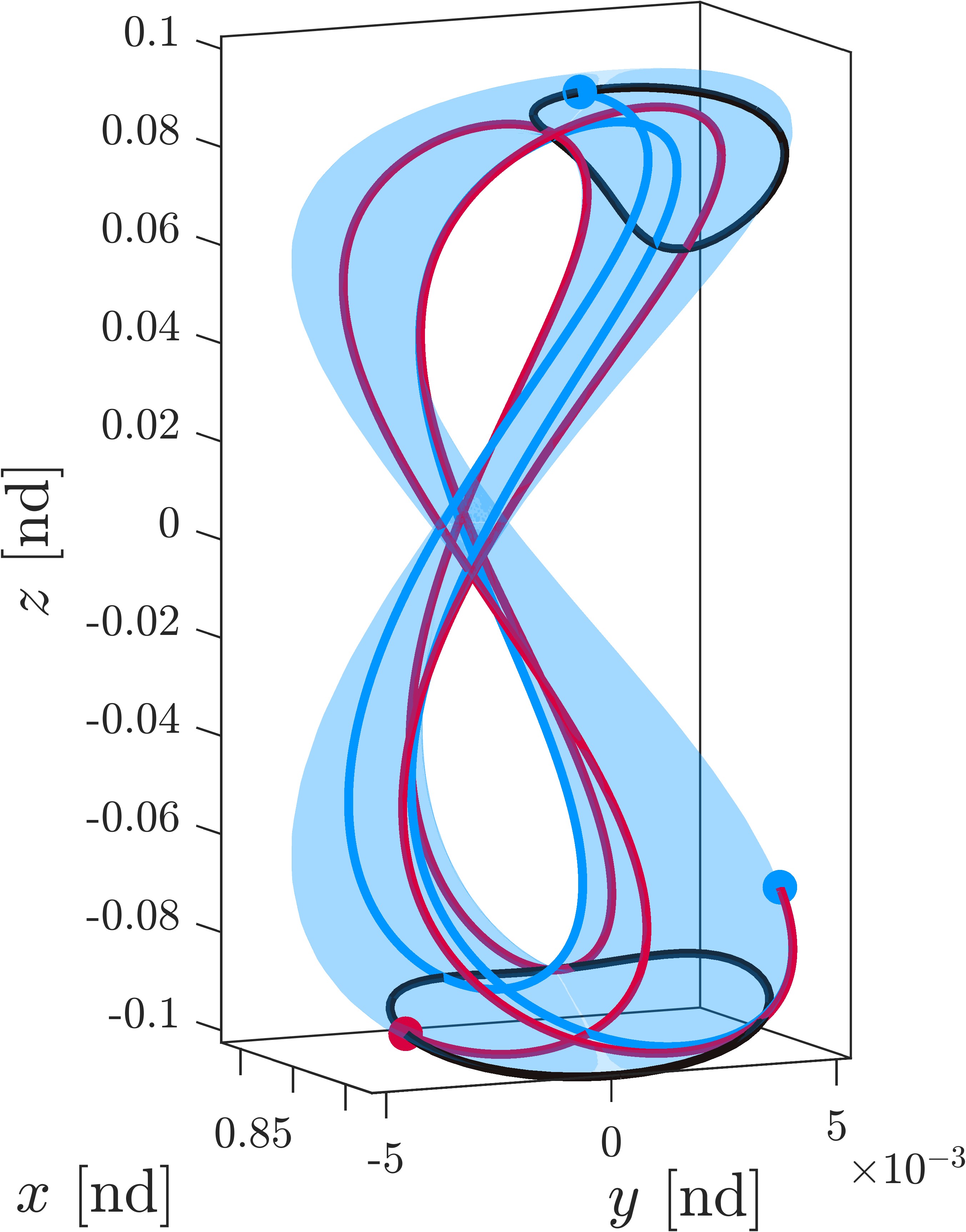}
     \end{subfigure}\hspace{1em}
     \begin{subfigure}{0.384\textwidth}
         \centering
         \includegraphics[width=1\textwidth]{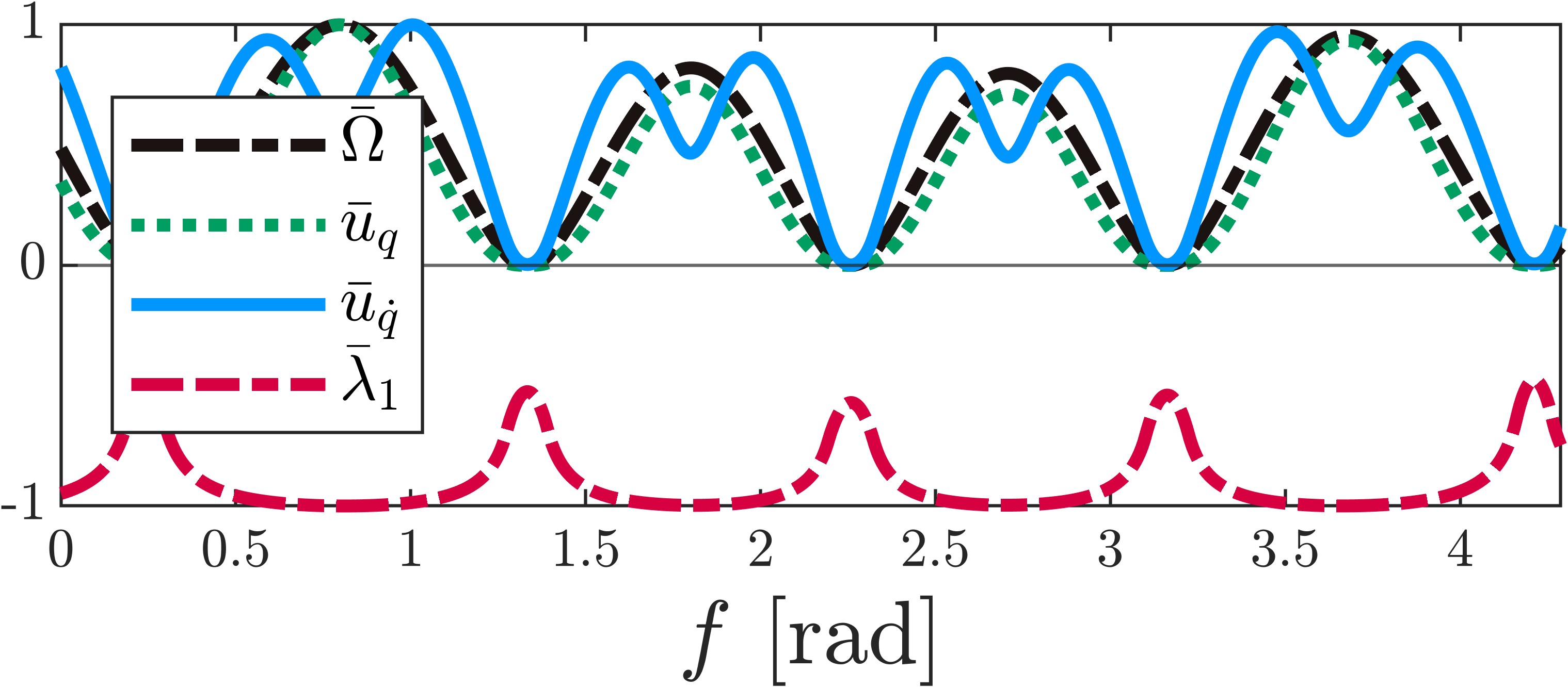}\\
         \includegraphics[width=1\textwidth]{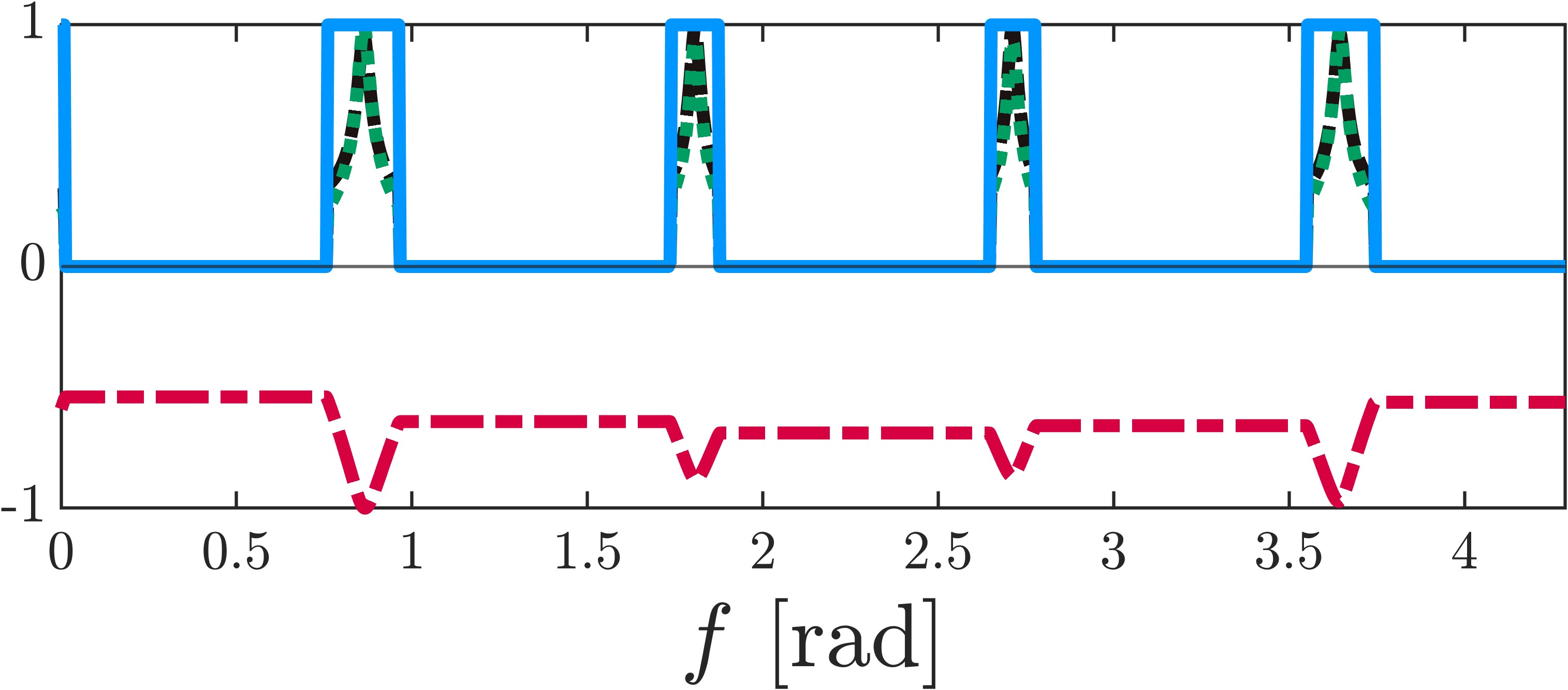}
     \end{subfigure}\hspace{1em}
     \begin{subfigure}{0.27\textwidth}
         \centering
         \includegraphics[width=1\textwidth]{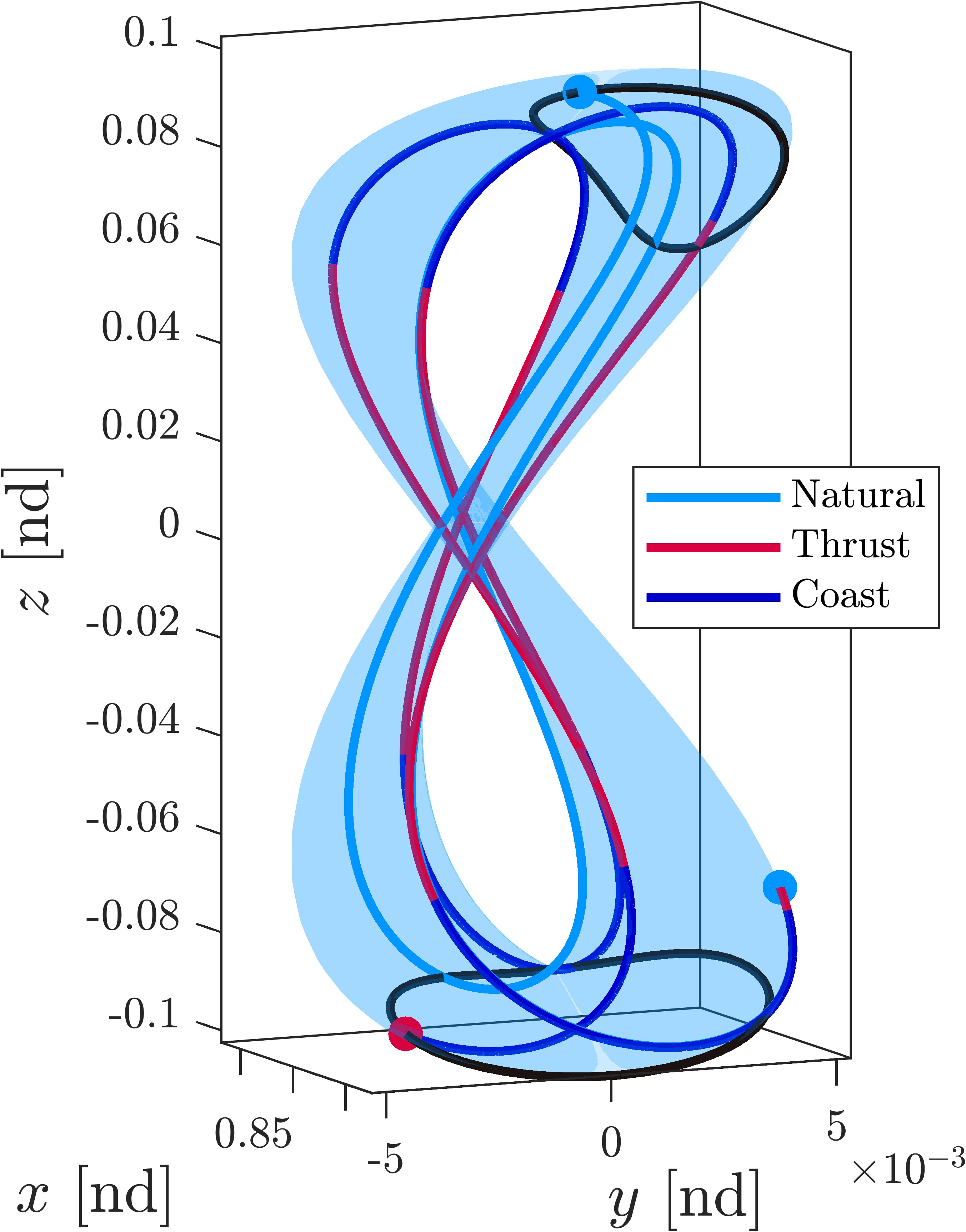}
     \end{subfigure}
        \caption{CMVE (left) and CMVT (right) torus space solutions to a fixed boundary condition rephasing maneuver on an $L_1$ quasi-vertical torus in the ER3BP.}
        \label{fig:ER3BPtorusspace}
\end{figure}

To demonstrate the proposed optimal control framework for this reduced degree-of-freedom torus, a fixed boundary condition rephasing problem is solved with $\boldsymbol{\theta}(f_0)=[1.3\,\,\pi/4]^\text{T}$, $f_0=\theta_{2,0}$, $\Delta f=3\pi/\omega_1$, and $\theta_{1,d}(f_0+\Delta f)=1.8048$. Results for the CMVE and CMVT solution to this boundary value problem are given in Fig. \ref{fig:ER3BPtorusspace}.
With the control frequency space reduction due to the resonance requirement for $\theta_2$, the goal now is to find optimal locations along the underlying periodic orbit direction over which to artificially change the torus angle. Both the CMVE and CMVT solutions find these locations to be along the vertical traverses of the torus, as the peak control frequency magnitudes nearly line up between the two solutions for each of their four local maxima. Thus, all coast arcs in the CMVT solution occur around the change in $z$-direction where the coordinate vector is changing most rapidly with $\boldsymbol{\theta}$. 

\begin{figure}[htbp!]
     \centering
     \begin{subfigure}{0.24\textwidth}
         \centering
         \includegraphics[width=1\textwidth]{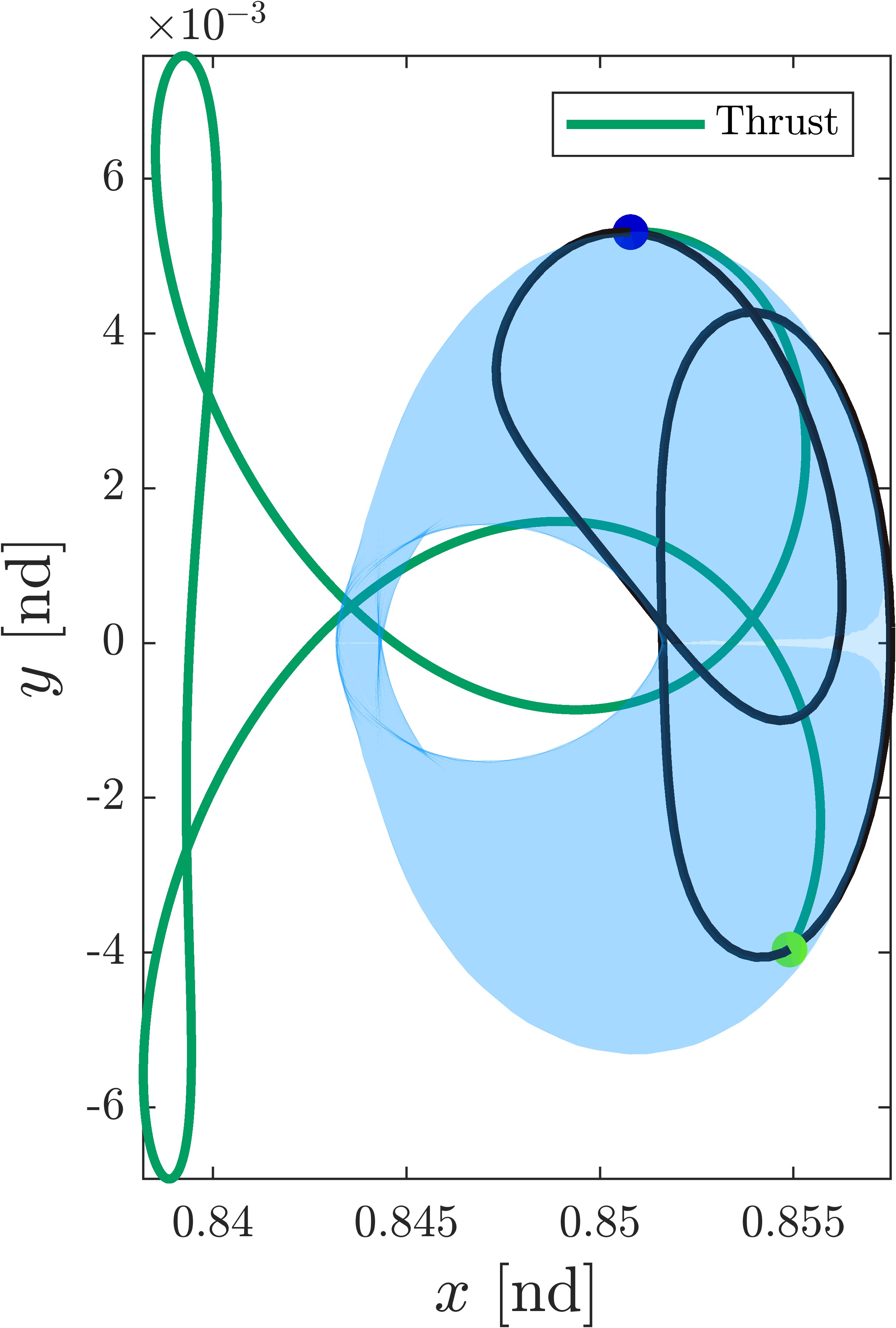}
         \caption{CME}
     \end{subfigure}\hfill
     \begin{subfigure}{0.24\textwidth}
         \centering
         \includegraphics[width=1\textwidth]{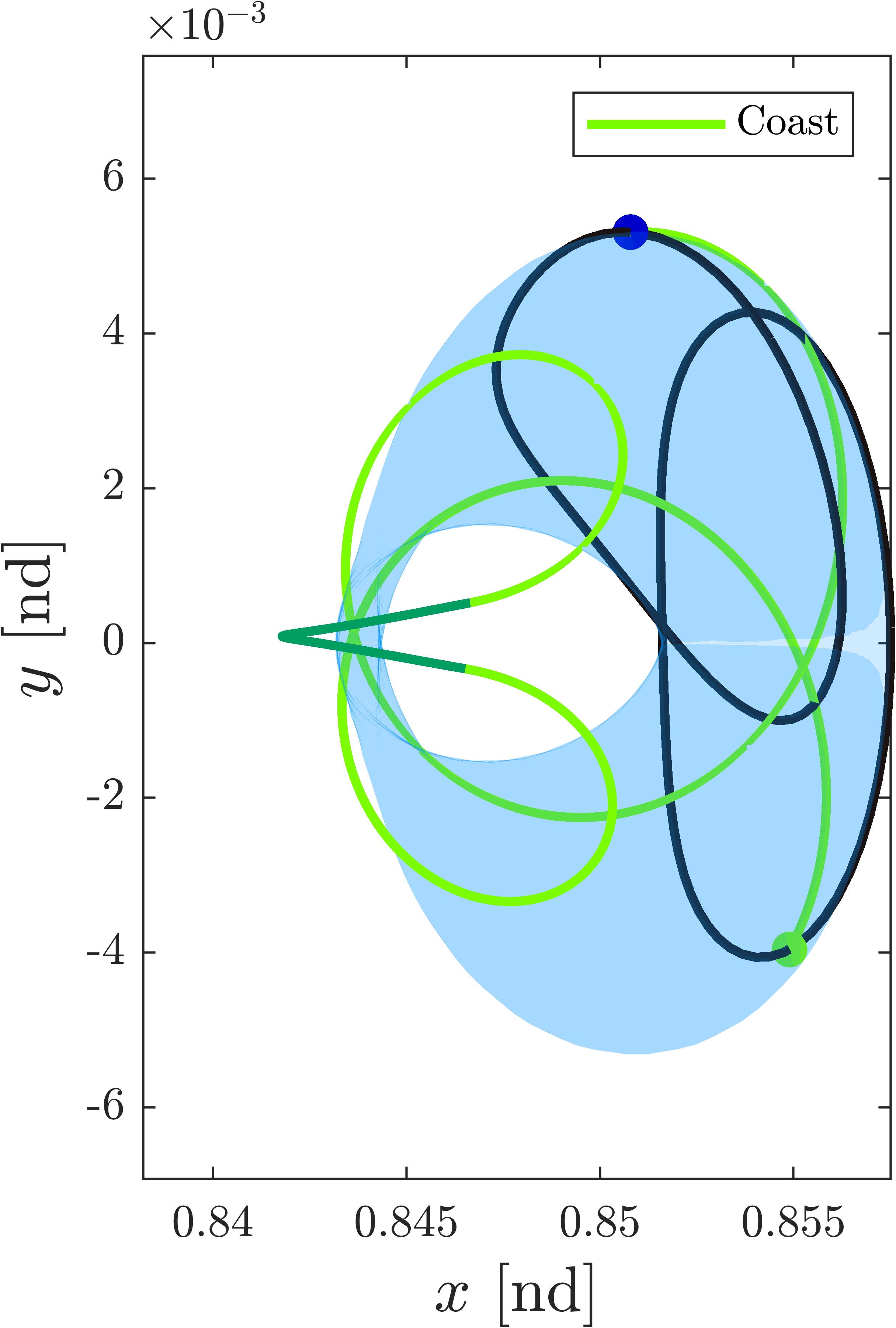}
         \caption{CMT}
     \end{subfigure}\hfill
     \begin{subfigure}{0.24\textwidth}
         \centering
         \includegraphics[width=1\textwidth]{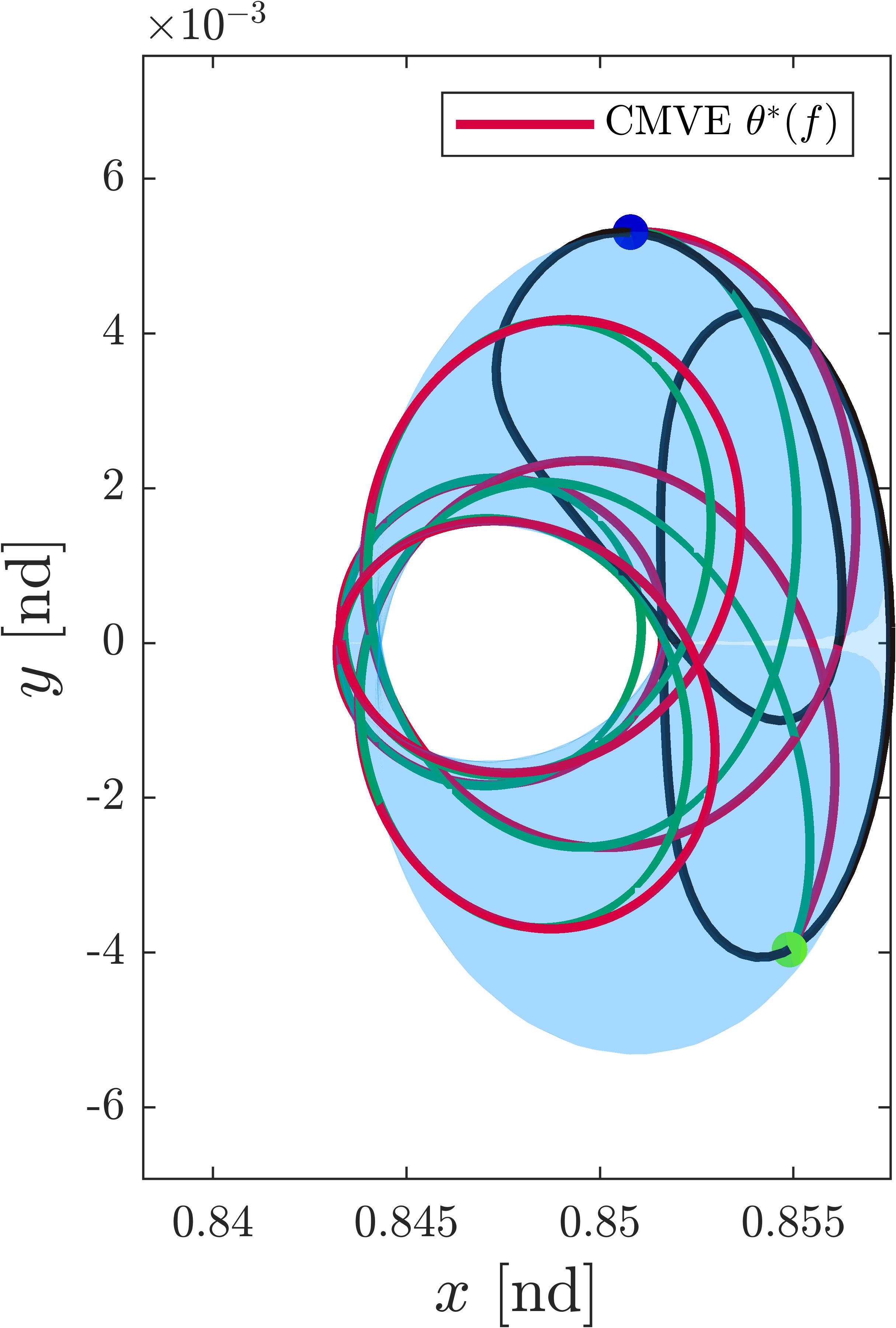}
         \caption{CMETEH $\varepsilon=0.9998$}
     \end{subfigure}\hfill
     \begin{subfigure}{0.24\textwidth}
         \centering
         \includegraphics[width=1\textwidth]{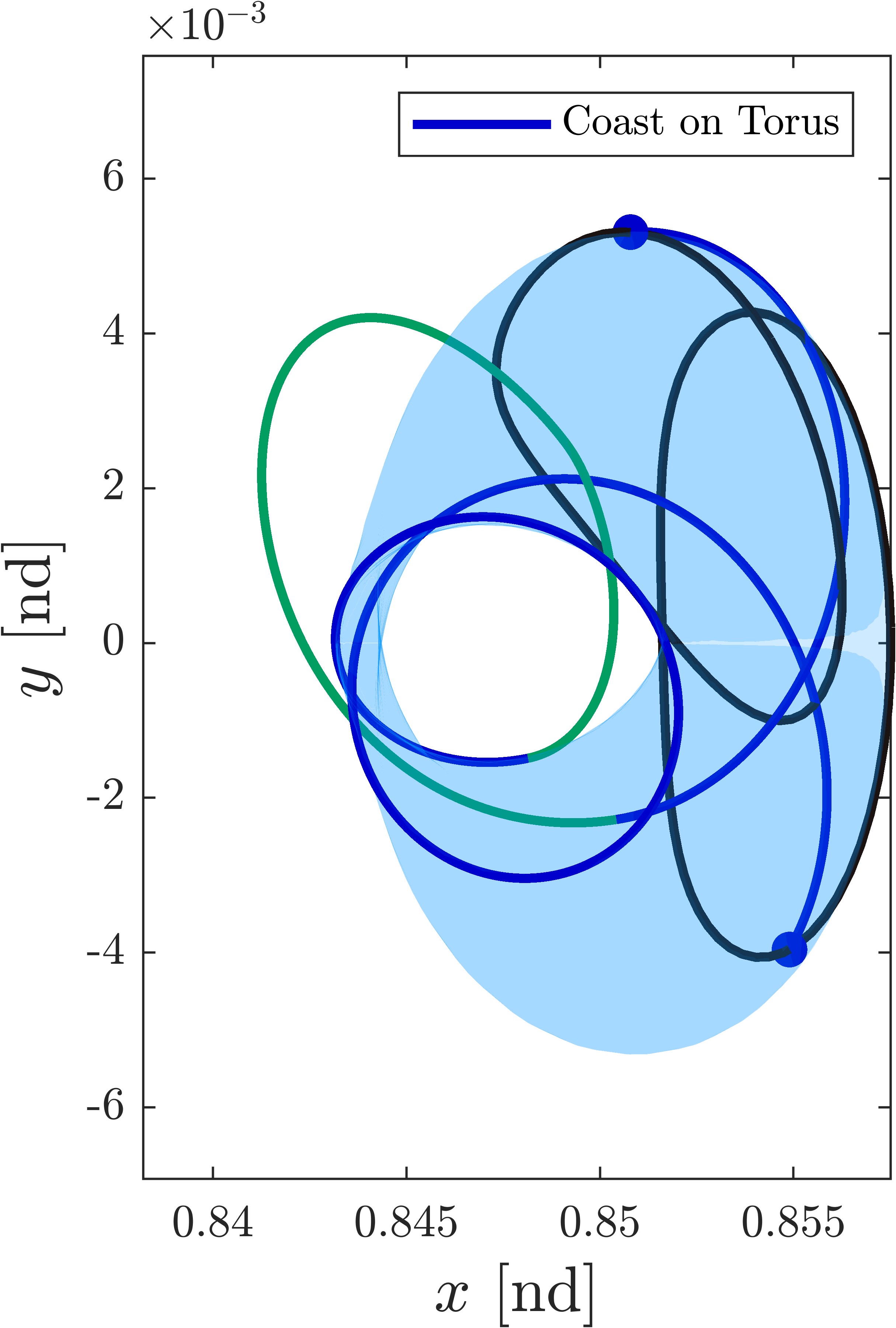}
         \caption{CMTP}
     \end{subfigure}
        \caption{Phase space solutions to a fixed boundary condition rephasing maneuver on an $L_1$ quasi-vertical torus in the ER3BP.}
        \label{fig:ER3BPphasespace}
\end{figure}

\begin{figure}[htbp!]
    \centering
    \includegraphics[width=0.4\linewidth]{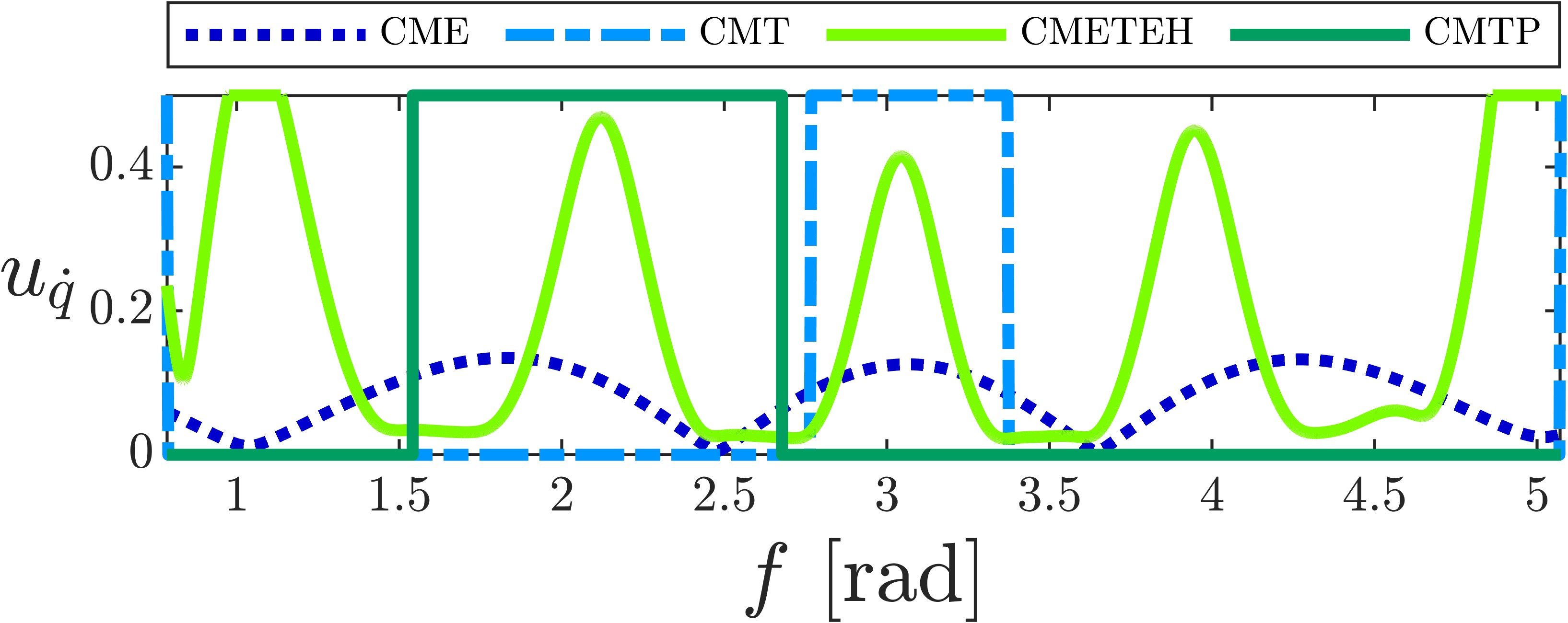}
    \caption{Control true anomaly histories for different solutions to the $L_1$ quasi-vertical torus rephasing problem.}
    \label{fig:ER3BPcontrol}
\end{figure}

Both torus agnostic phase space solutions, as well as those transitioned from the torus space are displayed in Fig. \ref{fig:ER3BPphasespace}. To better visualize the trajectories, axes are not scaled evenly, and hence thrust lines for each trajectory are omitted. 
The CME solution shows the most geometric variation in configuration space, while the CMT solution stays tighter to the torus surface. The CMT solution uses one finite duration thrust arc to connect trajectories on the torus' stable and unstable invariant manifolds. This can be seen in Fig. \ref{fig:ER3BPphasespace} as the thrust arc begins and ends well off the torus surface. The near-impulsive initial and final perturbing control inputs to ride these manifolds are shown in the control history plot of Fig. \ref{fig:ER3BPcontrol}.
The CMETEH solution was continued up to $\varepsilon=0.9998$. Neglecting the initial control impulse, the CMVT solution in the torus space revealed four thrust arcs. While each of these individual arcs can be successfully transitioned with time before the next, a better solution by measure of time spent on the torus and fuel consumption is computed by grouping all four coast arcs together. This is primarily due to the successively repeated thrust arc geometries for this particular torus and boundary value problem.

Performance metrics for these phase space solutions are presented in Table \ref{tab:level2er3bp}. The torus error measure of Eq. \ref{eq:cumerr} is modified to quantify the distance from available structures on the torus for fixed values $\theta_2=f$ (i.e invariant curves/surfaces) over the trajectory discretization. This is in contrast to the distance from the entire torus itself seen previously for tori in autonomous systems. The CME solution showed the largest cumulative torus error while the single patch CMTP solution showed the least. The CMT solution required the least fuel and naturally had a low torus error because of its excitation of local hyperbolic manifolds. The CMETEH solution required the most fuel in order to best track its reference trajectory. A near 2.5x increase in fuel resulted in a 2.5x decrease in cumulative torus error when compared to the CME solution with the same control structure. The CMTP solution traded a near 2x increase in fuel for a 1.5x decrease in torus error when compared to the CMT solution with the same control structure, but spent over three quarters of the transfer time on passively safe torus coast arcs. Thus, despite the loss in degree-of-freedom, the proposed bi-level optimal control framework is successful minimizing torus deviation for rephasing maneuvers at the cost of fuel expenditure. The CMTP solutions in particular for both examples highlight the utility of passive safety through time spent on the torus during maneuver.
\begin{table}[htbp!]
\caption{\label{tab:level2er3bp} rephasing performance metrics with $\varepsilon=0.9998$ used in the CMETEH solution.}
\centering
\begin{tabular}{llccc}
\hline
Torus Space & Phase Space & $\Delta V$ & $E$ & $f_{\text{on}}$ \\\hline
None & CME & 0.3439 & 0.0216 & 0\\
None & CMT& 0.3039 & 0.0055 & 0\\
\hline
CMVE & CMETEH & 0.8184 & 0.0088 & 0\\
CMVT & CMTP & 0.5679 & 0.0037 & 3.1518\\
\hline
\end{tabular}
\end{table}
\section{Minimum Time Recovery Maneuvers}\label{sec:recovery}

During long duration, low-thrust rephasing maneuvers, shifting mission priorities and the identification of new targets may require the abort and immediate return to the primary design structure such that nominal operations/observations may resume. Returning to a nearby center manifold structure (i.e. different periodic orbit or QPIT) is also possible, but ultimately requires more input and design from the satellite operator. Thus, return to the original torus is posed as the minimum time recovery problem in Eq. \ref{eq:recovprob} for autonomous systems, where the terminal state on the torus is left free.
\begin{align}
    \text{min}\quad &J = \int_{t_{0}}^{t_f}1\:\text{d}t,\quad t_f\,\,\text{free}\label{eq:recovprob}\\
    \text{s.t.}\quad &\dot{\boldsymbol{x}}=\boldsymbol{f}(\boldsymbol{x},\boldsymbol{\mu})+\boldsymbol{B}\boldsymbol{u}_{\dot{q}},\quad \left|\left|\boldsymbol{u}_{\dot{q}}\right|\right| \leq u_{\dot{q},\text{max}} \nonumber \\
    &\boldsymbol{x}(t_{0}) = \boldsymbol{x}_0\nonumber \\
    &\boldsymbol{x}(t_f) = \boldsymbol{\Gamma}(\boldsymbol{\theta}),\quad \boldsymbol{\theta}\,\,\text{free} \nonumber
\end{align}

This type of problem, where the desired terminal state is constrained to an orthogonally approximated manifold, is solved by Kelly \textit{et al}. for the low-thrust orbit transfer into a hyperbolic invariant manifold for capture in the CR3BP \cite{kelly2023orthogonal}. Here, it is applied to the free return to the original QPIT in the rephasing problem. The problem shares the same Hamiltonian, control expression, and costate dynamics as the CMTP problem in Sec. \ref{sec:bilevel}. However, free final time and torus angles require the following two transversality conditions to be enforced.
\begin{align}
    \mathcal{H}(t_f) = 0 \quad\quad &\\
    \boldsymbol{\lambda}_x^\text{T}(t_f)\left[\frac{\partial \boldsymbol{\Gamma}(\boldsymbol{\theta})}{\partial \boldsymbol{\theta}}\right] =\boldsymbol{0}&
\end{align}

\begin{figure}[htbp!]
     \centering
     \begin{subfigure}{0.4\textwidth}
         \centering
         \includegraphics[width=1\textwidth]{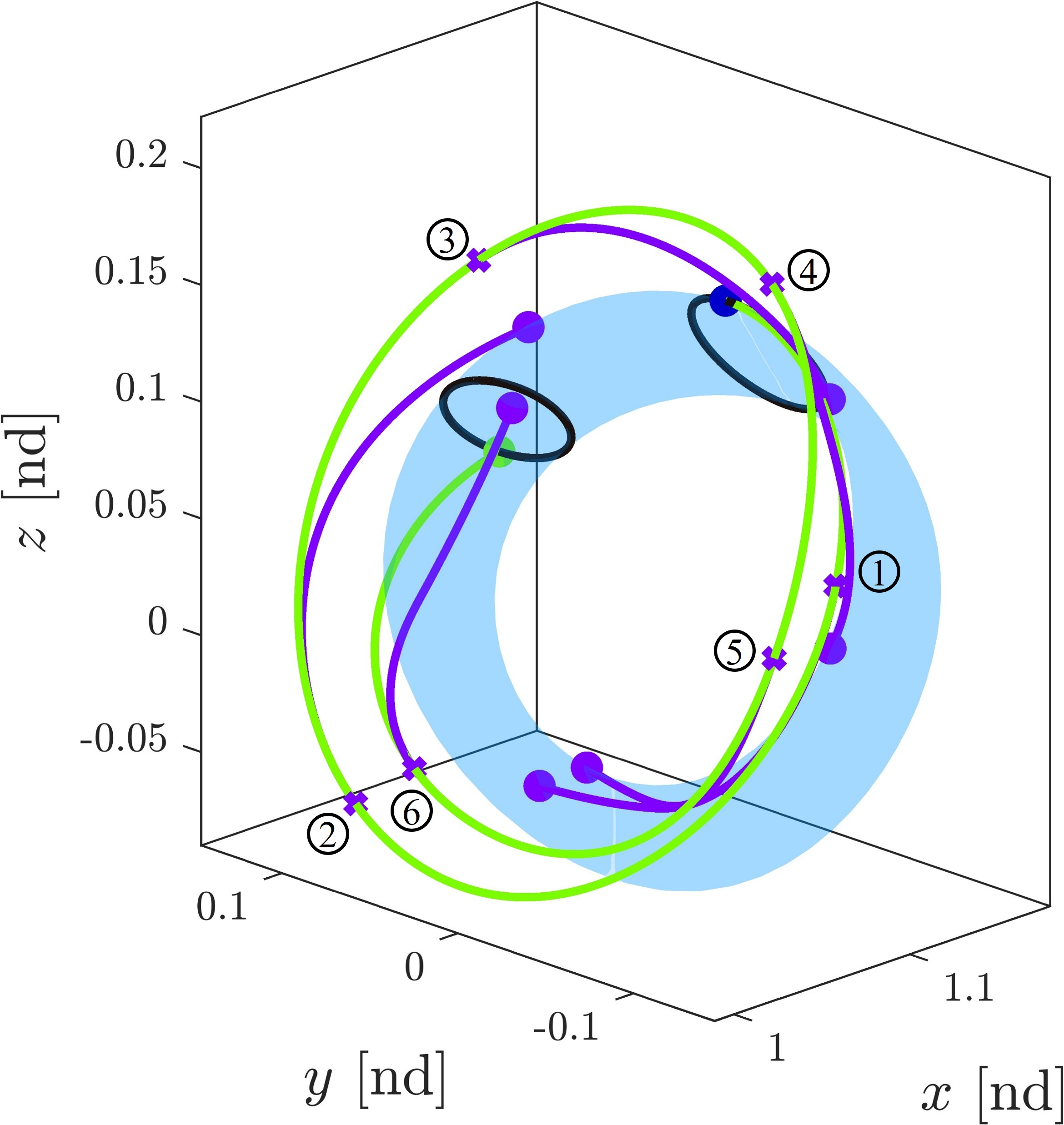}
         \caption{CME}
     \end{subfigure}\hspace{2em}
     \begin{subfigure}{0.4\textwidth}
         \centering
         \includegraphics[width=1\textwidth]{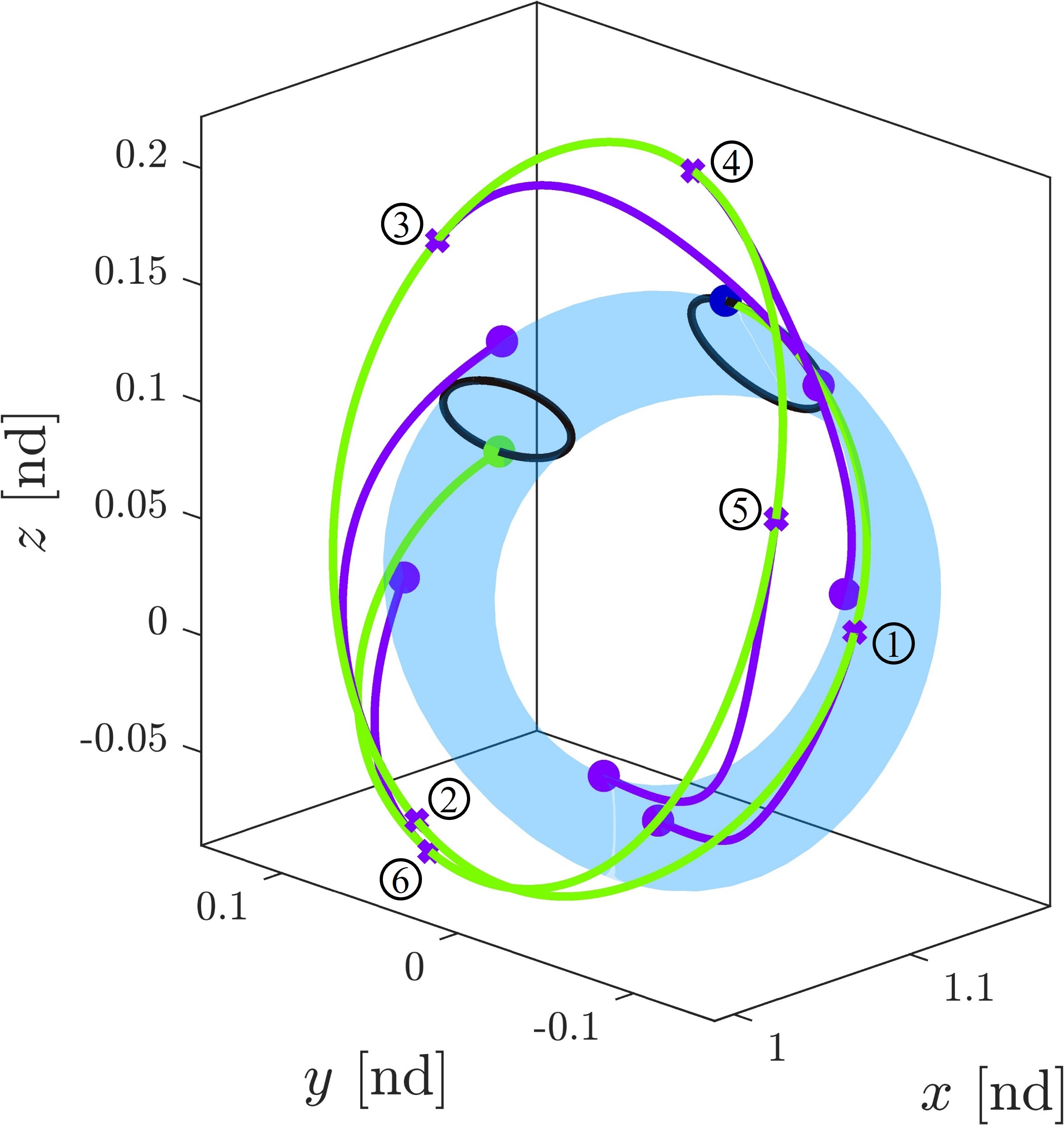}
         \caption{CMT}
     \end{subfigure}\\
     \begin{subfigure}{0.4\textwidth}
         \centering
         \includegraphics[width=1\textwidth]{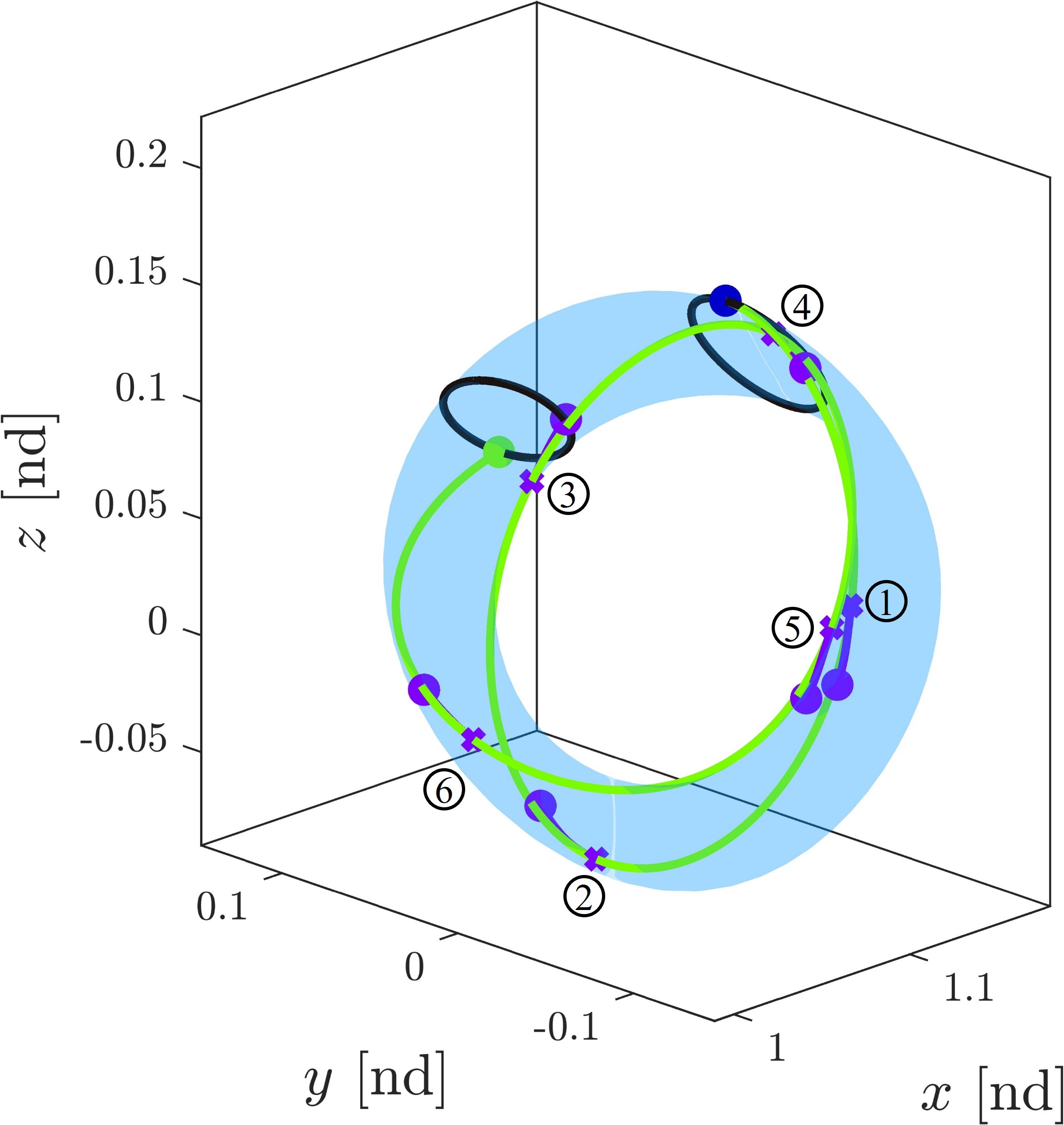}
         \caption{CMETEH}
     \end{subfigure}\hspace{2em}
     \begin{subfigure}{0.4\textwidth}
         \centering
         \includegraphics[width=1\textwidth]{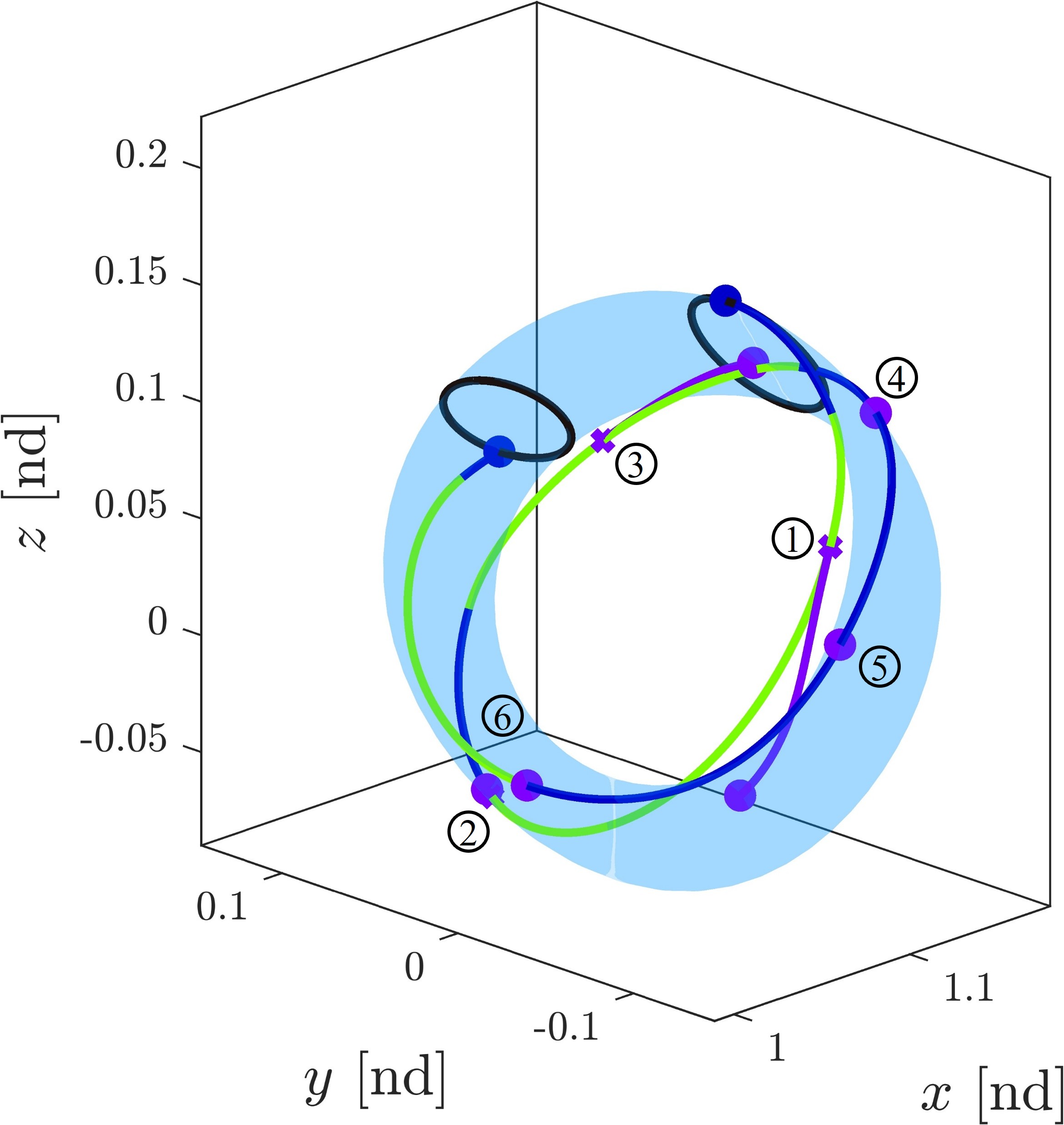}
         \caption{CMTP}
     \end{subfigure}
        \caption{Minimum time recovery trajectories for the fixed boundary condition rephasing case on a 2-dimensional QPIT in the CR3BP.}
        \label{fig:aborttraj}
\end{figure}

\begin{figure}[htbp!]
    \centering
    \includegraphics[width=0.4\linewidth]{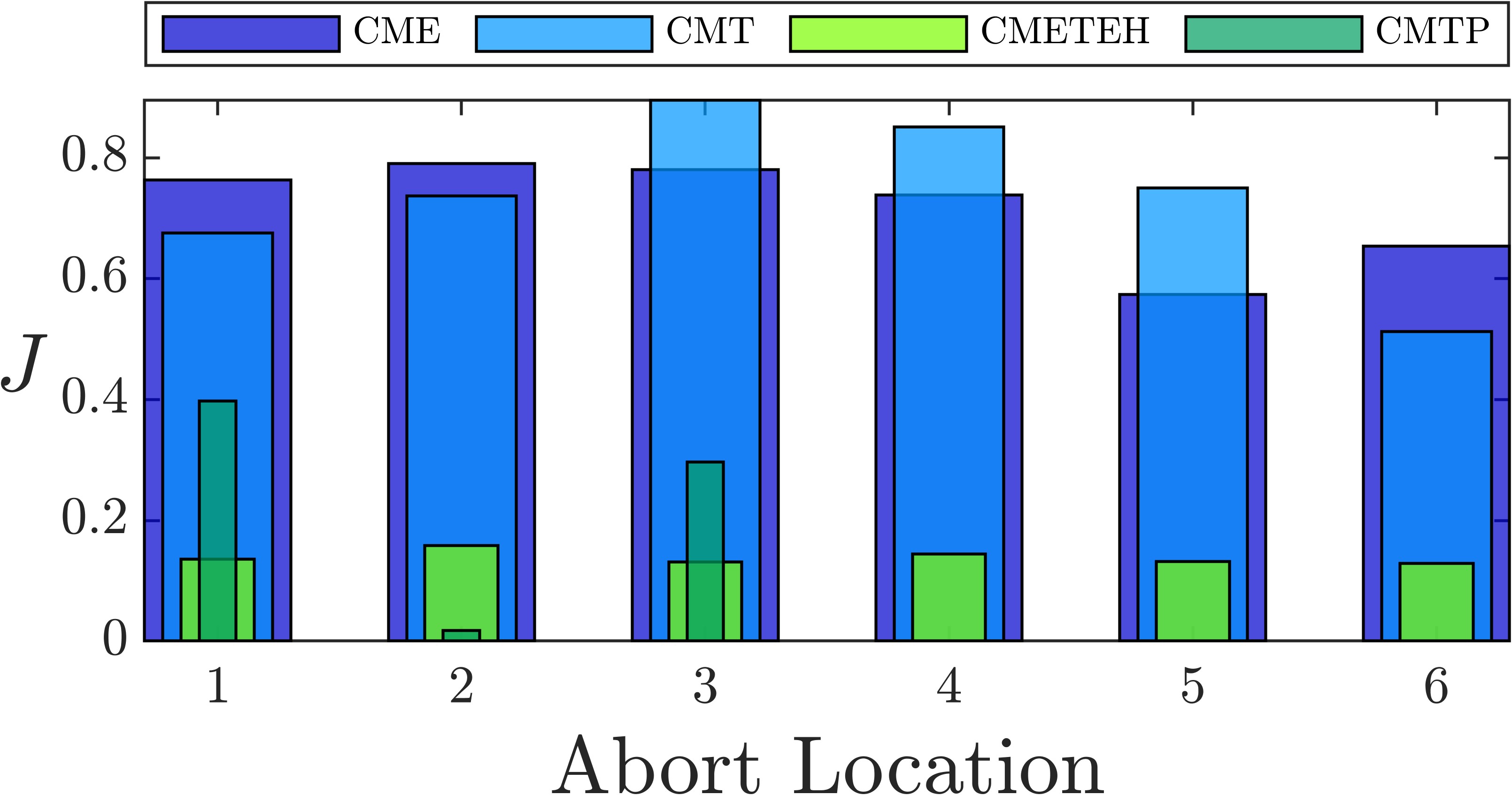}
    \caption{Recovery cost for each maneuver abort location by solution type.}
    \label{fig:abortJ}
\end{figure}

Intuitively, abort-to-recovery solutions generated from rephasing trajectories that minimize torus deviation will result in lower recovery times because they are already close to or on the torus at the abort time. To illustrate this concept, six equally spaced in time abort locations were selected across the rephasing trajectories on the quasi-halo torus example from Sec. \ref{sec:bilevel}. The resulting minimum time recovery trajectories are plotted in purple in Fig. \ref{fig:aborttraj} for each different solution to the rephasing problem.
The costs for each maneuver abort location by solution type are displayed in Fig. \ref{fig:abortJ}.
Note that $J=t_f\propto\Delta V$ for solutions of the minimum time variety because control input magnitude is set to a maximum at all times. From Figs. \ref{fig:aborttraj} and \ref{fig:abortJ} it is clear that the CMETEH and CMTP rephasing solutions result in a more agile system because the response time and hence required propellant needed to return to nominal operation is on average five times less than the CME and CMT torus agnostic solutions. Abort locations 4, 5, and 6 occur on torus coast arcs for the CMTP solution, hence no recovery maneuver is necessary. This analysis further demonstrates the utility in minimizing torus deviation during rephasing maneuvers on tori.

\section{Conclusion}\label{sec:conclusion}
This article introduced a methodology for designing rephasing maneuvers on quasi-periodic invariant tori in multi-body environments where deviation from the torus was minimized. Fourier approximations to the torus function and a modified torus invariance condition allowed this process to be split into two levels of optimization. First, reference trajectories were generated in the torus space by the minimization of mapped fictitious phase space control inputs. Second, these trajectories were transitioned to the phase space using physically available control through both a minimum tracking error homotopy and minimum time patches between torus space switching times. Results from 2-dimensional tori in the CR3BP and ER3BP of the proposed methodology compared to torus agnostic approaches show the method's success in torus conformity, as well as the tradeoff of increased fuel expenditure. This method is shown to generalize to $p$-dimensional tori in quasi-periodically forced dynamical systems with $m$ independent forcing frequencies, so long as $p>m$ and the torus can be approximated using a tractable amount of data points. This includes the simple case of the periodic orbit in an autonomous system ($p=1,\:m=0$). Lastly, the minimum time to torus recovery problem was posed and solved for rephasing maneuver abort locations along the paper's continued example. The results illustrate the disadvantage of traditional minimum fuel, torus agnostic solutions in responding to changing mission requirements and directives.

% The bi-level optimal control framework was formulated using an indirect optimal control approach because simple initial guess schemes coupled with multiple shooting techniques demonstrated reliable solution convergence over a wide range of problems and boundary conditions. The main practical drawback to this approach is the successive evaluation of the torus function within the numerical integration scheme. This resulted in longer computation times per solution iteration when compared to torus agnostic approaches, although the majority of cases converged within 30 seconds. To circumvent this issue, the CMVE and CMVT optimal control problems posed in this paper could be solved via direct transcription. Because the torus function only appears in the cost function for these torus space problems when the fundamental torus function is used, no function evaluation is required in the collocation of the torus dynamics, hence resulting in a reduction in computation time. Additional speed up is expected because analytical cost function gradients are readily available for use in a nonlinear program because of the Fourier parametrization of the torus function.

The trajectory design methodologies introduced herein are intended to enhance preliminary mission design and analysis tools for complex dynamical systems. Solutions produced from the bi-level optimal control framework in this paper must then be transitioned into a full ephemeris model to be used within a mission context. Towards this goal, there remains an option to the trajectory designer of whether to transition the reference trajectory from the first level of optimization, or the phase space trajectory from the second level of optimization. \rev1{Further analysis is required to understand the tradeoffs of this flexibility.}

\section*{Appendix}

\subsection{Two-Dimensional Torus Function Sensitivities} \label{sec:appA}
\vspace{-0.8cm}
\begin{align}
    \frac{\partial \boldsymbol{\Gamma}(\boldsymbol{\theta})}{\partial \theta_1} &= \boldsymbol{C}^\text{T}\left[\text{i}\cdot\text{diag}(\boldsymbol{k}_1)\cdot \text{e}^{\text{i}\cdot\boldsymbol{k}^\text{T}_1\theta_1}\otimes \text{e}^{\text{i}\cdot\boldsymbol{k}^\text{T}_2\theta_2}\right] \\
    \frac{\partial \boldsymbol{\Gamma}(\boldsymbol{\theta})}{\partial \theta_2} &= \boldsymbol{C}^\text{T}\left[\text{e}^{\text{i}\cdot\boldsymbol{k}^\text{T}_1\theta_1}\otimes \text{i}\cdot\text{diag}(\boldsymbol{k}_2)\cdot \text{e}^{\text{i}\cdot\boldsymbol{k}^\text{T}_2\theta_2}\right]\\
    \frac{\partial^2\boldsymbol{\Gamma}(\boldsymbol{\theta})}{\partial \theta_1^2} &= -\boldsymbol{C}^\text{T}\left[[\text{diag}(\boldsymbol{k}_1)]^2\cdot \text{e}^{\text{i}\cdot\boldsymbol{k}^\text{T}_1\theta_1}\otimes \text{e}^{\text{i}\cdot\boldsymbol{k}^\text{T}_2\theta_2}\right] \\
    \frac{\partial^2\boldsymbol{\Gamma}(\boldsymbol{\theta})}{\partial \theta_2^2} &= -\boldsymbol{C}^\text{T}\left[\text{e}^{\text{i}\cdot\boldsymbol{k}^\text{T}_1\theta_1}\otimes [\text{diag}(\boldsymbol{k}_2)]^2\cdot \text{e}^{\text{i}\cdot\boldsymbol{k}^\text{T}_2\theta_2}\right] \\
    \frac{\partial^2\boldsymbol{\Gamma}(\boldsymbol{\theta})}{\partial \theta_1\partial\theta_2} &= -\boldsymbol{C}^\text{T}\left[\text{diag}(\boldsymbol{k}_1)\cdot \text{e}^{\text{i}\cdot\boldsymbol{k}^\text{T}_1\theta_1}\otimes\text{diag}(\boldsymbol{k}_2)\cdot \text{e}^{\text{i}\cdot\boldsymbol{k}^\text{T}_2\theta_2}\right]
\end{align}
% \begin{equation}
%     \left[\frac{\partial \boldsymbol{\Gamma}(\boldsymbol{\theta})}{\partial \boldsymbol{\theta}}\right] = 
%         \left[\frac{\partial \boldsymbol{\Gamma}(\boldsymbol{\theta})}{\partial \theta_1} \quad \frac{\partial \boldsymbol{\Gamma}(\boldsymbol{\theta})}{\partial \theta_2}\right]
% \end{equation}
% \begin{align}
%     \frac{\partial^2\boldsymbol{\Gamma}(\boldsymbol{\theta})}{\partial \theta_1^2} &= -\boldsymbol{C}^\text{T}\left[[\text{diag}(\boldsymbol{k}_1)]^2\cdot \text{e}^{\text{i}\boldsymbol{k}^\text{T}_1\theta_1}\otimes \text{e}^{\text{i}\boldsymbol{k}^\text{T}_2\theta_2}\right] \\
%     \frac{\partial^2\boldsymbol{\Gamma}(\boldsymbol{\theta})}{\partial \theta_2^2} &= -\boldsymbol{C}^\text{T}\left[\text{e}^{\text{i}\boldsymbol{k}^\text{T}_1\theta_1}\otimes [\text{diag}(\boldsymbol{k}_2)]^2\cdot \text{e}^{\text{i}\boldsymbol{k}^\text{T}_2\theta_2}\right] \\
%     \frac{\partial^2\boldsymbol{\Gamma}(\boldsymbol{\theta})}{\partial \theta_1\partial\theta_2} &= -\boldsymbol{C}^\text{T}\left[\text{diag}(\boldsymbol{k}_1)\cdot \text{e}^{\text{i}\boldsymbol{k}^\text{T}_1\theta_1}\otimes\text{diag}(\boldsymbol{k}_2)\cdot \text{e}^{\text{i}\boldsymbol{k}^\text{T}_2\theta_2}\right]
% \end{align}

\subsection{Indirect Optimal Control Theory}\label{sec:appB}

Consider the optimal control problem of a continuous time dynamical system where the initial state and time are assumed fixed without loss of generality. Written in Bolza form, this problem is stated as
\begin{align}
    \text{min}\quad &J = \phi(\boldsymbol{x}(t_f),t_f) + \int_{t_0}^{t_f}\mathcal{L}(\boldsymbol{x}(t),\boldsymbol{u}(t),t)\text{d}t\label{eq:bolza}\\
    \text{s.t.}\quad &\dot{\boldsymbol{x}} = \boldsymbol{f}(\boldsymbol{x},\boldsymbol{u},t) \nonumber \\
    &\boldsymbol{x}(t_{0}) = \boldsymbol{x}_0\nonumber \\
    &\boldsymbol{\psi}(\boldsymbol{x}(t_f),t_f) = \boldsymbol{0} \nonumber
\end{align}
where $\phi$ is a terminal penalty, and $\mathcal{L}$ is a path penalty, also called the Lagrangian. A solution to Eq. \ref{eq:bolza} then determines control $\boldsymbol{u}(t)$ that minimizes $J$ while also satisfying the set of terminal constraints $\boldsymbol{\psi}$.

The indirect approach towards the solution of this optimal control problem sees its conversion to a boundary value problem first through the introduction of dynamic Lagrange multipliers $\boldsymbol{\lambda}$ in the augmented cost function
\begin{equation}
    J_a = \phi(\boldsymbol{x}(t_f),t_f) + \boldsymbol{\nu}^\text{T}\boldsymbol{\psi}(\boldsymbol{x}(t_f),t_f)+\int_{t_0}^{t_f}\left[\mathcal{H}(\boldsymbol{x}(t),\boldsymbol{u}(t),t)+\boldsymbol{\lambda}^\text{T}(t)\dot{\boldsymbol{x}}(t)\right]\text{d}t
\end{equation}
where $\boldsymbol{\nu}$ are a set of terminal constraint multipliers and $\mathcal{H}$ is the problem's Hamiltonian.
\begin{equation}
    \mathcal{H}(\boldsymbol{x}(t),\boldsymbol{u}(t),t) = \mathcal{L}(\boldsymbol{x}(t),\boldsymbol{u}(t),t) + \boldsymbol{\lambda}^\text{T}(t)\left(\boldsymbol{f}(\boldsymbol{x}(t),\boldsymbol{u}(t),t)\right)
\end{equation}
The total differential of $J$ is then taken using techniques from variational calculus as a function of differentials in $\boldsymbol{x}$, $\boldsymbol{u}$, $\boldsymbol{\lambda}$, $\boldsymbol{\nu}$, and $t$.
\begin{align}
    \text{d}J_a &= (\phi_x+\boldsymbol{\psi}_x^\text{T}\boldsymbol{\nu}-\boldsymbol{\lambda})^\text{T}\text{d}x|_{t_f} + (\phi_t + \boldsymbol{\psi}_t^\text{T}\boldsymbol{\nu}+\mathcal{H})\text{d}t|_{t_f} \nonumber\\
    &+\boldsymbol{\psi}^\text{T}|_{t_f}\text{d}\boldsymbol{\nu} - \mathcal{H}\text{d}t|_{t_0} + \int_{t_0}^{t_f}\left[(\mathcal{H}_x+\dot{\boldsymbol{\lambda}})^\text{T}\delta \boldsymbol{x} + \mathcal{H}_u^\text{T}\delta \boldsymbol{u} + (\mathcal{H}_\lambda-\dot{\boldsymbol{x}})^\text{T}\delta \boldsymbol{\lambda} \right]\text{d}t\label{eq:finalAugCF}
\end{align}
Subscripts represent partial derivatives with respect to that variable. The constrained local minimum of $J$ is found at the unconstrained local minimum of $J_a$, i.e. $\text{d}J_a=0$. This is accomplished by setting all arbitrary increments in Eq. \ref{eq:finalAugCF} to zero, leading to the following set of first-order necessary conditions.
\begin{align*}
    \text{State Equation}&:\quad \dot{\boldsymbol{x}}=\frac{\partial \mathcal{H}}{\partial \boldsymbol{\lambda}}=\boldsymbol{f}(\boldsymbol{x},\boldsymbol{u},t) \\
    \text{Costate Equation}&:\quad\dot{\boldsymbol{\lambda}}=-\left[\frac{\partial \mathcal{H}}{\partial \boldsymbol{x}}\right]^{\text{T}}\\
    \text{Stationary Equation}&:\quad\boldsymbol{0}=\frac{\partial \mathcal{H}}{\partial \boldsymbol{u}}\\
    \text{Boundary Conditions}&:\quad\boldsymbol{0}=\boldsymbol{\psi}\\
    &\quad\:\:\:\,\boldsymbol{0}=(\phi_x+\boldsymbol{\psi}_x^\text{T}\boldsymbol{\nu}-\boldsymbol{\lambda})|_{t_f}\\
    &\quad\:\:\:\,\boldsymbol{0}=(\phi_t + \boldsymbol{\psi}_t^\text{T}\boldsymbol{\nu}+\mathcal{H})|_{t_f}\\
    &\quad\:\:\:\,0 = \mathcal{H}|_{t_0}
\end{align*}
The state, costate, and stationary equations are applied to all indirect optimal control problems, regardless of boundary conditions. Boundary conditions other than the original terminal boundary constraint $\boldsymbol{\psi}$ are known as \textit{transversality} conditions. These \textit{transversality} conditions are only applied to a problem if their associated increment is allowed. For example, if an optimal control problem has a fixed initial time, then $\text{d}t|_{t_0}=0$ and the condition $\mathcal{H}|_{t_0} = 0$ is ignored.

The optimal control problem has now been transformed into a boundary value problem that can be solved via differential corrections procedures such as forward shooting where initial costate values $\boldsymbol{\lambda}(t_0)$ and terminal constraint multipliers $\boldsymbol{\nu}$ must be found to drive the system to the terminal constraint surface $\boldsymbol{\psi}$ at the final time. For a well-defined and correct problem formulation, this boundary value problem is square with a \textit{locally} unique solution.

\section*{Funding Sources}
AFOSR Space University Research Initiative project "Multi-Phenomenological, Autonomous, and Understandable SDA and XDA Decision Support" (Grant No: FA9550-23-1-0726), as well as the Hagler Institute of Advanced Study at Texas A\&M University and the Texas A\&M University Department of Aerospace Engineering National EXcellence in Aerospace Sciences Fellowship (NEXAS) are all acknowledged for their support of the research herein.

\section*{Acknowledgments}
The authors would like to thank Dr. Srinivas Vadali of Texas A\&M University for helpful discussions on optimal control theory. Dr. James McElreath and Dr. David van Wijk are acknowledged for their feedback on the construction of this document.

\bibliography{11_references}

\end{document}